
\catcode`\^^J=10
\magnification=\magstep1

\newcount\sectionpageno
\def\initsectionpageno#1 {\sectionpageno=#1\pageno=#1}

\output={\plainoutput}

 1


\def\qed{\hfill\hbox{\vrule width 4pt height 6pt depth 1.5 pt}}

\def\today{\ifcase\month\or January \or February \or March
\or April  \or
May \or June \or July \or August \or September \or October \or
November \or December \fi\space
\oldnos{\number\day}, \oldnos{\number\year}}

\hyphenation{dif-fer-en-tial di-men-sion-al
Helm-holtz Cum-mings  Czech-o-sla-vak }

\newif\ifproofmode    \proofmodefalse
\newif\ifbibmakemode  \bibmakemodefalse
\newif\ifbibcitemode  \bibcitemodefalse
\newif\ifmultisection \multisectionfalse

for

\def\ikedocument{\ifbibcitemode\input bibfilen.tex\fi
\ifmultisection\immediate\openout2=\jobname.eqn\fi  }

\def\ikeenddocument{\ifbibmakemode
\leftappenditem{zzzzz99z}\to\biblist
\openout1=\jobname.bib
{\edef\\##1{\write1{##1}}\biblist}\closeout1\fi
\ifmultisection\closeout2\fi }


\def\strutdepth{\dp\strutbox}
\def\margintagleft#1{\strut\vadjust{\kern-\strutdepth
{\vtop to \strutdepth{\baselineskip\strutdepth\vss\llap{\sevenrm
#1\quad}\null}}}}

\newcount\equationnumber
\newcount\equationnumberlet
\newcount\sectionnumber
\newcount\statementnumber

\def\sectionno#1 {\sectionnumber=#1 \equationnumber=1
\statementnumber=0}

\def\equationtag#1#2{\ifproofmode\margintagleft{#1}\fi
\expandafter\ifx\csname
s\expandafter\romannumeral\the\sectionnumber
eq\romannumeral#1\endcsname\relax\else
\message{Eqtag \the\sectionnumber.#1.#2 already exists} \fi
\ifmultisection
\immediate\write2{
\noexpand\expandafter\def
\noexpand\csname
s\expandafter\romannumeral\the\sectionnumber
eq\romannumeral#1 \noexpand\endcsname{\the\equationnumber}}\fi
\eqno(\the\sectionnumber.\the\equationnumber#2)\expandafter\xdef\csname
s\expandafter\romannumeral\the\sectionnumber
eq\romannumeral#1\endcsname{\the\equationnumber}
\global\equationnumberlet=\equationnumber
\global\advance\equationnumber by 1}
\def\aligneqtag#1#2{\ifproofmode\margintagleft{#1}\fi
\expandafter\ifx\csname
s\expandafter\romannumeral\the\sectionnumber
eq\romannumeral#1\endcsname\relax\else
\message{Eqtag \the\sectionnumber.#1.#2 already exists} \fi
\ifmultisection
\immediate\write2{
\noexpand\expandafter\def
\noexpand\csname
s\expandafter\romannumeral\the\sectionnumber
eq\romannumeral#1 \noexpand\endcsname{\the\equationnumber}}\fi
&{(\the\sectionnumber.\the\equationnumber#2)}\cr
\noalign{\expandafter\xdef\csname
s\expandafter\romannumeral\the\sectionnumber
eq\romannumeral#1\endcsname{\the\equationnumber}
\global\equationnumberlet=\equationnumber
\global\advance\equationnumber by 1}}

\def\aleqtag#1 {\aligneqtag{#1}{\null}}
\def\aligneqtagletter#1 {&(\the\sectionnumber.\the\equationnumberlet#1)}

\def\eqtag#1 {\equationtag{#1}{\null}}
\def\eqtagletter#1 {\eqno(\the\sectionnumber.\the\equationnumberlet#1)}

\def\equationlabel#1#2#3{\if#1X XXX\else
{\xdef\cmmm{\csname
 s\romannumeral#1eq\romannumeral#2\endcsname}\expandafter\ifx
 \csname s\romannumeral#1eq\romannumeral#2\endcsname
\relax\message{Equation #1.#2.#3 not defined ...}\fi(#1.\cmmm#3)}\fi}

\def\eq#1{\equationlabel{\the\sectionnumber}{#1}{}}


\def\statementtag#1 {
\expandafter\ifx\csname
s\expandafter\romannumeral\the\sectionnumber
stat\romannumeral#1\endcsname\relax\else
\message{Statementtag \the\sectionnumber.#1 already exists} \fi
\ifproofmode
\margintagleft{#1}\fi\global\advance\statementnumber by 1
\expandafter\xdef\csname
s\expandafter\romannumeral\the\sectionnumber
 stat\romannumeral#1\endcsname{\the\statementnumber}\ifmultisection
\immediate\write2{\noexpand\expandafter\def\noexpand\csname
s\expandafter\romannumeral\the\sectionnumber
stat\romannumeral#1
\noexpand\endcsname{\the\statementnumber}}
\fi\the\sectionnumber.\the\statementnumber}

\def\statementlabel#1#2{\if#1X XXX \else\xdef\cmmm{\csname
s\romannumeral#1stat\romannumeral#2\endcsname}\expandafter\ifx
 \csname
s\romannumeral#1stat\romannumeral#2\endcsname\relax\message{Stateme
nt
 #1.#2 not defined ...}\fi#1.\cmmm\fi}

\def\statement#1{\statementlabel{\the\sectionnumber}{#1}}




\newcount\addcitation
\let\biblist=\empty

\toksdef\ta=0  \toksdef\tb=2
\long\def\leftappenditem#1\to#2{\ta={\\{#1}}\tb=\expandafter{#2}
\edef#2{\the\ta\the\tb}}

\def\bibyear#1#2#3#4#5#6#7#8 {\edef
\romanyear{#1#2#3#4#5\romannumeral#6#7 #8}}

\font\slr=cmsl10
\def\cite#1{\ifbibmakemode\addcitation=1
\def\\##1{\def\citea{#1}\def\citeb{##1}\ifx\citea\citeb\addcitation=0
\fi}\biblist\ifnum\addcitation=1\leftappenditem#1\to\biblist\fi[{\bf00}]
\else
\ifbibcitemode
\bibyear#1
\expandafter\ifx\csname\romanyear
bibnumber\endcsname\relax\message{#1 not in
bibfile}\fi
[{\slr \csname\romanyear bibnumber\endcsname}\hbox{\kern
1pt}]\else[{\bf00}]\fi\fi}

\def\Ex#1 {\par\medskip\noindent{\smc Example#1.} }
\def\endEx{\qed\par\medskip}



\def\hook{\mathbin{\raise2.5pt\hbox{\hbox{{\vbox{\hrule height.4pt
width6pt depth0pt}}}\vrule height3pt width.4pt depth0pt}\,}}

\def\mypartial{\partial{\hbox{\kern 1 pt}}}

\def\mygamma{{\hbox{\raise 1pt\hbox{$\gamma$}}}}

\def\Ev{\CE\hbox{\kern - 1pt}v}

  
\def\CE{\Cal E}

\def\body                       
  {\beginparmode}               

\def\refto#1{[#1]}           
\def\subhead#1{                 
  \vskip 0.25truein             
  \noindent{{\it {#1}} \par}
   \nobreak\vskip 0.15truein\nobreak}
\def\beginparmode{\endmode
  \begingroup \def\endmode{\par\endgroup}}
\let\endmode=\par
{\obeylines\gdef\
{}}

\def\references                 
  {\subhead{\bf References}         
commas)
   \beginparmode
   \frenchspacing \parindent=0pt \leftskip=1truecm
   \everypar{\hangindent=\parindent}}
\def\endreferences{\body}

\gdef\refis#1{\indent\hbox to 0pt{\hss#1.~}}    

\gdef\journal#1, #2, #3, 1#4#5#6{               
set
    {\sl #1~}{\bf #2}, #3 (1#4#5#6)}           

\def\refstylenp{                
  \gdef\refto##1{ [##1]}                                
  \gdef\refis##1{\indent\hbox to 0pt{\hss##1)~}}        
  \gdef\journal##1, ##2, ##3, ##4 {                     
     {\sl ##1~}{\bf ##2~}(##3) ##4 }}

\def\refstyleprnp{              
  \gdef\refto##1{ [##1]}                                
  \gdef\refis##1{\indent\hbox to 0pt{\hss##1)~}}        
  \gdef\journal##1, ##2, ##3, 1##4##5##6{               
    {\sl ##1~}{\bf ##2~}(1##4##5##6) ##3}}

\def\prd{\journal Phys. Rev. D, }

\def\prl{\journal Phys. Rev. Lett., }

\def\jmp{\journal J. Math. Phys., }

\def\np{\journal Nucl. Phys., }

\def\ann{\journal Ann. Phys., }

\def\grg{\journal Gen. Rel. Grav., }

\def\ref#1{Ref. #1}                     
\def\Ref#1{Ref. #1}                     

\catcode`@=11
\newcount\r@fcount \r@fcount=0
\newcount\r@fcurr
\immediate\newwrite\reffile
\newif\ifr@ffile\r@ffilefalse
\def\w@rnwrite#1{\ifr@ffile\immediate\write\reffile{#1}\fi\message{#1}}

\def\writer@f#1>>{}
\def\referencefile{
  \r@ffiletrue\immediate\openout\reffile=\jobname.ref%
  \def\writer@f##1>>{\ifr@ffile\immediate\write\reffile%
    {\noexpand\refis{##1} = \csname r@fnum##1\endcsname = %
     \expandafter\expandafter\expandafter\strip@t\expandafter%
     \meaning\csname r@ftext\csname
r@fnum##1\endcsname\endcsname}\fi}%
  \def\strip@t##1>>{}}

\def\citeall#1{\xdef#1##1{#1{\noexpand\cite{##1}}}}
\def\cite#1{\each@rg\citer@nge{#1}}	
separated by

\def\each@rg#1#2{{\let\thecsname=#1\expandafter\first@rg#2,\end,}}
\def\first@rg#1,{\thecsname{#1}\apply@rg}	
purpose
\def\apply@rg#1,{\ifx\end#1\let\next=\relax
macro.
\else,\thecsname{#1}\let\next=\apply@rg\fi\next}
commas

\def\citer@nge#1{\citedor@nge#1-\end-}	
and N numbers)
\def\citer@ngeat#1\end-{#1}
\def\citedor@nge#1-#2-{\ifx\end#2\r@featspace#1 
  \else\citel@@p{#1}{#2}\citer@ngeat\fi}	
\def\citel@@p#1#2{\ifnum#1>#2{\errmessage{Reference range #1-
#2\space is bad.}%
    \errhelp{If you cite a series of references by the notation M-N, then M
and
    N must be integers, and N must be greater than or equal to M.}}\else%
 {\count0=#1\count1=#2\advance\count1
by1\relax\expandafter\r@fcite\the\count0,
  \loop\advance\count0 by1\relax
    \ifnum\count0<\count1,\expandafter\r@fcite\the\count0,%
  \repeat}\fi}

\def\r@featspace#1#2 {\r@fcite#1#2,}	
end of arg
\def\r@fcite#1,{\ifuncit@d{#1}
    \newr@f{#1}%
    \expandafter\gdef\csname r@ftext\number\r@fcount\endcsname%
                     {\message{Reference #1 to be supplied.}%
                      \writer@f#1>>#1 to be supplied.\par}%
 \fi%
 \csname r@fnum#1\endcsname}
\def\ifuncit@d#1{\expandafter\ifx\csname r@fnum#1\endcsname\relax}%
\def\newr@f#1{\global\advance\r@fcount by1%
    \expandafter\xdef\csname r@fnum#1\endcsname{\number\r@fcount}}

\let\r@fis=\refis			
\def\refis#1#2#3\par{\ifuncit@d{#1}
blank
   \newr@f{#1}%
   \w@rnwrite{Reference #1=\number\r@fcount\space is not cited up to
now.}\fi%
  \expandafter\gdef\csname r@ftext\csname
r@fnum#1\endcsname\endcsname%
  {\writer@f#1>>#2#3\par}}

\def\ignoreuncited{
   \def\refis##1##2##3\par{\ifuncit@d{##1}%
    \else\expandafter\gdef\csname r@ftext\csname
r@fnum##1\endcsname\endcsname%
     {\writer@f##1>>##2##3\par}\fi}}

\def\r@ferr{\endreferences\errmessage{I was expecting to see
\noexpand\endreferences before now;  I have inserted it here.}}
\let\r@ferences=\references
\def\references{\r@ferences\def\endmode{\r@ferr\par\endgroup}}

\let\endr@ferences=\endreferences
\def\endreferences{\r@fcurr=0
redefine
  {\loop\ifnum\r@fcurr<\r@fcount
produce text
    \advance\r@fcurr by
1\relax\expandafter\r@fis\expandafter{\number\r@fcurr}%
    \csname r@ftext\number\r@fcurr\endcsname%
  \repeat}\gdef\r@ferr{}\endr@ferences}


\let\r@fend=\endpaper\gdef\endpaper{\ifr@ffile
\immediate\write16{Cross References written on
[]\jobname.REF.}\fi\r@fend}

\catcode`@=12

\citeall\refto		
\citeall\ref		%
\citeall\Ref		%
%
%
%

%
%
{\nopagenumbers
\line{}

\vskip 1 true in
\centerline{\bf CLASSIFICATION OF GENERALIZED SYMMETRIES }
\vskip 9pt
\centerline{\bf FOR THE }
\vskip 9pt
\centerline{\bf  VACUUM EINSTEIN EQUATIONS}

\vskip 45pt

\hbox{\hfil\vbox{\hsize 150pt
\centerline{Ian M. Anderson}
\centerline{Department of Mathematics}
\centerline{Utah State University}
\centerline{Logan, UT 84322--3900}
\centerline{USA} }
\raise 20 pt \vbox{\hsize 50pt \centerline{and}}
\vbox{\hsize 150pt
\centerline{Charles G. Torre}
\centerline{Department of Physics}
\centerline{Utah State University}
\centerline{Logan, UT 84322--4415 }
\centerline{USA} }
\hfill}
\vskip 1 true in

\noindent{\bf Abstract:}

A generalized symmetry of a system of differential equations is an
infinitesimal transformation depending locally upon the fields and their
derivatives which carries solutions to solutions.  We classify all generalized
symmetries of the vacuum Einstein equations in
four spacetime dimensions.
To begin, we analyze symmetries that can be built from the metric,
curvature,
and covariant derivatives of the curvature to any order;  these are called
natural symmetries and are globally defined on any spacetime manifold.
We next classify first-order generalized symmetries, that is, symmetries
that depend on the metric and its first
derivatives.  Finally, using
results from the classification of natural symmetries, we reduce the
classification of all higher-order generalized symmetries to the first-order
case.  In each case we find
that the generalized symmetries are infinitesimal generalized
diffeomorphisms and constant metric scalings.  There are no non-trivial
conservation laws associated with these symmetries.
A novel feature of our analysis is the use of a fundamental set of spinorial
coordinates on the infinite jet space of Ricci-flat metrics,
which are derived from Penrose's ``exact set of fields'' for the vacuum
equations.
\vfill\eject}
\pageno1
%
\multisectiontrue
\ikedocument
\sectionno1

\noindent{\bf 1. Introduction}

Symmetry plays an important role throughout
theoretical physics and one of central importance in field theory
\refto{Elliot1979}, \refto{Gourdin1969}.  Indeed, in the
construction of a field theory physical considerations usually demand that
the field equations (or the Lagrangian) possess certain
symmetries.  These symmetries include Poincar\' e symmetry, gauge
symmetry, diffeomorphism symmetry, various discrete symmetries, and a
host of
specialized symmetries needed to ensure the conservation of appropriate
quantum numbers.
Symmetries also play an important role in the mathematical analysis of
differential equations \refto{Olver1993}, \refto{Bluman1989}.
Originating with the work of Lie, symmetry group
methods and their recent generalizations have proved useful in
understanding
conservation laws, in constructing exact solutions, and in establishing
complete integrability of certain systems of differential equations.

The symmetries encountered in field theory are usually of the type
commonly referred to as point symmetries.  A point symmetry of a system
of differential equations is a 1-parameter group of transformations of the
underlying space of independent and dependent variables that carries
any solution of the equations to another solution.  If a point symmetry
preserves
an underlying Lagrangian for the system of equations, then there is a
corresponding conservation law.  However, not all conservation laws stem
from point symmetries.  To account for all conservation laws in
Lagrangian field theory one must
enlarge the notion of symmetry to include generalized symmetries
\refto{Noether1918}.  A
{\it generalized symmetry} is an infinitesimal transformation, constructed
locally from the independent variables, the dependent variables, and the
{\it derivatives} of the dependent variables, that carries solutions of the
differential equations to nearby solutions.  The importance of generalized
symmetries is underscored by their role in completely integrable
systems of non-linear differential equations.  In particular, when
a system of differential equations is integrable, it invariably admits
``hidden'' generalized
symmetries  \refto{Olver1993}, \refto{Fokas1987},
\refto{Mikhailov1991}.

In recent years considerable attention has been devoted to
applications of symmetry group methods to a variety of non-linear
partial
differential equations, but relatively few complete results have been
obtained for
the Einstein equations.  It is, of course, natural to inquire whether or not
the Einstein equations admit any hidden generalized symmetries, but the
apparent
complexity of the ensuing analysis has, to date, precluded substantive
progress.  The existence of
hidden symmetries of the Einstein equations would lead to solution
generating--classification techniques, and perhaps even information about
the general
solution to the Einstein equations.  There are hints that such symmetries
may exist. The two Killing field reduction of the Einstein equations leads to
an
integrable system of partial differential equations \refto{Belinsky1979},
\refto{Hauser1981}, \refto{Husain1994}; the self-dual Einstein equations
exhibit an infinite number of
symmetries and can be integrated using twistor methods
\refto{Husain1994}, \refto{Penrose1976},
\refto{Winternitz1989}, \refto{Grant1993} .  A complete
generalized symmetry analysis provides a systematic and rigorous way to
unravel some aspects of the integrable behavior of the gravitational field
equations.  In particular, such an analysis indicates whether the rich
structure of special reductions of the Einstein equations extends to the full
theory.

An equally important consequence of a generalized symmetry analysis
stems from the fact that the existence of generalized symmetries of the
Einstein
equations is a necessary condition for the existence of local differential
conservation laws for
the gravitational field.  If such conservation laws could be
found, they would
lead to observables for the gravitational field \refto{CGT1993b}.  It has
long been an
open problem in relativity theory to exhibit such observables, and the lack
thereof currently hampers progress in canonical quantization of
general relativity \refto{Smolin1990}.

Recently, Gurses \refto{Gurses1993} proposed
infinite-dimensional families of generalized symmetries for the vacuum
Einstein equations.  Subsequent investigations showed that a subset of the
proposed symmetry transformations were in fact infinitesimal
diffeomorphism symmetries \refto{Hauser1993a}.  The remaining
transformations proposed in \refto{Gurses1993} fail to be symmetries in
the sense that the transformations are not infinitesimal maps from {\it any}
solution of the vacuum equations to another solution \refto{Hauser1993b},
\refto{Capovilla1994}.

In this paper we will give a {\it complete}
classification of all arbitrary-order generalized symmetries for the vacuum
Einstein equations in four spacetime
dimensions.  We shall show that the only generalized symmetries admitted
by the
vacuum Einstein equations consist of the diffeomorphism symmetry that is
inherent in the Einstein equations and a trivial scaling symmetry.
More precisely, we will prove the following theorem.

\proclaim Theorem.
Let $$h_{ab}=h_{ab}(x^i,g_{ij},g_{ij,h_1},\ldots,g_{ij,h_1\cdots h_k})$$
be the components of a $k^{th}$-order generalized symmetry of the
vacuum Einstein equations $R_{ij}=0$ in four spacetime dimensions.
Then there is a constant $c$ and a generalized vector field
$$X^i=X^i(x^i,g_{ij},g_{ij,h_1},\ldots,g_{ij,h_1\cdots h_{k-1}})$$ such
that, modulo the Einstein equations,
$$
h_{ab}=c g_{ab} + \nabla_a X_b + \nabla_b X_a.
$$

\noindent This result was announced in \refto{CGT1993}.

Because the existence of generalized symmetries is necessary for the
existence of (local differential) conservation laws, it is natural to ask what
is conserved by virtue of the symmetries of the Einstein equations.  It is
straightforward to show that there are no conservation laws associated with
the scaling symmetry.  This is because the Hilbert Lagrangian $\sqrt{g}R$
is not preserved (even up to a divergence) under metric re-scalings.  The
diffeomorphism symmetries do lead to conservation laws in the form of the
contracted Bianchi identities, but of course the conserved quantities all
vanish when the field equations are satisfied.

The plan of this paper is as follows.  In \S2 we begin with a summary of
the theory of generalized symmetries.  We then present elementary
applications of this theory to the Einstein equations.  The technical
machinery needed for our analysis is then summarized.  In \S3 we classify
natural symmetries, which are symmetries built from the metric, curvature
and covariant derivatives of the curvature to any order.  In \S4 we classify
first-order generalized symmetries, which require a considerably more
intricate analysis than needed for natural generalized symmetries.  In \S5
we extend the analysis of \S3 to obtain a classification of all generalized
symmetries.  The analysis of \S5 uses an induction argument to reduce the
classification to that of first-order generalized symmetries.

We believe the
methods that are used to prove these results are of no less importance than
the results themselves.  In classifying the
generalized symmetries of the Einstein equations we
have developed an effective spinor--jet bundle formalism for analyzing
mathematical properties of the Einstein equations and related equations
\refto{CGT1994a}.  By far, the most important ingredient in this
formalism is the use of what Penrose calls an ``exact set of fields'' for the
field equations \refto{Penrose1960}, \refto{Penrose1984}. These are
spinor fields which allow us to parametrize the jet space of vacuum
Einstein metrics.  In future work we will
apply these spinor--jet techniques to related aspects of general relativity.
Specifically, our methods can be used to classify systematically (i) all
closed $p$-forms that are built locally from a Ricci-flat metric, (ii) all
symplectic forms for the Einstein equations, and (iii) all divergence-free
symmetric tensors built locally from Einstein metrics.  Finally, it is worth
pointing out that the existence of an exact set of fields is not limited to the
Einstein equations.  For example, preliminary computations show that the
generalized symmetries of the Yang-Mills equations are amenable to
analysis using these techniques.
\vfill\eject

%
\sectionno2
\abovedisplayskip=12pt plus 1pt minus 3pt
\belowdisplayskip=12pt plus 1pt minus 3pt
\def\ss{\scriptscriptstyle}
\def\proof{\par\noindent {\bf Proof:\ }}
\def\eqskip{\noalign{\vskip\jot}}
\def\K{{\cal K}}
\def\s{{\cal S}}
\def\Psibar{\overline\Psi{}}
\def\psibar{\overline\psi{}}
\def\phibar{\overline\phi{}}
\def\betabar{\overline\beta{}}
\def\alphabar{\overline\alpha{}}
\def\chibar{\overline\chi{}}

\def\hab{h^{\ss A\kern 0.5pt B}_{\ss A^{\kern -.8pt\prime}\kern -1.8pt
B^{\kern -.8pt\prime}}}
\def\dab{d^{\ss A\kern 0.5pt B}_{\ss A^{\kern -.8pt\prime}\kern -1.8pt
B^{\kern -.8pt\prime}}}
\def\kab{k^{\ss A\kern 0.5pt B}_{\ss A^{\kern -.8pt\prime}\kern -1.8pt
B^{\kern -.8pt\prime}}}
\def\Q{{\cal G}}
\overfullrule=0pt
\def\lineq{\equationlabel{2}{2001}{}}

\noindent{\bf 2. Preliminaries.}

In \S2A we briefly review the geometric theory of generalized
symmetries
for differential equations and their role in constructing local conservation
laws.  For more on generalized symmetries and their applications, see
\refto{Olver1993}.  In \S2B  we derive the defining equations for the
generalized symmetries of the vacuum Einstein equations and present some
preliminary results concerning solutions to these equations.  We then
present in sections \S2C and \S2D the
technical machinery needed to compute the generalized symmetries of the
Einstein equations.  A complete presentation of the results in these latter
two sections can be found in \refto{CGT1994a}.

\vskip0.2truein
\noindent{\bf 2A. Generalized Symmetries for Classical Field Theories.}

In classical field theory, the
fields are usually identified with sections $\varphi\colon M\to E$ of a fiber
bundle $\pi\colon E\to M$.  In general
relativity, $M$ is
a 4-dimensional manifold and $\pi$ is the bundle $\pi\colon \Q\to M$ of
quadratic
forms on the tangent space $TM$ with signature $(-+++)$.  A section
$g\colon M\to
\Q$ is a choice of Lorentz metric on $M$.

Let $\pi^k_M\colon J^k(E)\to M$ be the bundle of $k$-th order jets of
local
sections of
$E$.  A point $\sigma\in J^k(E)$ is, by definition, an equivalence class of
local
sections defined in a neighborhood $U$ of the point $x=\pi^k_M(\sigma)$;
two
local sections $\varphi_1,\varphi_2\colon U\to E$ are equivalent if
$\varphi_1$ and
$\varphi_2$ and all their partial derivatives to order $k$ agree at $x$. If
$\varphi\colon U\to E$ is a local section of $E$, then the canonical lift
$$
j^k(\varphi)\colon U\to J^k(E)
$$
is the map that assigns to each point $x\in U$ the
$k$-jet $j^k(\varphi)(x)$ represented by $\varphi$ at $x$.  There are
also canonical
projections
$$
\pi^k_l\colon J^k(E)\to J^l(E),
$$
defined for all $k\geq l$.  When $l=0$, we write $\pi_E^k\colon J^k(E)\to
E$.
The
infinite jet bundle $\pi^\infty_M\colon J^\infty(E)\to M$ is similarly
defined.  For a more detailed presentation of jet bundles, see
\refto{Olver1993}, \refto{Saunders1989}.

A differential form $\omega$ on $J^\infty(E)$ is called a {\it contact
form} if,
for every local section $\varphi\colon U\to E$,
$$
[j^\infty(\varphi)]^*(\omega)=0.
$$
The set of all contact forms on $J^\infty(E)$ is a differential ideal in the
ring $\Omega^*(J^\infty(E))$ of all differential forms on
$J^\infty(E)$, and we denote this ideal by ${\cal C}(J^\infty(E))$.

A {\it generalized vector field} $Z$ on $E$ is a vector field along the map
$\pi^\infty_E$, that is, for each point $\sigma\in J^\infty(E)$, $Z_\sigma$
is
a tangent vector in $T_p(E)$, where $p=\pi^\infty_E(\sigma)$.  If $Z$ is a
generalized vector field on $E$, then there is a unique vector field ${\rm
pr}\,Z$
on
$J^\infty(E)$, called the {\it infinite prolongation} of $Z$ such that

\medskip
\itemitem{(i)} for
each $\sigma\in J^\infty(E)$,
$(\pi^\infty_E)_*[{\rm (pr}\, Z)_\sigma]=Z_{\pi^\infty_E(\sigma)}$, and
\smallskip
\itemitem{(ii)} ${\rm pr}\,Z$ preserves the contact ideal, that is, under
Lie differentiation
$
{\cal L}_{{\rm pr}\,Z}\,{\cal C}(J^\infty(E))\subset {\cal C}(J^\infty(E)).
$

\medskip
\noindent We shall give local expressions for $Z$ and ${\rm pr}\,Z$
shortly.
A generalized vector field $Y$ on $E$ that is $\pi$-vertical, {\it i.e.,}
$$
\pi_*(Y_\sigma)=0,
$$
for all $\sigma\in J^\infty(E)$, is called an {\it evolutionary vector field}.
Evolutionary vector fields determine ``infinitesimal field variations'', and
their prolongations determine the induced variations in the derivatives of
the fields.
Finally, a {\it generalized vector field $X$ on $M$} is a vector field along
the
map $\pi^\infty_M$, and a {\it generalized tensor field $A$ of type $(p,q)$
on $M$} is a smooth map
$$
A\colon J^\infty(E)\to T^p_q(M)
$$
along $\pi^\infty_M$, where $T^p_q(M)$ is the bundle of tensors of type
$(p,q)$ over $M$.  Note that if $Z$ is a generalized vector field on $E$,
then
$Z_M=\pi_*(Z)$ is a
generalized vector field on $M$.

Every generalized vector field $X$ on $M$ defines a
unique
vector field tot$X$ on $J^\infty(E)$, called the {\it total vector field} of
$X$, with the
properties

\medskip
\itemitem{(i)}
$
(\pi^\infty_M)_*[({\rm
tot}X)_\sigma]=X_{\pi^\infty_M(\sigma)},\quad{\rm and}
$\smallskip
\itemitem{(ii)} tot$X$ annihilates all contact 1-forms, that is, if $\omega$
is a
contact 1-form, then\break
${\rm tot}\,X\hook\omega=0.
$

\medskip
The following theorem is easily established from the local formulas for
pr$Z$ and tot$X$ that we shall give momentarily.

\proclaim Theorem{\statementtag204 }.
Let Z be a generalized vector field on $E$.  Then there exists a
unique evolutionary vector field $Z_{\rm ev}$ such that
$$
{\rm pr}\,Z={\rm tot}\,Z_M+ {\rm pr}\,Z_{\rm ev},\eqtag1
$$
where $Z_M=\pi_*(Z)$.

If $Z_1$ and $Z_2$ are generalized vector fields on $E$, then there exists
a generalized vector field $Z_3$ such that $[{\rm pr}\,Z_1,{\rm
pr}\,Z_2]={\rm pr}\,Z_3$.  We call $Z_3$ the generalized Lie bracket of
$Z_1$ and $Z_2$ and write
$$
[Z_1,Z_2]=Z_3.
$$

We remark that if $X$ is a generalized vector field on $M$ and
$X_E=(\pi^\infty_E)_*({\rm tot}\,X)$, then
$$
{\rm pr}\,X_E={\rm tot}\,X.
$$
In other words, tot$X$ is also a prolongation of a vector field and
therefore tot$X$ preserves the contact ideal.  It is straightforward to verify
that if tot$X_1$ and tot$X_2$ are two total vector fields, then $[{\rm
tot}\,X_1,{\rm tot}\,X_2]$ is also a total vector field, $[{\rm
tot}\,X_1,{\rm tot}\,X_2]={\rm tot}\,X_3$.  (Hence the set of all total
vector fields on $J^\infty(E)$ is a connection of general type on
$J^\infty(E)\longrightarrow M$.)

Now suppose a system of differential equations for the sections of $E$ is
given.  These are the field equations for the classical field theory.  If these
equations are of order $k$ (typically $k=2$), then they determine a smooth
subbundle
$$
{\cal R}^k\hookrightarrow J^k(E)
$$
with projection $\pi^k_M\colon {\cal R}^k\to M$.  We call ${\cal R}^k$
the {\it
equation manifold} for the classical field theory.  The derivatives of the
field equations to order $l$ then define the $l$-th {\it prolonged equation
manifold}
$$
{\cal R}^{k+l}\hookrightarrow J^{k+l}(E).
$$
The field equations, together with all their derivatives, determine the
{\it infinite prolonged equation manifold}
$$
{\cal R}^\infty\hookrightarrow J^\infty(E).
$$
It is customary to assume \refto{Tsujishita1982}, \refto{Tsujishita1989}
that the maps
$$
\pi^{l+1}_{l}\colon {\cal R}^{l+1}\to {\cal R}^{l}
$$
are surjective for all $l\geq k$ and have constant rank.  The fiber
dimension of $\pi^{l+1}_l$ represents the number of ``degrees of
freedom'' available in constructing a formal power series solution for the
field equations to order $l+1$  from a given solution to order $l$.
Roughly speaking, equations that are not ``over-determined'' will satisfy
the surjectivity assumption.  As we shall see, the vacuum Einstein equations
also satisfy these surjectivity and constant rank assumptions
\refto{CGT1994a}.

\proclaim Definition{\statementtag205 }.
A generalized vector field $Z$ on $E$ is called a generalized
symmetry of the given field equations if pr$\,Z$ is tangent to the infinitely
prolonged equation manifold ${\cal R}^\infty$, that is, for all
$\sigma\in{\cal R}^\infty$
$$
({\rm pr}\,Z)_\sigma\in T_\sigma({\cal R}^\infty).
$$

Generalized symmetries are sometimes called ``Lie-B\" acklund
symmetries''. If $Z_1$ and $Z_2$ are two generalized symmetries for
${\cal R}^\infty$, then the generalized Lie bracket $[Z_1,Z_2]$ is also a
generalized symmetry.

It is easy to see from our local coordinate formulas, given below, that if
$X$ is a generalized vector field on $M$, then
tot$X$ (or more precisely $X_E=\pi^\infty_E({\rm tot}X)$)
is always a generalized symmetry for any system of equations.  Total
vector fields are therefore viewed as trivial symmetries.  A generalized
symmetry $Z$ is also considered trivial if $Z$ vanishes on the prolonged
equation manifold ${\cal R}^\infty$.  Two generalized symmetries are said
to be
equivalent if their difference is a trivial symmetry.  Theorem
\statement{204}
implies that {\it every generalized symmetry $Z$ of a given system of
equations is
equivalent to a generalized symmetry $Y$ which is $\pi$-vertical, that is,
to an evolutionary generalized symmetry.}

We now give local coordinate descriptions of these various notions.  If
$(x^i,\varphi^\alpha)$, $i=1,2,\dots, n$ and $\alpha=1,2,\dots,m$, are local
coordinates on $E$, then the standard local coordinates for $J^\infty(E)$
are
$$
(x^i,\varphi^\alpha,\varphi^\alpha_{i_1},\varphi^\alpha_{i_1i_2},\ldots,
\varphi^\alpha_{i_1i_2\cdots i_k},\ldots),
$$
where, for a given local section $\varphi^\alpha=\varphi^\alpha(x^i)$,
$$
\varphi^\alpha_{i_1\cdots
i_k}(j^\infty(\varphi)(x))={\partial^k\varphi^\alpha(x)\over\partial
x^{i_1}\cdots\partial x^{i_k}}.
$$
The contact ideal ${\cal C}(J^\infty(E))$ is spanned locally by the contact
1-forms
$$
\theta^\alpha_{i_1\cdots i_k}=d\varphi^\alpha_{i_1\cdots
i_k}-\,\varphi^\alpha_{i_1\cdots i_kj}dx^j
$$
for $k=0,1,2,\dots$.  These forms satisfy the structure equations
$$
d\,\theta^\alpha_{i_1\cdots i_k}=dx^j\wedge\theta^\alpha_{i_1\cdots i_kj}.
$$

A generalized vector field $Z$ on $E$ assumes the form
$$
Z=A^i{\partial\hfill\over\partial
x^i}+B^\alpha{\partial\hfill\over\partial\varphi^\alpha},
$$
where
$$
A^i=A^i(x^j,\varphi^\beta,\varphi^\beta_{i_1},\dots,
\varphi^\beta_{i_1\cdots i_k}),\qquad
{\rm and}\qquad
B^\alpha=B^\alpha(x^j,\varphi^\beta,\varphi^\beta_{i_1},\dots,\varphi^\beta
_{i_1\cdots i_k}).
$$
A generalized vector field $X$ on $M$ and an
evolutionary vector field $Y$ on $E$ take the form
$$
X=A^i{\partial\hfill\over\partial x^i}\quad{\rm and}\quad
Y=B^\alpha{\partial\hfill\over\partial\varphi^\alpha},
$$
where, again, the coefficients $A^i$ and $B^\alpha$ are functions of $x^i$,
$\varphi^\alpha$ and the derivatives $\varphi^\alpha_{i_1\cdots i_k}$ to
some
arbitrary but finite order.
The vector field tot$X$ is given by
$$
{\rm tot}\,X=A^iD_i,
$$
where $D_i$ is the total derivative operator
$$
D_i={\partial\hfill\over\partial x^i} +
\varphi^\alpha_i{\partial\hfill\over\partial \varphi^\alpha}+
\varphi^\alpha_{ii_1}{\partial\hfill\over\partial\varphi^\alpha_{i_1}}
+\varphi^\alpha_{ii_1i_2}{\partial\hfill\over\partial\varphi^\alpha_{i_1i_2
}}
+\cdots.
$$
We write
$$
D_{i_1i_2\cdots i_k}=D_{i_1}D_{i_2}\cdots D_{i_k}.
$$

The prolongation of $Z$ is given by the prolongation formula
\refto{Olver1993}
$$
{\rm pr}\,Z=A^iD_i + \sum_{k=0}^\infty D_{i_1i_2\cdots
i_k}(B^\alpha-\varphi^\alpha_iA^i){\partial\hfill\over
\partial\varphi^\alpha_{i_1i_2\cdots i_k}}.\eqtag26
$$
Note that, in particular, the prolongation of the evolutionary vector field
$Y=B^\alpha{\displaystyle{\partial\hfill\over\partial\varphi^\alpha}}$
is
$$
{\rm pr}\,Y=\sum_{k=0}^\infty (D_{i_1i_2\cdots
i_k}B^\alpha){\partial\hfill
\over\partial\varphi^\alpha_{i_1i_2\cdots i_k}}.\eqtag27
$$
We now remark that \eq{26} and \eq{27} together prove Theorem
\statement{204}, with
$$
Z_{\rm ev}=(B^\alpha-
\varphi^\alpha_iA^i){\partial\hfill\over\partial\varphi^\alpha}.\eqtag8
$$

If $X_1=A^i_1{\displaystyle{\partial\hfill\over\partial x^i}}$ and
$X_2=A^i_2{\displaystyle{\partial\hfill\over\partial x^i}}$ are generalized
vector fields on $M$, then
$$
[X_1,X_2]=[A^i_1(D_iA^j_2)-A_2^i(D_iA^j_1)]{\partial\hfill\over\partial
x^j}.
$$
If
$Y_1=B^\alpha_1{\displaystyle{\partial\hfill\over\partial\varphi^\alpha}}$
and
$Y_2=B^\alpha_2{\displaystyle{\partial\hfill\over\partial\varphi^\alpha}}$
are evolutionary vector fields on $E$, then
$$
[Y_1,Y_2]=[{\rm pr}\,Y_1(B_2^\alpha)-{\rm
pr}\,Y_2(B_1^\alpha)]{\partial\hfill\over\partial\varphi^\alpha}.
$$

An evolutionary vector field
$Y=B^\alpha{\displaystyle{\partial\hfill\over\partial\varphi^\alpha}}$
defines ``infinitesimal field variations'' $\delta\varphi^\alpha_{i_1\cdots
i_l}$, $l=0,1,\ldots$, which depend locally on the fields and their
derivatives.  Explicitly, $\delta\varphi^\alpha_{i_1\cdots i_l}$ is defined by
letting the prolonged vector field ${\rm pr}\,Y$ act on the coordinates
$\varphi^\alpha_{i_1\cdots i_l}$, which are viewed as functions on
$J^\infty(E)$:
$$
\delta\varphi^\alpha_{i_1\cdots i_l}={\rm pr}\,Y(\varphi^\alpha_{i_1\cdots
i_l})=(D_{i_1i_2\cdots
i_l}B^\alpha)(x^i,\varphi^\alpha,\varphi^\alpha_i,\ldots,\varphi^\alpha_{i_1
\cdots i_{l+k}}).
$$

If
$$
\Delta_\beta(x^i,\varphi^\alpha,\varphi^\alpha_{i_1},\ldots,
\varphi^\alpha_{i_1\cdots i_k})=0,\quad\beta=1,\ldots,m\eqtag28
$$
is a system of field equations for the fields $\varphi^\alpha$, then ${\cal
R}^k\subset J^k(E)$ is the manifold defined by these equations.  The
infinite
prolonged equation manifold ${\cal R}^\infty$ is defined by the equations
\eq{28}
together with the equations
$$
D_{i_1i_2\cdots i_l}\Delta_\beta=0
$$
for $l=1,2,\dots$.  The evolutionary vector field $Y=B^\alpha
{\displaystyle{\partial\hfill\over\partial\varphi^\alpha}}$ is, according to
the tangency condition in Definition \statement{205}, a generalized
symmetry of
\eq{28} if and only if the coefficient functions $B^\alpha$ satisfy the linear
total
differential equation
$$
\sum_{l=0}^k {\partial\Delta_\beta\over\partial\varphi^\alpha_{i_1\cdots
i_l}}
[D_{i_1\cdots i_l}B^\alpha]=0\qquad {\rm on}\quad {\cal R}^\infty.\eqtag2
$$
This equation is called the {\it formal linearization}
of \eq{28}, or the {\it defining equation} for the generalized symmetry
$Y$.

Let us remark that when $Z$ is an ordinary vector field on $E$, that is,
$$
Z=A^i(x^j,\varphi^\beta){\partial\hfill\over\partial
x^i}+B^\alpha(x^j,\varphi^\beta)
{\partial\hfill\over\partial\varphi^\alpha},
$$
and $({\rm pr}\,Z)(\Delta_\beta)=0$ on the equation manifold
$\Delta_\beta=0$,
then $Z$ is called a {\it point symmetry} of the equations.  Point
symmetries
are in
one-to-one correspondence with first-order evolutionary symmetries
$$
Y=B^\beta(x^i,\varphi^\alpha,\varphi^\alpha_i){\partial\hfill\over
\partial\varphi^\beta},
$$
with $B^\alpha$ a collection of affine linear functions of the first
derivatives $\varphi^\alpha_i$.

Finally, we cite a version of Noether's theorem as it applies to
generalized
symmetries \refto{Olver1993}. Recall that a {\it local differential
conservation law} $V$ for the field equations $\Delta_\beta=0$ is a
generalized vector density
$$
V=V^i(x^i,\varphi^\alpha,\varphi^\alpha_{i_1},\dots,\varphi^\alpha_{i_1\c
dots i_k})
{\partial\hfill\over\partial x^i}
$$
on $M$ such that the total divergence
$$
{\rm Div}V=D_iV^i=0\qquad {\rm on}\quad {\cal R}^\infty.
$$
A conservation law $V$ is said to be trivial if there is a generalized skew-
symmetric tensor density
$$
S^{ij}=S^{ij}(x^k,\varphi^\alpha,\varphi^\alpha_{i_1},\varphi^\alpha_{i_1
i_2},\ldots,\varphi^\alpha_{i_1\cdots i_l})
$$
such that
$$
V^i=D_jS^{ij}\qquad{\rm on\ }{\cal R}^\infty.
$$
Two conservation laws are said to be equivalent if their difference is a
trivial conservation law.  Following Olver \refto{Olver1993}, an
evolutionary vector field
$Y=B^\alpha{\displaystyle{\partial\hfill\over\partial\varphi^\alpha}}$ is
called a
{\it characteristic vector field} for the conservation law $V$ if
$$
{\rm Div}V=B^\alpha\Delta_\alpha\eqtag281
$$
{\it identically}.  Under mild regularity conditions on the equations
$\Delta_\beta=0$, it can be shown that every conservation law $V^\prime$
is equivalent to a conservation law $V$ whose divergence satisfies
\eq{281}.  It is a simple result from the variational calculus that if
$\Delta_\alpha$ are the components of the Euler-Lagrange operator
$E_\alpha(L)$ for some
Lagrangian,
$L=L(x^i,\varphi^\alpha,\varphi^\alpha_{i_1},
\dots,\varphi^\alpha_{i_1\cdots i_k})$,
$$
E_\alpha(L)= {\partial L\over\partial\varphi^\alpha}-D_{i_1}{\partial
L\over\partial\varphi^\alpha_{i_1}}+\cdots\pm D_{i_1\cdots i_k}{\partial
L\over\partial\varphi^\alpha_{i_1\cdots i_k}},
$$
then {\it
every characteristic vector field Y for a local differential conservation law
for the equations $\Delta_\alpha=0$ defines a generalized symmetry}.  The
converse need not be true.  For example, scaling symmetries of Euler-
Lagrange equations typically will not lead to conservation laws.

\vskip0.2truein
\noindent{\bf 2B. The Formal Linearization of the Einstein Equations.}

To study the generalized symmetries of the Einstein field equations, we let
$\pi\colon \Q\to M$ be the bundle of Lorentz metrics over the spacetime
manifold $M$.
Standard
local coordinates for $J^k(\Q)$ are
$$
(x^i, g_{ij}, g_{ij},_{i_1},\dots, g_{ij},_{i_1i_2\cdots i_k}).
$$
The Christoffel symbols $\Gamma_{ij}^k$, the curvature tensor
$R_{i\phantom{h}jk}^{\phantom{i}h}$, and their derivatives are
all
considered now as functions on $J^k(\Q)$.  The covariant derivatives of a
generalized tensor field on $M$ are defined in terms of total derivatives.
For example, if
$$
A_a=A_a(x^i, g_{ij},
g_{ij},_{i_1},g_{ij},_{i_1i_2},\dots,g_{ij},_{i_1i_2\cdots i_k})
$$
are the components of a generalized 1-form on $M$, then
$$
\eqalign{
\nabla_bA_a&=D_bA_a-\Gamma_{ab}^cA_c\cr
&={\partial A_a\over\partial x^b}+ {\partial A_a\over\partial
 g_{ij}}g_{ij},_b
+{\partial A_a\over\partial g_{ij},_{i_1}}g_{ij},_{i_1b}
+\cdots+{\partial A_a\over\partial g_{ij},_{i_1\cdots
i_k}}g_{ij},_{i_1\cdots i_kb}-\Gamma_{ab}^cA_c.}
$$

We now compute the formal linearization \eq{2} of the vacuum Einstein
equations.

\proclaim Proposition{\statementtag206 }.
Let
$$
Y=h_{ab}(x^i,g_{ij},g_{ij},_{i_1},\ldots,g_{ij},_{i_1i_2\cdots
i_k}){\partial\hfill\over\partial g_{ab}}
$$
be an evolutionary vector field on the bundle of Lorentz metrics.  Then
$Y$ is a generalized symmetry of the Einstein equations $R_{ij}=0$ if and
only if
$$
\left[-g^{cd}\delta^a_i\delta^b_j-g^{ab}\delta^c_i\delta^d_j
+g^{ac}\left(\delta^b_i\delta^d_j+\delta^b_j\delta^d_i\right)\right]\nabla_c
\nabla_dh_{ab}=0\eqtag29
$$
whenever $R_{ij}$ and its covariant derivatives to order $k$ vanish.

\proof
This is an easy computation based upon the identities
$$
({\rm pr}Y)(\Gamma_{ij}^l)
={1\over2}g^{lm}[\nabla_ih_{mj}+\nabla_jh_{mi}
-\nabla_mh_{ij}],\eqtag30
$$
and
$$
({\rm pr}Y)(R_{i\phantom{l}jk}^{\phantom{i}l})
=\nabla_k({\rm pr}Y(\Gamma_{ij}^l))-
\nabla_j({\rm pr}Y(\Gamma_{ik}^l)).\eqtag31
$$
These formulas are, of course, familiar from the variational calculus.
We emphasize that now \eq{30} and \eq{31} are to be viewed as
identities on $J^k(\Q)$, where they are direct consequences of the
prolongation formula \eq{27}.
\qed

We remark that Proposition \statement{206} could also be formulated in
terms of the Einstein tensor $G_{ij}$ and its derivatives.  The symmetry
conditions so-obtained are equivalent to \eq{29}.

Let $X=X^i(x){\displaystyle{\partial\hfill\over\partial x^i}}$ be a vector
field on $M$ with local flow $\phi_t\colon M\to M$.  Then $\phi_t$
induces, by pull-back, a local flow on ${\cal G}$ with corresponding
vector field $\widetilde X$ on ${\cal G}$ given by
$$
\widetilde X=X^a{\partial\hfill\over\partial x^a}-({\partial X^a\over\partial
x^i}g_{aj}+{\partial X^a\over\partial x^j}g_{ai})
{\partial\hfill\over\partial g_{ij}}.
$$
The associated evolutionary vector field is, by \eq{8},
$$
\widetilde X_{\rm ev}=-
(\nabla_iX_j+\nabla_jX_i){\partial\hfill\over\partial g_{ij}}.
$$
It is well-known \refto{Ibragimov1985} that $\widetilde X$, or
equivalently  $\widetilde X_{\rm ev}$, represents a point symmetry of the
Einstein equations corresponding to the diffeomorphism invariance of the
Einstein equations.   This observation motivates the following definition.
\proclaim Definition{\statementtag2061 }.
Let
$$
X=X^a(x^i,g_{ij},g_{ij},_{i_1},\ldots,g_{ij},_{i_1i_2\cdots
i_l}){\partial\hfill\over\partial x^a}
$$
be a generalized vector field on $M$.  We call the evolutionary vector field
$$
\K_X=(\nabla_iX_j + \nabla_jX_i){\partial\hfill\over\partial g_{ij}},
$$
where $X_i=g_{ij}X^j$, the associated generalized diffeomorphism
vector field on $\Q$.

We remark that if $X_1$ and $X_2$ are generalized vector fields on $M$,
then
$$
[\K_{X_1},\K_{X_2}]=\K_{[X_1,X_2]}.
$$

\proclaim Proposition{\statementtag207 }.
For any generalized vector field $X$ on $M$, the associated generalized
diffeomorphism vector field $\K_X$ is a generalized symmetry of the
vacuum Einstein equations.

\proof
By virtue of \eq{30}, we find that
$$
({\rm pr}\,\K_X)(\Gamma_{ij}^l)=\nabla_j\nabla_iX^l
+R_{i\phantom{l}jp}^{\phantom{i}l}X^p,
$$
and hence, by \eq{31},
$$
({\rm pr}\,\K_X)R_{ij}=(\nabla_pR_{ij})X^p
+R_{pj}\nabla_iX^p+R_{ip}\nabla_jX^p,
$$
which vanishes when $R_{ij}=0$ and $\nabla_kR_{ij}=0$.\qed\medskip

We call the symmetry $\K_X$ a {\it generalized
diffeomorphism symmetry} of the Einstein equations.  Note that the
generalized diffeomorphism
vector fields $\K_X$ will be symmetries for any generally covariant set of
field equations on $\Q$.  In particular, Proposition \statement{207}
generalizes to the Einstein equations with cosmological constant.

There is one more obvious symmetry of the vacuum Einstein equations
$R_{ij}=0$.

\proclaim Proposition{\statementtag208 }.
For any constant $c$, the vector field
$$
\s_c=c\,g_{ij}{\partial\hfill\over\partial g_{ij}}\eqtag311
$$
is a point symmetry of the vacuum Einstein equations $R_{ij}=0$.

\proof
This proposition follows from the fact that
$$
({\rm pr}\,\s_c)(\Gamma_{ij}^k)=0
$$
and hence
$$({\rm
pr}\,\s_c)(R_{ij})=0.
$$
Alternatively, $h_{ij}=c\,g_{ij}$ clearly satisfies \eq{29}.  \qed\medskip

On a 4-dimensional manifold $M$ we have
$$
({\rm pr}\,\s_c)(\sqrt{g}R)= c \sqrt{g}R,
$$
so the scaling symmetry $\s_c$ of the Einstein equations does not preserve
the Hilbert Lagrangian (even up to a divergence) and therefore does not
generate a conservation law.  The generalized diffeomorphism symmetry
${\cal K}_X$ is a characteristic for a conservation law for the Einstein
equations, namely,
$$
\nabla_j(2X_iG^{ij})=(\nabla_iX_j+\nabla_jX_i)G^{ij}.
$$
But the conserved vector field $V^j=2X_iG^{ij}$ is trivial.

We remark that the scaling symmetry $\s_c$ and the point diffeomorphism
symmetry $\widetilde X$ are the only {\it point} symmetries of the
vacuum Einstein equations \refto{Ibragimov1985}.

\vskip0.2truein
\noindent{\bf 2C. Spinor Coordinates for Prolonged Einstein Equation
Manifolds.}

Let ${\cal E}^k\subset J^k(\Q)$ be the set of $k$-jets that satisfy the
Einstein equations and the covariant derivatives of the Einstein equations to
order $k-2$,
$$
{\cal E}^k=\{\,j^k(g)(x_0)\in J^k(\Q)\,|\,G_{ij}=0,
G_{ij|i_1}=0,\dots,G_{ij|i_1\cdots i_{k-2}}=0 \ {\rm at}\  j^k(g)(x_0)\,\}.
$$
In what follows, we will either use the vertical bar or $\nabla$ to indicate
covariant differentiation.

If $h_{ab}=h_{ab}(x^i,g_{ij},g_{ij,j_1},\ldots,g_{ij,j_1\cdots j_k})$ is a
generalized symmetry of the vacuum Einstein equations, then the linearized
equations \eq{29} must hold identically at each point of ${\cal E}^{k+2}$.
To solve these equations we shall construct explicit coordinates for these
prolonged equation manifolds.  To this end, we let $\Gamma^i_{jk}$ be
the Christoffel symbols of the metric $g_{ij}$ and inductively define {\it
higher-order Christoffel symbols} by
$$
\Gamma^i_{j_0j_1\cdots j_k}=\Gamma^i_{(j_0j_1\cdots j_{k-1},j_k)}
-(k-1)\Gamma^i_{m(j_1\cdots j_{k-2}}\Gamma^m_{j_{k-
1}j_k)},\eqtag3102
$$
for $k\geq1$.
These higher-order symbols arise naturally from the prolongations of the
geodesic equations and play a prominent role in T. Y. Thomas' theory of
normal extensions \refto{Thomas1934}.  We will denote the generalized
Christoffel symbols \eq{3102} by $\Gamma^k$.   Note that
$\Gamma^i_{j_0j_1\cdots j_k}$ is completely symmetric in the indices
$j_0j_1\cdots j_k$ and depends on the metric and its first $k$ derivatives.

Next, let \refto{Penrose1960}
$$
Q_{ij},_{j_1\cdots
j_k}=g_{ir}g_{js}R^{r\phantom{(j_1}s}_{\phantom{r}(j_1\phantom{s}j_
2|j_3\cdots j_k)},\eqtag312
$$
for $k\geq2$.
This tensor is a generalized tensor on $M$ of order $k$,  which we denote
by $Q^k$.  Note that $Q_{ij},_{j_1\cdots j_k}$ is symmetric in $ij$ and
$j_1\cdots j_k$, and satisfies the cyclic identity
$$
Q_{i(j},_{j_1\cdots j_k)}=0.\eqtag3
$$
It is possible to prove, for example, by applying T.Y. Thomas'
Replacement Theorem \refto{Thomas1934}, that
$$
\nabla_{j_{k+1}}Q_{ij,j_1\cdots j_k}=
Q_{ij,j_1\cdots j_{k+1}} + {2\over k+2}Q_{j_{k+1}(i,j)j_1\cdots j_k}
+{k\over k+2}Q_{(j_1j_2,j_3\cdots j_k)ij}
+L_{ij,j_1\cdots j_{k+1}},\eqtag3101
$$
where
$$
L_{ij,j_1\cdots j_{k+1}}=L_{ij,j_1\cdots j_{k+1}}(g_{ab},
Q_{ab,c_1c_2},\ldots,Q_{ab,c_1c_2\cdots c_{k-1}}).
$$
A straightforward calculation, starting with the expression for $R_{ijkl}$
in terms of the derivatives of the metric, shows that
$$
g_{ij,j_1\cdots j_k}=-2{k-1\over k+1}Q_{ij,j_1\cdots j_k} +
g_{il}\Gamma^l_{jj_1\cdots j_k} + g_{jl}\Gamma^l_{ij_1\cdots j_k}
+P_{ij,j_1\cdots j_k},
$$
where
$$
P_{ij,j_1\cdots j_k}=P_{ij,j_1\cdots
j_k}(g_{hk},g_{hk,j_1},\ldots,g_{hk,j_1\cdots j_{k-1}}).
$$
{}From this equation it is then possible to prove \refto{CGT1994a} that the
variables
$$
(x^i,g_{ij},\Gamma^i_{j_0j_1},\ldots,\Gamma^i_{j_0j_1\cdots
j_k},Q_{ij},_{j_1j_2},\ldots,Q_{ij},_{j_1\cdots j_k})\eqtag32
$$
can be used as coordinates for the bundle $J^k(\Q)$.  In particular, suppose
we are given quantities $q_{ij}$, $X^i_{j_0j_1\cdots j_l}$, and
$Y_{ij,j_1\cdots j_m}$, for $l=1,2,\dots,k$ and $m=2,3,\dots,k$, where
$q_{ij}$, $X^i_{j_0j_1\cdots j_l}$ and $Y_{ij,j_1\cdots j_m}$ have the
symmetries of $g_{ij}$, $\Gamma^i_{j_0j_1\cdots j_l}$ and
$Q_{ij,j_1\cdots j_m}$.  Then the $k$-jet $j^k(g)(x_0)$ defined
inductively by
$$
g_{ij}(x_0)=q_{ij},
$$
and
$$
\eqalign{
g_{ij,j_1\cdots j_l}(x_0)=&-2{k-1\over k+1}Y_{ij,j_1\cdots j_m} +
g_{il}(x_0)X^l_{jj_1\cdots j_l} + g_{jl}(x_0)X^l_{ij_1\cdots j_l}\cr
&+P_{ij,j_1\cdots j_l}(g_{ij}(x_0),g_{ij,j_1}(x_0),\ldots,g_{ij,j_1\cdots
j_{l-1}}(x_0))}
$$
for $l=1,2,\ldots,k$ and $m=2,3,\dots,k$, satisfies
$$
\Gamma^i_{j_0j_1\dots j_l}(j^l(g)(x_0))=X^i_{j_0j_1\cdots j_l},
$$
and
$$
Q_{ij,j_1\cdots j_m}(j^m(g)(x_0))=Y_{ij,j_1\cdots j_m}
$$
for $l=1,2,\ldots,k$ and $m=2,3,\dots,k$.

The coordinates \eq{32} are well-suited for describing the prolonged
Einstein equation manifold ${\cal E}^k$.  If we let
$$
{\rm [tr}_1Q]_{j_1j_2\cdots j_k}=g^{ij}Q_{ij},_{j_1j_2\cdots j_k},\quad
{\rm [tr}_2Q]_{j,j_2\cdots j_k}=g^{ij_1}Q_{ij},_{j_1j_2\cdots j_k},\quad
{\rm [tr}_3Q]_{ij,j_3\cdots j_k}=g^{j_1j_2}Q_{ij},_{j_1j_2\cdots j_k},
$$
and
$$
{\rm [tr}_{13}Q]_{j_3\cdots j_k}=g^{ij}g^{j_1j_2}Q_{ij},_{j_1j_2\cdots
j_k},\quad
{\rm [tr}_{22}Q]_{j_3\cdots j_k}=g^{ij_1}g^{jj_2}Q_{ij,j_1j_2\cdots
j_k},
$$
then it is not difficult to prove that
$$
\eqalign{
G_{ij|(j_3\cdots j_k)}
&=R_{ij|(j_3\cdots j_k)}-{1\over2}g_{ij}R_{|(j_3\cdots j_k)}\cr
&={k-1\over k+1}\Big(\,{\rm [tr}_3Q]_{ij,j_3j_4\cdots j_k}-{\rm
[tr}_2Q]_{i,jj_3\cdots j_k}-\ {\rm [tr}_2Q]_{j,ij_3\cdots j_k}
+\ {\rm [tr}_1Q]_{ijj_3j_4\cdots j_k}\cr
&\quad\ -\ g_{ij}({\rm [tr}_{13}Q]_{j_3j_4\cdots j_k}-\ {\rm
[tr}_{22}Q]_{j_3j_4\cdots j_k})\,\Big)+\ F_{ij,j_3j_4\cdots j_k},}
\eqtag33
$$
where $F_{ij,j_3j_4\cdots j_k}$ is a tensor of order $k-2$ that is
symmetric in $ij$ and in $j_3j_4\cdots j_k$.  By contracting this equation
with $g^{jj_3}$ we deduce that
$$
g^{jj_3}F_{ij,j_3j_4\cdots j_k}(j^{k-2}(g)(x_0))=0\quad{\rm
whenever}\quad j^{k-2}(g)(x_0)\in {\cal E}^{k-2}.\eqtag330
$$

In \refto{CGT1994a} we carefully analyze \eq{33} to prove the following
two theorems.  These are purely algebraic results.

\proclaim Theorem{\statementtag209 }.
Let $j^k(g^1)(x_0)$ and $j^k(g^2)(x_0)$ be any two points in ${\cal E}^k$
and let $Q^l_1$, $Q^l_2$ denote the values of the tensors $Q_{ij,j_1\cdots
j_l}$, $l=2,\dots,k$ at $j^k(g^1)(x_0)$ and $j^k(g^2)(x_0)$.  If
$g^1_{ij}(x_0)=g^2_{ij}(x_0)$, and if the completely trace-free parts of
the tensors $Q^l_1$ and $Q^l_2$ agree, that is,
$$
[Q^l_1]_{\ss\rm tracefree}=[Q^l_2]_{\ss\rm tracefree},
$$
for each $l=2,\dots,k$, then
$$
Q^l_1=Q^l_2
$$
for each $l=2,\dots,k$.

\proclaim Theorem{\statementtag210 }.
Let $j^{k-1}(g)(x_0)\in{\cal E}^{k-1}$, and let $S_{ab,j_1\cdots j_k}$,
which is denoted $S^k$, be a given tensor with the following
properties:\hfill\break\smallskip
\line{\indent(i) $S_{ab,j_1\cdots j_k}$ is symmetric in $ab$ and $j_1\cdots
j_k$,\hfill}\smallskip
\line{\indent(ii) $S_{ab,j_1\cdots j_k}$ is trace-free on any pair of indices,
and\hfill}\smallskip
\line{\indent(iii) $S_{a(b,j_1\cdots j_k)}=0$.\hfill}
\medskip
\noindent Then there exists a metric $k$-jet $j^k(\widetilde g)(x_0)$ such
that $j^k(\widetilde g)(x_0)\in{\cal E}^k$, and
$$
j^{k-1}(\widetilde g)(x_0)=j^{k-1}(g)(x_0),
$$
and
$$
[Q^k(j^k(g)(x_0))]_{\ss \rm tracefree}=S^k.
$$

\noindent Together, Theorems \statement{209}, \statement{210} show that
local coordinates for ${\cal E}^k$ are given by
$$
(x^i, g_{ij},\Gamma^i_{j_0j_1\cdots j_l}, [Q_{ij,j_1\cdots j_l}]_{\ss\rm
tracefree})\quad{\rm for}\quad l\leq k.\eqtag36
$$
In particular, Theorem \statement{210} shows that the projection map
$\pi^{k}_{k-1}\colon{\cal E}^{k}\to{\cal E}^{k-1}$ is surjective.

The tensor $[Q_{ij,j_1\cdots j_l}]_{\ss\rm tracefree}$ has a remarkable
characterization in terms of two-component spinors.  This spinor
characterization is based on the following theorem \refto{CGT1994a}.

\proclaim Theorem{\statementtag211 }.
Let $S_{ab,j_1\cdots j_k}$ be a complex tensor which satisfies the
properties (i)--(iii) of Theorem \statement{210},
and let the tensor $S_{ab,j_1\cdots j_k}$ have the spinor representation
$$
S_{ab,j_1\cdots j_k}\longleftrightarrow S{}_{\ss A}^{\ss
A^\prime}{}_{\ss B}^{\ss B^\prime}{}_{\ss J_1}^{\ss
J_1^\prime}{}^{\cdots}_{\cdots}{}_{\ss J_k}^{\ss J_k^\prime}.
$$
Then there exist unique spinors $U{}_{\ss J_1\cdots J_{k+2}}^{\ss
J_1^\prime \cdots J_{k-2}^\prime}$ and $V{}_{\ss J_1\cdots J_{k-
2}}^{\ss J_1^\prime \cdots J_{k+2}^\prime}$, both completely symmetric
in their primed and unprimed indices, such that
$$
S{}_{\ss A}^{\ss A^\prime}{}_{\ss B}^{\ss B^\prime}{}_{\ss J_1}^{\ss
J_1^\prime}{}^{\cdots}_{\cdots}{}_{\ss J_k}^{\ss
J_k^\prime}=\epsilon^{\ss A^\prime (J_1^\prime}\epsilon^{\ss |B^\prime|
J_2^\prime}U{}_{\ss ABJ_1\cdots J_{k}}^{\ss J_3^\prime \cdots
J_{k}^\prime)}+\epsilon_{\ss A (J_1}\epsilon_{\ss |B| J_2}V{}^{\ss
A^\prime B^\prime J_1^\prime \cdots J_{k}^\prime }_{\ss J_3 \cdots
J_{k})}.
$$

Now we consider the spinor representation of the curvature tensor
\refto{Penrose1984},
$$
R_{abcd}\longleftrightarrow R{}_{\ss A}^{\ss A^\prime}{}_{\ss B}^{\ss
B^\prime}{}_{\ss C}^{\ss C^\prime}{}_{\ss D}^{\ss D^\prime},
$$
where
$$
\eqalign{
R{}_{\ss A}^{\ss A^\prime}{}_{\ss B}^{\ss B^\prime}{}_{\ss C}^{\ss
C^\prime}{}_{\ss D}^{\ss D^\prime}
&=\Psi_{\ss ABCD}\epsilon^{\ss A^\prime B^\prime}\epsilon^{\ss
C^\prime D^\prime}+\Psibar^{\ss A^\prime B^\prime C^\prime D^\prime
}\epsilon_{\ss AB}\epsilon_{\ss CD}\cr
&+\Phi{}_{\ss A}^{\ss C^\prime}{}_{\ss B}^{\ss D^\prime}\epsilon_{\ss
CD}\epsilon^{\ss A^\prime B^\prime}
+\Phi{}_{\ss C}^{\ss A^\prime}{}_{\ss D}^{\ss B^\prime}\epsilon_{\ss
AB}\epsilon^{\ss C^\prime D^\prime}\cr
&+2\Lambda(\epsilon_{\ss AC}\epsilon_{\ss BD}\epsilon^{\ss A^\prime
C^\prime}\epsilon^{\ss B^\prime D^\prime}
-\epsilon_{\ss AD}\epsilon_{\ss BC}\epsilon^{\ss A^\prime
D^\prime}\epsilon^{\ss B^\prime C^\prime}).}\eqtag34
$$
The totally symmetric spinors $\Psi_{\ss ABCD}$ and $\Psibar^{\ss
A^\prime B^\prime C^\prime D^\prime}$ correspond to the spinor
representation of the Weyl tensor. The symmetric spinor $\Phi_{\ss
A\,B}^{\ss C^\prime D^\prime}$ corresponds to the trace-free Ricci
tensor, and the scalar $\Lambda$ corresponds to the scalar curvature.  If
we set
$$
[Q_{ab,j_1\cdots j_k}]_{\ss\rm tracefree}\longleftrightarrow
Q{}_{\ss A}^{\ss A^\prime}{}_{\ss B}^{\ss B^\prime}{}_{\ss J_1}^{\ss
J_1^\prime}{}^{\cdots}_{\cdots}{}_{\ss J_k}^{\ss J_k^\prime},
$$
then it is not too difficult to show, using Theorem \statement{211} and
\eq{34}, that
$$
Q{}_{\ss A}^{\ss A^\prime}{}_{\ss B}^{\ss B^\prime}{}_{\ss J_1}^{\ss
J_1^\prime}{}^{\cdots}_{\cdots}{}_{\ss J_k}^{\ss
J_k^\prime}=\epsilon^{\ss A^\prime (J_1^\prime}\epsilon^{\ss |B^\prime|
J_2^\prime}\Psi{}_{\ss ABJ_1\cdots J_{k}}^{\ss J_3^\prime \cdots
J_{k}^\prime)}+\epsilon_{\ss A (J_1}\epsilon_{\ss |B| J_2}\Psibar{}^{\ss
A^\prime B^\prime J_1^\prime \cdots J_{k}^\prime }_{\ss J_3 \cdots
J_{k})},\eqtag38
$$
where
$$
\Psi{}_{\ss J_1\cdots J_{k+2}}^{\ss J_1^\prime \cdots J_{k-2}^\prime}
=\nabla_{\ss (J_1}^{\ss( J_1^\prime}\cdots\nabla_{\ss J_{k-2}}^{\ss
J^\prime_{k-2})}\Psi_{\ss J_{k-1}J_kJ_{k+1}J_{k+2})},
$$
and
$$
\Psibar{}_{\ss J_1\cdots J_{k-2}}^{\ss J_1^\prime \cdots J_{k+2}^\prime}
=\nabla_{\ss (J_1}^{\ss( J_1^\prime}\cdots\nabla_{\ss J_{k-2})}^{\ss
J^\prime_{k-2}}\Psibar^{\ss J_{k-1}^\prime J_k^\prime J_{k+1}^\prime
J_{k+2}^\prime )}.
$$

In summary, {\it the natural spinor coordinates for the prolonged Einstein
equation manifold ${\cal E}^k$ are}
$$
(x^i,g_{ij},\Gamma^i_{j_0j_1},\ldots,\Gamma^i_{j_0\cdots j_k},\Psi_{\ss
J_1J_2J_3J_4},\Psibar^{\ss J_1^\prime J_2^\prime J_3^\prime J_4^\prime
},\ldots,\Psi_{\ss J_1\cdots J_{k+2}}^{\ss J_1^\prime \cdots J_{k-
2}^\prime},\Psibar_{\ss J_1\cdots J_{k-2}}^{\ss J_1^\prime \cdots
J_{k+2}^\prime}). \eqtag35
$$
As an example, the natural spinor coordinates for ${\cal E}^2$ and ${\cal
E}^3$ are
$$
(x^i,g_{ij},\Gamma^i_{j_0j_1},\Gamma^i_{j_0j_1j_2},\Psi_{\ss
J_1J_2J_3J_4},\Psibar^{\ss J_1^\prime J_2^\prime J_3^\prime J_4^\prime
}),
$$
and
$$
(x^i,g_{ij},\Gamma^i_{j_0j_1},\Gamma^i_{j_0j_1j_2},\Gamma^i_{j_0j_1
j_2j_3},\Psi_{\ss J_1J_2J_3J_4},\Psibar^{\ss J_1^\prime J_2^\prime
J_3^\prime J_4^\prime },\Psi^{\ss J_1^\prime}_{\ss
J_1J_2J_3J_4J_5},\Psibar_{\ss J_1}^{\ss J_1^\prime J_2^\prime J_3^\prime
J_4^\prime J_5^\prime}).
$$

The symmetrized covariant derivatives $\Psi_{\ss J_1\cdots J_{k+2}}^{\ss
J_1^\prime \cdots J_{k-2}^\prime}$ and $\Psibar_{\ss J_1\cdots J_{k-
2}}^{\ss J_1^\prime \cdots J_{k+2}^\prime}$ derive from Penrose's
notion of an exact set of fields for the vacuum Einstein equations
\refto{Penrose1960}.  Henceforth we refer to these spinors as the {\it
Penrose fields} for the vacuum Einstein equations, and we denote them by
$\Psi^k$ and $\Psibar^k$.  We remark that to pass between the coordinates
\eq{35} and \eq{36} we use any soldering form $\sigma_a^{\ss
AA^\prime}$ such that
$$
g_{ij}=\sigma_i^{\ss AA^\prime}\sigma_{j{\ss AA^\prime}}.
$$

We have the following important structure equation for the Penrose fields
\refto{Penrose1960}.

\proclaim Proposition{\statementtag213 }.
The spinorial covariant derivative of $\Psi_{\ss J_1\cdots J_{k+2}}^{\ss
J_1^\prime \cdots J_{k-2}^\prime}$, when evaluated on ${\cal E}^{k+1}$,
is given by
$$
\nabla_{\ss A}^{\ss A^\prime}\Psi_{\ss J_1\cdots J_{k+2}}^{\ss
J_1^\prime \cdots J_{k-2}^\prime}=\Psi_{\ss A\ J_1\cdots J_{k+2}}^{\ss
A^\prime J_1^\prime \cdots J_{k-2}^\prime}+\{\star\},\eqtag39
$$
where $\{\star\}$ denotes a spinor-valued function of the Penrose fields
$\Psi^2,\Psibar^2,\ldots,\Psi^{k-1},\Psibar^{k-1}$.

{\it The fact that the lower-order terms $\{\star\}$ are of order less than
or equal to $k-1$ is essential to much of our symmetry analysis.}  Equation
\eq{39} can be viewed as a special case of \eq{3101}

We now turn our attention to {\it natural tensors}, which we can define
precisely using the language of jets.  If $f\colon M\to M$ is a
diffeomorphism, we let
$$f_*\colon T^p_q(M)\to T^p_q(M)
$$
be the induced map on tensors.  By definition, if $T_x\in T^p_q(M)_x$ and
$y=f(x)$, then for covectors $\alpha_1,\ldots,\alpha_p\in T^*(M)_y$ and
vectors $X_1,\ldots,X_q\in T(M)_y$, $(f_*T)_y$ is the type $(p,q)$ tensor
at $y$ given by
$$
(f_*T)_y(\alpha_1,\ldots,\alpha_p,X_1,\ldots,X_q)
=T_x(f^*\alpha_1,\ldots,f^*\alpha_p,(f^{-1})_*X_1,\ldots,(f^{-1})_*X_q),
$$
where $f^*\alpha_i$ is the pull-back of $\alpha_i$ to $T^*(M)_x$.
Next, we let
$$
\widetilde f\colon \Q\to \Q
$$
be the lift of $f$ to a bundle morphism on $\Q$, that is,
$$
\widetilde f(x,g)=(f(x),f_*g).
$$
The map $\widetilde f$ can then, in turn, be lifted by prolongation to a
bundle morphism on $J^k(\Q)$ \refto{Olver1993}:
$$
{\rm pr}^k\widetilde f\colon J^k(\Q)\to J^k(\Q).
$$

\proclaim Definition{\statementtag214 }.
A natural tensor of type $(p,q)$ and order $k$ on the bundle of quadratic
forms $\Q$ is a smooth bundle map
$$
T\colon J^k(\Q)\to T^p_q(M)
$$
such that for each diffeomorphism $f\colon M\to M$ and every point ${\bf
g}=j^k(g)(x_0)\in J^k(\Q)$,
$$
T\big(({\rm pr}^k\widetilde f)({\bf g})\big)=(f_*T)({\bf g}).
$$

By appealing to the Replacement Theorem of T. Y. Thomas
\refto{Thomas1934}, it can be shown that the restriction to ${\cal E}^k$
of any natural tensor on $J^k(\Q)$, say
$$
T_{a_1\cdots a_p}(g_{ij},g_{ij,j_1},\ldots,g_{ij,j_1\cdots j_k}),
$$
may be uniquely expressed as a function of the Penrose fields, that is,
$$
T_{a_1\cdots a_p}\longleftrightarrow T_{\ss A_1\cdots A_p}^{\ss
A_1^\prime\cdots A_p^\prime}(\Psi_{\ss J_1J_2J_3J_4},\Psibar^{\ss
J_1^\prime J_2^\prime J_3^\prime J_4^\prime },\ldots,\Psi_{\ss J_1\cdots
J_{k+2}}^{\ss J_1^\prime \cdots J_{k-2}^\prime},\Psibar_{\ss J_1\cdots
J_{k-2}}^{\ss J_1^\prime \cdots J_{k+2}^\prime}).\eqtag40
$$
Under an arbitrary $SL(2,{\bf C})$ transformation $\Lambda^{\ss
A}_{\ss B}$, the spinor $T$ satisfies the identity
$$
T_{\ss A_1\cdots A_p}^{\ss A_1^\prime\cdots A_p^\prime}[\Lambda\cdot
\Psi]=\Lambda_{\ss A_1}^{\ss B_1}\cdots \Lambda_{\ss A_p}^{\ss B_p}
\overline\Lambda^{\ss A_1^\prime}_{\ss B_1^\prime}\cdots
\overline\Lambda^{\ss A_p^\prime}_{\ss B_p^\prime}
T_{\ss B_1\cdots B_p}^{\ss B_1^\prime\cdots B_p^\prime}[ \Psi],\eqtag41
$$
where $\Lambda\cdot \Psi$ denotes the action of $SL(2,{\bf C})$ on the
Penrose fields, for example,
$$
[\Lambda\cdot\Psi]_{\ss ABCD} = \Lambda_{\ss A}^{\ss J}\Lambda_{\ss
B}^{\ss K}\Lambda_{\ss C}^{\ss L}\Lambda_{\ss D}^{\ss M}\Psi_{\ss
JKLM}.
$$
We call spinors \eq{40} that satisfy \eq{41} {\it natural spinors of the
Penrose fields} $\Psi^2,\Psibar^2,\ldots,\Psi^k,\Psibar^k$.

\smallskip
We let $\partial_{\Psi}{}^{\ss J_1\cdots J_{k+2}}_{\ss J_1^\prime \cdots
J_{k-2}^\prime}$, $\partial_{\Psibar}{}^{\ss J_1\cdots J_{k-2}}_{\ss
J_1^\prime \cdots J_{k+2}^\prime}$, and
$\partial_{\Gamma}{}_i^{j_0\cdots j_k}$ denote the (symmetrized) partial
differential operators with respect to the coordinates $\Psi{}_{\ss J_1\cdots
J_{k+2}}^{\ss J_1^\prime \cdots J_{k-2}^\prime}$, $\Psibar{}_{\ss
J_1\cdots J_{k-2}}^{\ss J_1^\prime \cdots J_{k+2}^\prime}$, and
$\Gamma^i_{j_0\cdots j_k}$.  For example,
$$
\partial_{\Psi}{}^{\ss J_1J_2J_3J_4}(\Psi_{\ss ABCD})=\delta^{\ss
(J_1}_{\ss A}\delta^{\ss J_2}_{\ss B}\delta^{\ss J_3}_{\ss C}\delta^{\ss
J_4)}_{\ss D}.
$$

As a consequence of \eq{41} we have the following result
\refto{CGT1994a}.

\proclaim Proposition{\statementtag215 }.
Let $T{}_{\ss A_1\cdots A_q}^{\ss A_1^\prime\cdots A_p^\prime}$ be a
natural spinor of the fields $\Psi^2,\Psibar^2,\ldots,\Psi^k,\Psibar^k$.  The
spinorial covariant derivative of $T{}_{\ss A_1\cdots A_q}^{\ss
A_1^\prime\cdots A_p^\prime}$ is a natural spinor of the Penrose fields
$\Psi^2,\Psibar^2,\ldots,\Psi^{k+1},\Psibar^{k+1}$, and is given by
$$
\nabla_{\ss B}^{\ss B^\prime}T{}_{\ss A_1\cdots A_q}^{\ss
A_1^\prime\cdots A_p^\prime}
=\sum_{l=2}^k\big[\partial^{\phantom{\Psi}\ss J_1\cdots J_{l+2}}_{\Psi
\ss J_1^\prime \cdots J_{l-2}^\prime }T{}_{\ss A_1\cdots A_q}^{\ss
A_1^\prime\cdots A_p^\prime}\big]\nabla_{\ss B}^{\ss B^\prime}\Psi_{\ss
J_1\cdots J_{l+2}}^{\ss J_1^\prime\cdots J_{l-2}^\prime}
+\sum_{l=2}^k\big[\partial^{\phantom{\Psibar}\ss J_1\cdots J_{l-
2}}_{\Psibar\ss J_1^\prime \cdots J_{l+2}^\prime }T{}_{\ss A_1\cdots
A_q}^{\ss A_1^\prime\cdots A_p^\prime}\big]\nabla_{\ss B}^{\ss
B^\prime}\Psibar_{\ss J_1\cdots J_{l-2}}^{\ss J_1^\prime\cdots
J_{l+2}^\prime}.
$$

We close this section by deriving a spinor expression for the linearized
Einstein equations \eq{29} that we shall use to compute generalized
symmetries.
Starting from \eq{29}, and using the spinor correspondence
$$
\nabla_c\nabla_d\longleftrightarrow\nabla_{\ss CC^\prime}\nabla_{\ss
DD^\prime}
$$
$$
h_{ab}\longleftrightarrow h_{\ss ABA^\prime B^\prime}
$$
$$
g^{cd}\longleftrightarrow \epsilon^{\ss CD}\epsilon^{\ss C^\prime
D^\prime},
$$
the defining equation \eq{29} takes the form
$$
\eqalign{
[&-\epsilon^{\ss CD}\epsilon^{\ss C^\prime D^\prime }
\delta^{\ss A}_{\ss M}\delta^{\ss A^\prime}_{\ss M^\prime}
\delta^{\ss B}_{\ss N}\delta^{\ss B^\prime}_{\ss N^\prime}
-\epsilon^{\ss AB}\epsilon^{\ss A^\prime B^\prime }
\delta^{\ss C}_{\ss M}\delta^{\ss C^\prime}_{\ss M^\prime}
\delta^{\ss D}_{\ss N}\delta^{\ss D^\prime}_{\ss N^\prime}\cr
&+\epsilon^{\ss AC}\epsilon^{\ss A^\prime C^\prime }
(\delta^{\ss B}_{\ss M}\delta^{\ss B^\prime}_{\ss M^\prime}
\delta^{\ss D}_{\ss N}\delta^{\ss D^\prime}_{\ss N^\prime}
+\delta^{\ss D}_{\ss M}\delta^{\ss D^\prime}_{\ss M^\prime}
\delta^{\ss B}_{\ss N}\delta^{\ss B^\prime}_{\ss N^\prime})]
\nabla_{\ss CC^\prime}\nabla_{\ss DD^\prime}
h_{\ss ABA^\prime B^\prime}=0.}\eqtag43
$$
Since $h_{\ss ABA^\prime B^\prime}=h_{\ss BAB^\prime A^\prime}$, we
have that
$$
h_{\ss ABA^\prime B^\prime}=h_{\ss BAA^\prime B^\prime}
+{1\over2}\epsilon_{\ss AB}\epsilon_{\ss A^\prime B^\prime }h,\eqtag42
$$
where the trace $h$ of $h_{\ss ABA^\prime B^\prime}$ is given by
$$
h=\epsilon^{\ss AB}\epsilon^{\ss A^\prime B^\prime }h_{\ss ABA^\prime
B^\prime}.
$$
Substituting \eq{42} into the last two terms of \eq{43}, we find that all the
trace terms cancel leaving us with
$$
[-\epsilon^{\ss CD}\epsilon^{\ss C^\prime D^\prime }
\delta^{\ss A}_{\ss M}\delta^{\ss A^\prime}_{\ss M^\prime}
\delta^{\ss B}_{\ss N}\delta^{\ss B^\prime}_{\ss N^\prime}
+\epsilon^{\ss BC}\epsilon^{\ss A^\prime C^\prime }
\delta^{\ss A}_{\ss M}\delta^{\ss B^\prime}_{\ss M^\prime}
\delta^{\ss D}_{\ss N}\delta^{\ss D^\prime}_{\ss N^\prime}
+\epsilon^{\ss BC}\epsilon^{\ss A^\prime C^\prime }
\delta^{\ss D}_{\ss M}\delta^{\ss D^\prime}_{\ss M^\prime}
\delta^{\ss A}_{\ss N}\delta^{\ss B^\prime}_{\ss N^\prime}]
\nabla_{\ss CC^\prime}\nabla_{\ss DD^\prime}
h_{\ss ABA^\prime B^\prime}=0.
$$
We now multiply this expression with arbitrary spinors $\alpha^{\ss M}$,
$\alphabar^{\ss M^\prime}$, $\beta^{\ss N}$, $\betabar^{\ss N^\prime}$
to get our final spinor form of the linearized equations.

\proclaim Proposition{\statementtag217 }.
If $\hab$ are the spinor components of a generalized symmetry of the
vacuum Einstein equations, then for all spinors $\alpha^{\ss M}$,
$\alphabar^{\ss M^\prime}$, $\beta^{\ss N}$, $\betabar^{\ss N^\prime}$
$$
\eqalign{
[-\epsilon_{\ss CD}\epsilon^{\ss C^\prime D^\prime}\alpha_{\ss A}
\beta_{\ss B}\alphabar^{\ss A^\prime}\betabar^{\ss B^\prime}
&+\epsilon_{\ss BC}\epsilon^{\ss A^\prime C^\prime}\alpha_{\ss A}
\beta_{\ss D}\alphabar^{\ss B^\prime}\betabar^{\ss D^\prime}\cr
&+\epsilon_{\ss BC}\epsilon^{\ss A^\prime C^\prime}\alpha_{\ss D}
\beta_{\ss A}\alphabar^{\ss D^\prime}\betabar^{\ss B^\prime}]
\nabla_{\ss C^\prime}^{\ss C}\nabla_{\ss D^\prime}^{\ss D}\hab
=0\qquad{\rm on}\  {\cal E}^{k+2}.}\eqtag44
$$
In general $\hab$ is a function of the coordinates \eq{35}, that is,
$$
\hab=\hab(x^i,\sigma^a_{\ss
AB},\Gamma^i_{j_0j_1},\ldots,\Gamma^i_{j_0\cdots j_k},\Psi_{\ss
J_1J_2J_3J_4},\Psibar^{\ss J_1^\prime J_2^\prime J_3^\prime J_4^\prime
},\ldots,\Psi_{\ss J_1\cdots J_{k+2}}^{\ss J_1^\prime \cdots J_{k-
2}^\prime},\Psibar_{\ss J_1\cdots J_{k-2}}^{\ss J_1^\prime \cdots
J_{k+2}^\prime}).
$$
When $\hab$ is a natural generalized symmetry,
$$
\hab=\hab(\Psi_{\ss J_1J_2J_3J_4},\Psibar^{\ss J_1^\prime J_2^\prime
J_3^\prime J_4^\prime },\ldots,\Psi_{\ss J_1\cdots J_{k+2}}^{\ss
J_1^\prime \cdots J_{k-2}^\prime},\Psibar_{\ss J_1\cdots J_{k-2}}^{\ss
J_1^\prime \cdots J_{k+2}^\prime}).
$$
In both cases, $\hab$ satisfies the $SL(2,{\bf C})$ invariance properties
$$
\Lambda_{\ss A}^{\ss C}\Lambda_{\ss B}^{\ss D}\Lambda^{\ss
A^\prime}_{\ss C^\prime}\Lambda^{\ss B^\prime}_{\ss D^\prime}
\hab(x,\sigma,\Gamma,\Psi)
=h^{\ss C\kern 0.5pt D}_{\ss C^{\kern -.8pt\prime}\kern -1.8pt D^{\kern
-.8pt\prime}}(x,\Lambda\cdot\sigma,\Gamma,\Lambda\cdot\Psi),
$$
where $\Lambda\cdot\sigma$, and $\Lambda\cdot\Psi$ denote the action of
$SL(2,{\bf C})$ on the soldering form and Penrose fields.

\vskip0.2truein
\noindent{\bf 2D. Results from Tensor and Spinor Algebra.}

Here we gather together a number of key algebraic results which we shall
use repeatedly in our study of the generalized symmetries of the Einstein
equations.  Following the standard algebraic treatment of tensors, we
consider spinors as multi-linear maps on complex 2-dimensional vector
spaces.
For notational convenience, we
separate groups of symmetric spinor (or tensor) arguments with a comma
and we use
no delimiters between arguments within a symmetric set.  As an example,
$A(\alpha\,\beta,\gamma,\delta)$ denotes a rank 4 spinor that is symmetric
in
$\alpha$ and $\beta$,
$$
A(\alpha\,\beta,\gamma,\delta)=A(\beta\,\alpha,\gamma,\delta),
$$
but otherwise has no symmetries.  Repeated symmetric arguments of a
spinor (or tensor) will be abbreviated using an exponential notation.  For
example, if
$T$ is a spinor of type $(k,1)$ that is totally symmetric in its first $k$
arguments, we will write
$$
T(\psi^k,\alphabar)=T(\underbrace{\psi,\dots,\psi}_{k\ times},\alphabar).
$$
It is important to note that the values of
$T(\psi_1\psi_2\cdots\psi_k,\alphabar)$, where $\psi_1$, $\psi_2$, \dots,
$\psi_k$ are arbitrary spinors, are completely determined by the values of
$T(\psi^k,\alphabar)$.
In addition, if $\alpha_{\ss B}$ and $\beta^{\ss A}$ are spinors of type
$(0,1)$ and type $(1,0)$ respectively, we set
$$
\beta_{\ss B}=\epsilon_{\ss AB}\beta^{\ss A}\quad{\rm and}\quad
\alpha^{\ss A}=\epsilon^{\ss AB}\alpha_{\ss B}.
$$
The skew-symmetric inner product between $\alpha_{\ss B}$ and
$\beta^{\ss A}$ is given by
$$
<\alpha,\beta>=\alpha_{\ss A}\beta^{\ss A}=\epsilon^{\ss AB}\alpha_{\ss
A}\beta_{\ss B} = -<\beta,\alpha>.
$$
We denote by $<X,Y>$ the metric inner product between two vectors $X$
and $Y$.

The following propositions are all elementary facts which we shall use
repeatedly.  See \refto{CGT1994a} for proofs.
\proclaim Proposition{\statementtag201 }.
Let $P=P(\psi^k,\alpha)$ be a rank $(k+1)$ spinor that is
symmetric in its first $k$ arguments.  Then there are unique, totally
symmetric spinors $P^*$ and $Q$, of rank $k+1$ and $k-1$
respectively, such that
$$
P(\psi^ k,\alpha)=P^*(\psi^k\alpha)\ + <\psi,\alpha>Q(\psi^{k-1}).
$$
If $P$ is a natural spinor of the Penrose fields $\Psi^2$, $\Psibar^2$,
\dots,$\Psi^k$, $\Psibar^k$, then so are $P^*$ and $Q$.

\proclaim Proposition{\statementtag202 }.
Let $P=P(\psi^k,\alpha)$ be a rank $(k+1)$ spinor that is
symmetric in its first $k$ arguments.  If $P(\psi^k,\alpha)$ satisfies
$$
P(\psi^k,\psi)=0,\eqtag202
$$
then there is a totally symmetric spinor $Q=Q(\psi^{k-1})$ such that
$$
P(\psi^k,\alpha)=<\psi,\alpha>Q(\psi^{k-1}).\eqtag203
$$
If $P$ is a natural spinor, then so is $Q$.

\noindent We note for future use that \eq{203} is equivalent to
$$
P(\psi^1\cdots\psi^k,\alpha)={1\over
k}\sum_{i=1}^k<\psi^i,\alpha>Q(\psi^1\cdots\psi^{i-
1}\psi^{i+1}\cdots\psi^k).\eqtag2031
$$

\proclaim Proposition{\statementtag203 }.
Let $P=P(\psi^k,\alpha)$ be a rank $(k+1)$ spinor that is
symmetric in its first $k$ arguments.  If $P(\psi^k,\alpha)$ satisfies
$$
<\psi,\alpha>P(\psi^k,\beta)=<\psi,\beta>P(\psi^k,\alpha),\eqtag24
$$
then there is a unique totally symmetric spinor $Q$ of rank $k-1$ such that
$$
P(\psi^k,\alpha)=<\psi,\alpha>Q(\psi^{k-1}).\eqtag25
$$
The spinor $Q$ is natural if $P$ is natural.
If $P(\psi^k,\alpha)$ satisfies
$$
<\psi,\alpha>P(\psi^k,\beta)=-<\psi,\beta>P(\psi^k,\alpha),
\eqtag2051
$$
then $P=0$.

\proclaim Proposition{\statementtag2044 }.
Let $T$ be a symmetric rank-$k$ tensor, and suppose that
$$
T(X^k)=0
$$
whenever $X$ is a null vector.  Then there exists a symmetric tensor $P$
of rank $k-2$ such that, for any vector $X$,
$$
T(X^k) = <X,X>P(X^{k-2}).\eqtag4100
$$

\proclaim Proposition{\statementtag 2041 }.
Let $T(Y^{p},X)$ be a tensor that vanishes whenever $<Y,X>=0$.  Then
there is a unique tensor $U(Y^{p-1})$ such that
$$
T(Y^{p},X)=<Y,X>U(Y^{p-1}).\eqtag1110
$$

We close this section with a characterization of spinors with certain
symmetries which arise in our symmetry analysis of the Einstein equations.
The proof of this theorem is rather lengthy; for details, see
\refto{CGT1994a}.

\proclaim Theorem{\statementtag405 }.
Let $P(\psi^{k+2},\psibar^{k-2},\alpha,\beta,\alphabar,\betabar)$ be a
spinor that is symmetric in its first $k+2$ and next $k-2$ arguments.
The spinor $P(\psi^{k+2},\psibar^{k-2},\alpha,\beta,\alphabar,\betabar)$
enjoys the two symmetry properties
$$
P(\psi^{k+2},\psibar^{k-2},\alpha,\beta,\alphabar,\betabar)=
P(\psi^{k+2},\psibar^{k-2},\beta,\alpha,\betabar,\alphabar)\eqtag4102
$$
and
$$
P(\psi^{k+2},\psibar^{k-2},\psi,\alpha,\betabar,\psibar)=0\eqtag4103
$$
if and only if there are spinors,
$$
A=A(\psi^{k},\psibar^{k}),\quad
B=B(\psi^{k+4},\psibar^{k-4}),\quad
W=W(\psi^{k+1},\psibar^{k-3},\alpha,\alphabar),\eqtag3041
$$
such that
$$\eqalign{
&P(\psi^{k+2},\psibar^{k-2},\alpha,\beta,\alphabar,\betabar)\cr\eqskip
&=<\psi,\alpha><\psi,\beta>A(\psi^{k},\psibar^{k-2}\alphabar\betabar)
+<\psibar,\alphabar><\psibar,\betabar>B(\psi^{k+2}\alpha\beta,\psibar^{k-
4})\cr\eqskip
&+<\psi,\alpha><\alphabar,\psibar>W(\psi^{k+1},\psibar^{k-
3},\beta,\betabar) \ + <\psi,\beta><\betabar,\psibar>
W(\psi^{k+1},\psibar^{k-3},\alpha,\alphabar).}\eqtag305
$$
The spinor $A$ is symmetric in its first $k$ and last $k$ arguments; the
spinor $B$ is symmetric in its first $k+4$ and last $k-4$ arguments; and
the spinor $W$ is symmetric in its first $k+1$ and following $k-3$
arguments.  With these symmetries, the spinors $A, B, W$ are uniquely
determined by $P$.  When $k=3$, \eq{305} is valid with $B=0$ and
$W=W(\psi^{4},\alpha,\alphabar)$.  When $k=2$, \eq{305} holds with
$B=0$ and $W=0$.

\vfill\eject

%
\sectionno3
\abovedisplayskip=12pt plus 1pt minus 3pt
\belowdisplayskip=12pt plus 1pt minus 3pt
\def\ss{\scriptscriptstyle}
\def\proof{\par\noindent {\bf Proof:\ }}
\def\eqskip{\noalign{\vskip6pt}}
\def\Psibar{{\overline\Psi}{}}
\def\psibar{\overline\psi{}}
\def\phibar{\overline\phi{}}
\def\betabar{\overline\beta{}}
\def\alphabar{\overline\alpha{}}
\def\chibar{\overline\chi{}}

\def\hab{h^{\ss A\kern 0.5pt B}_{\ss A^{\kern -.8pt\prime}\kern -1.8pt
B^{\kern -.8pt\prime}}}
\def\dab{d^{\ss A\kern 0.5pt B}_{\ss A^{\kern -.8pt\prime}\kern -1.8pt
B^{\kern -.8pt\prime}}}
\def\kab{k^{\ss A\kern 0.5pt B}_{\ss A^{\kern -.8pt\prime}\kern -1.8pt
B^{\kern -.8pt\prime}}}
\overfullrule=0pt
\def\lineq{\eq{300}}
\noindent{\bf 3. Natural Generalized Symmetries of the Vacuum Einstein
Equations.}

In this section we obtain a complete classification of all natural generalized
symmetries
of the vacuum Einstein equations, that is, we find all solutions to the
linearized equations
$$
\eqalign{
[-\epsilon_{\ss CD}\epsilon^{\ss C^\prime D^\prime}\alpha_{\ss A}
\beta_{\ss B}\alphabar^{\ss A^\prime}\betabar^{\ss B^\prime}
&+\epsilon_{\ss BC}\epsilon^{\ss A^\prime C^\prime}\alpha_{\ss A}
\beta_{\ss D}\alphabar^{\ss B^\prime}\betabar^{\ss D^\prime}\cr
&+\epsilon_{\ss BC}\epsilon^{\ss A^\prime C^\prime}\alpha_{\ss D}
\beta_{\ss A}\alphabar^{\ss D^\prime}\betabar^{\ss B^\prime}]
\nabla_{\ss C^\prime}^{\ss C}\nabla_{\ss D^\prime}^{\ss D}\hab
=0,}\eqtag300
$$
where
$$
\hab=\hab(\Psi^2,\Psibar^2,\Psi^3,\Psibar^3,\dots,\Psi^k,\Psibar^k)
$$
is a natural spinor depending upon the Penrose fields to order $k$.
Equation \eq{300} and all subsequent equations in this section hold by
virtue of the Einstein equations and their derivatives.

Before beginning the detailed analysis of \lineq, let us outline the principal
steps.  Since $\hab$ is assumed to be of order $k$, the linearized equation
is an identity to order $k+2$ in the Penrose fields.  It is easy to see that
this
identity can be written symbolically as
$$
\alpha \Psi^{k+2} + \beta\Psibar^{k+2} + \gamma\Psi^{k+1}\Psi^{k+1}
+\delta\Psi^{k+1}\Psibar^{k+1} + \epsilon\Psibar^{k+1}\Psibar^{k+1}
+\rho\Psi^{k+1}+\tau\Psibar^{k+1}+\upsilon=0,\eqtag3000
$$
where the coefficients $\alpha$, $\beta$, \dots, $\upsilon$ are complicated
expressions of order $k$ involving $\hab$ and its repeated derivatives with
respect to $\Psi^2$, $\Psibar^2$, \dots $\Psi^k$, $\Psibar^k$.  Each of the
coefficients $\alpha$, $\beta$, \dots, $\upsilon$ must vanish identically
because the fields $\Psi^{k+2}$,
$\Psibar^{k+2}$,$\Psi^{k+1}$,$\Psibar^{k+1}$ may be freely specified on
${\cal E}^{k+2}$.  As is standard practice in symmetry group analysis, we
analyze this complicated identity beginning with the highest-order
conditions $\alpha=0$ and $\beta=0$.

Let $\partial^k_\Psi h$ and $\partial^k_\Psibar h$ denote the partial
derivatives of $\hab$ with respect to $\Psi^k$ and $\Psibar^k$.  The
conditions $\alpha=0$ and $\beta=0$ impose certain algebraic conditions on
the spinors $\partial^k_\Psi h$ and $\partial^k_\Psibar h$ which, when
carefully analyzed, lead to unique spinor decompositions that we shall write
symbolically as
$$
\partial^k_\Psi h=A+B+W\qquad{\rm and}\qquad
\partial^k_\Psibar h=D+E+U.\eqtag3001
$$
This we do in \S3A; see Propositions 3.4 and 3.5.  Each term $A$,
$B$,\dots, $U$ in these decompositions separately satisfies the algebraic
conditions arising from $\alpha=0$ and $\beta=0$.  In \S3B we show that
the vanishing of the coefficients $\gamma$, $\delta$, $\epsilon$ force
$\hab$ to be linear in the highest-order Penrose fields $\Psi^k$ and
$\Psibar^k$, so that the spinors $A$, $B$,\dots, $U$ in the representation
\eq{3001} are all at most of order $k-1$.  The analysis of the conditions
$\rho=0$ and $\tau=0$ is accomplished in two steps.  In \S3C we prove that
$A$, $B$, $D$, $E$ must actually be of order $k-2$, and that there is a
generalized natural vector field
$$
X^{\ss A}_{\ss A^\prime}=X^{\ss A}_{\ss
A^\prime}(\Psi^2,\Psibar^2,\dots,\Psi^{k-1},\Psibar^{k-1})
$$
such that
$$
W=\partial^{k-1}_\Psi X\quad{\rm and}\quad U=\partial^{k-1}_\Psibar X.
$$
We let
$$
\kab=\hab-(\nabla^{\ss A}_{\ss A^\prime}X^{\ss B}_{\ss B^\prime} +
\nabla^{\ss B}_{\ss B^\prime}X^{\ss A}_{\ss A^\prime}).
$$
Then $\kab$ satisfies \eq{3000} and \eq{3001} with $W=0$ and $U=0$.
In \S3D we find that the remaining coefficients $A$, $B$, $D$, $E$ in
\eq{3001} now satisfy certain covariant constancy conditions, from which
it readily follows that $A=B=D=E=0$.  The classification of the natural
generalized symmetries of the Einstein equations is then completed by a
simple induction argument.

\vskip0.2truein
\noindent{\bf  Notation and Commutation Rules.}

We begin by fixing some notation.  If
$$
T{}^{\ss C_1\dots C_p}_{\ss C_1^\prime\dots C_q^\prime}=T{}^{\ss
C_1\dots C_p}_{\ss C_1^\prime\dots C_q^\prime}(\Psi^2,
\Psibar^2,\dots,\Psi^k,\Psibar^k)
$$
is a natural spinor of type $(p,q)$ and order $k$, then the partial
derivative of
$T{}^{\ss C_1\dots C_p}_{\ss C_1^\prime\dots C_q^\prime}$ with respect
to $\Psi^l$ is a natural spinor of type $(p+l+2, q+l-2)$.  We shall write
$$
[\partial^l_\Psi T{}^{\ss C_1\cdots C_p}_{\ss C_1^\prime\dots
C_q^\prime}](\psi^1\cdots\psi^{l+2},\psibar_1\cdots\psibar_{l-2})
=[\partial_\Psi{}_{\ss A_1^\prime\dots A_{l-2}^\prime}^{\ss A_1\dots
A_{l+2}}\,T{}^{\ss C_1\dots C_p}_{\ss C_1^\prime\dots C_q^\prime}]
\psi^1_{\ss A_1}\cdots\psi^{l+2}_{\ss A_{l+2}}\psibar_1^{\ss
A^\prime_1}\cdots\psibar_{l-2}^{\ss A^\prime_{l-2}}.\eqtag301
$$

Further, let $\phi^1,\dots,\phi^p$ and $\phibar_1,\dots,\phibar_q$ be
arbitrary spinors of type $(1,0)$ and $(0,1)$ respectively; we shall write
$$
[\partial^l_\Psi T](\psi^{l+2},\psibar^{l-
2};\phi^1,\dots,\phi^p,\phibar_1,\dots,
\phibar_q)=[\partial^l_\Psi T{}^{\ss C_1\dots C_p}_{\ss C_1^\prime\dots
C_q^\prime}](\psi^{l+2},\psibar^{l-2}) \phi^1_{\ss C_1}\cdots\phi^p_{\ss
C_p}
\phibar_1^{\ss C^\prime_1}\cdots\phibar_q^{\ss C^\prime_q}.
$$
A semi-colon will always be used to separate arguments corresponding to
derivatives with respect to the coordinates \equationlabel{2}{35}{}.
Partial derivatives with respect to $\Psibar^{\ss A_1\dots A_{l-2}}_{\ss
A_1^\prime\dots A_{l+2}^\prime}$ will be similarly denoted.

We shall repeatedly need certain commutation relations between the partial
derivative operators $\partial_\Psi{}^{\ss A_1\dots A_{m+2}}_{\ss
A^\prime_1\dots A^\prime_{m-2}}$ and $\partial_\Psibar{}^{\ss A_1\dots
A_{m-2}}_{\ss A^\prime_1\dots A^\prime_{m+2}}$ and the covariant
derivative operator $\nabla^{\ss C}_{\ss C^\prime}$.

\proclaim Proposition{\statementtag301 }.
Let
$$
T{}^{\cdots}_{\cdots} =
T{}^{\cdots}_{\cdots}(\Psi^2,\Psibar^2,\dots,\Psi^m,\Psibar^m)
$$
be a natural spinor of order $m$.  Then
$$
[\partial_\Psi^{m+1}\nabla^{\ss C}_{\ss C^\prime}T{}^{\cdots}_{\cdots}]
(\psi^{m+3},\psibar^{m-1})
=\psi^{\ss C}\psibar_{\ss C^\prime}[\partial_\Psi^mT{}^{\cdots}_{\cdots}]
(\psi^{m+2},\psibar^{m-2}),\eqtag3011
$$
and
$$
[\partial_\Psi^{m}\nabla^{\ss C}_{\ss C^\prime}T{}^{\cdots}_{\cdots}]
(\psi^{m+2},\psibar^{m-2})=
[\nabla^{\ss C}_{\ss C^\prime}\partial_\Psi^{m}T{}^{\cdots}_{\cdots}]
(\psi^{m+2},\psibar^{m-2})+
\psi^{\ss C}\psibar_{\ss C^\prime}[\partial_\Psi^{m-
1}T^{\cdots}_{\cdots}]
(\psi^{m+1},\psibar^{m-3}),\eqtag3012
$$
and similarly,
$$
[\partial^{m+1}_\Psibar\nabla^{\ss C}_{\ss
C^\prime}T{}^{\cdots}_{\cdots}]
(\psi^{m-1},\psibar^{m+3})
=\psi^{\ss C}\psibar_{\ss C^\prime}[\partial^m_\Psibar
T{}^{\cdots}_{\cdots}]
(\psi^{m-2},\psibar^{m+2}),\eqtag3013
$$
and
$$
[\partial^{m}_\Psibar\nabla^{\ss C}_{\ss C^\prime}T{}^{\cdots}_{\cdots}]
(\psi^{m-2},\psibar^{m+2})
=[\nabla^{\ss C}_{\ss C^\prime}\partial^{m}_\Psibar
T{}^{\cdots}_{\cdots})]
(\psi^{m-2},\psibar^{m+2})+
\psi^{\ss C}\psibar_{\ss C^\prime}[\partial^{m-1}_\Psibar
T^{\cdots}_{\cdots}]
(\psi^{m-3},\psibar^{m+1}).\eqtag3014
$$

\proof These formulas follow directly from Proposition
\statementlabel{2}{215} and the structure equations
\equationlabel{2}{39}{}.\endEx

As an application of these commutation relations, we prove a proposition
that we shall need later.

\proclaim Proposition\statementtag308 .
Let
$$
P{}^{\ss A_1\cdots A_r}_{\ss B_1^\prime \cdots B_s^\prime}=
P{}^{\ss A_1\cdots A_r}_{\ss B_1^\prime \cdots
B_s^\prime}(\Psi^2,\Psibar^2,\dots,\Psi^k,\Psibar^k)
$$
be a natural spinor that is completely symmetric in the indices $A_1\dots
A_r$ and $B_1^\prime\dots B_s^\prime$.  If
$$
\nabla^{\ss (C}_{\ss (C^\prime}
P{}^{\ss A_1\cdots A_r)}_{\ss B_1^\prime \cdots B_s^\prime)}
=0\qquad{\rm on}\ {\cal E}^{k+1},\eqtag353
$$
then $P$ vanishes.

\proof
Equation \eq{353} is equivalent to
$$
[{\rm Grad}\ P](\alpha,\alphabar;\alpha^r,\alphabar^s)=0,\eqtag1
$$
where we have introduced the notation
$$
[{\rm Grad}\  P](\beta,\betabar;\alpha^r,\alphabar^s)=
\beta_{\ss A}\betabar^{\ss A^\prime}[\nabla^{\ss A}_{\ss A^\prime}
\,P](\alpha^r,\alphabar^s).\eqtag35001
$$

We differentiate \eq{1} with respect to $\Psi^{k+1}$ and use the
commutation relation \eq{3011} to deduce that
$$
[\partial_\Psi^kP](\psi^{k+2},\psibar^{k-
2};\alpha^r,\alphabar^s)=0.\eqtag354
$$
Similarly, if we differentiate with respect to $\Psibar^{k+1}$ we find that
$$
[\partial^k_\Psibar P](\psi^{k-
2},\psibar^{k+2};\alpha^r,\alphabar^s)=0.\eqtag355
$$
Equations \eq{354} and \eq{355} show $P$ to be independent of $\Psi^k$
and $\Psibar^k$.  A simple induction argument proves that $P$ is
independent of all the Weyl spinors
$\Psi^k,\Psibar^k,\dots,\Psi^2,\Psibar^2$.

The expansion of \eq{353} in terms of the spinor connection coefficients
$\gamma_{\ss C^\prime B}^{\ss CA}$ and $\gamma_{\ss C^\prime
B^\prime}^{\ss CA^\prime}$ now leads to
$$
\gamma^{\ss (CA_1}_{\ss (C^\prime|D|}
P^{\ss |D|A_2\dots A_r)}_{\ss B_1^\prime B_2^\prime\dots B^\prime_s)}
-\overline\gamma^{\ss (C|D^\prime|}_{\ss (C^\prime B_1^\prime}
P^{\ss A_1A_2\dots A_r)}_{\ss |D^\prime| B_2^\prime\dots
B^\prime_s)}=0.
$$
This is an identity that must hold for all spinor connection coefficients and
therefore, taking into account the identity
$$
\gamma^{\ss CA}_{\ss C^\prime D}\epsilon_{\ss AB}
+\gamma^{\ss CA}_{\ss C^\prime B}\epsilon_{\ss DA}=0,
$$
we conclude that
$$
<\alpha,\beta>P(\gamma\alpha^{r-1},\alphabar^s)
+<\alpha,\gamma>P(\beta\alpha^{r-1},\alphabar^s)
=0.
$$
Setting $\beta=\gamma$ we conclude that
$$
P(\alpha^r,\alphabar^s)=0.
$$
Alternatively, one may conclude that $P=0$ from the fact that there are no
completely symmetric natural spinors of order zero.
\endEx

\vskip0.2truein
\noindent{\bf 3A. The $\Psi^{k+2}$ and $\Psibar^{k+2}$ Analysis.}

We suppose that $\hab$ is a natural generalized symmetry of the vacuum
Einstein equations of order $k$:
$$
\hab = \hab(\Psi^2,\Psibar^2,\dots,\Psi^k,\Psibar^k).
$$
In this section we derive necessary and sufficient conditions for the
vanishing of the coefficients $\alpha$ and $\beta$ in \eq{3000}, and we
analyze these conditions in detail.

We have, by two applications of \eq{3011}
$$
\eqalign{
[\partial_\Psi^{k+2}\nabla^{\ss C}_{\ss C^\prime}\nabla^{\ss D}_{\ss
D^\prime}\hab](\psi^{k+4},\psibar^{k})
&=\psi^{\ss C}\psibar_{\ss C^\prime}[\partial_\Psi^{k+1}\nabla^{\ss
D}_{\ss D^\prime}\hab](\psi^{k+3},\psibar^{k-1})\cr
\eqskip
&=\psi^{\ss C}\psi^{\ss D}\psibar_{\ss C^\prime}\psibar_{\ss
D^\prime}[\partial_\Psi^{k}\hab](\psi^{k+2},\psibar^{k-2}).}
$$
Therefore, if we differentiate equation \eq{300} with respect to
$\Psi^{k+2}$ it follows that
$$
\eqalignno{
&<\beta,\psi><\betabar,\psibar>[\partial^k_\Psi h](\psi^{k+2},\psibar^{k-
2};\alpha,
\psi,\psibar,\alphabar)\cr
\eqskip
+&<\alpha,\psi><\alphabar,\psibar>[\partial_\Psi^kh](\psi^{k+2},\psibar^{k
-2};\beta,
\psi,\psibar,\betabar)=0.\aleqtag302 }
$$
Similarly, we differentiate the linearized equations \eq{300} with respect
to $\Psibar^{k+2}$ and use \eq{3013} to find
$$
\eqalignno{
&<\beta,\psi><\betabar,\psibar>[\partial^k_\Psibar h](\psi^{k-
2},\psibar^{k+2};\alpha,
\psi,\psibar,\alphabar)\cr \eqskip+
&<\alpha,\psi><\alphabar,\psibar>[\partial^k_\Psibar h](\psi^{k-
2},\psibar^{k+2};\beta,
\psi,\psibar,\betabar)=0.\aleqtag3002 }
$$

\proclaim Proposition{\statementtag302 }.
If $\hab$ is a natural generalized symmetry of order $k$ for the vacuum
Einstein equations, then
$$
[\partial_\Psi^kh](\psi^{k+2},\psibar^{k-2};\psi,
\alpha,\alphabar,\psibar)=0\eqtag303
$$
and
$$
[\partial^k_\Psibar h](\psi^{k-2},\psibar^{k+2};\psi,
\alpha,\alphabar,\psibar)=0.\eqtag304
$$

\proof
In equation \eq{302} we set $\alpha=\beta$ and $\alphabar=\betabar$ to
deduce that
$$
[\partial_\Psi^kh](\psi^{k+2},\psibar^{k-2};\alpha,\psi,
\psibar,\alphabar)=0.
$$
The symmetry $h_{\ss ABA^\prime B^\prime}=h_{\ss BAB^\prime
A^\prime}$ then leads to \eq{303}.
In equation \eq{3002} we set $\alpha=\beta$ and $\alphabar=\betabar$, and
then use the symmetry of $h_{\ss ABA^\prime B^\prime}$ to arrive at
\eq{304}.  Note that \eq{303} and \eq{304}  are necessary and sufficient
for \eq{302} and \eq{3002} to hold.
\endEx
Theorem \statementlabel{2}{405} allows us to explicitly characterize all
natural spinors that satisfy \eq{303} and \eq{304}.

\proclaim Proposition{\statementtag303 }.
The spinor $[\partial_\Psi^kh](\psi^{k+2},\psibar^{k-
2};\alpha,\beta,\alphabar,\betabar)$ satisfies the symmetry conditions
\eq{303} if and only if there are natural spinors,
$$
A=A(\psi^{k},\psibar^{k}),\quad
B=B(\psi^{k+4},\psibar^{k-4}),\quad
W=W(\psi^{k+1},\psibar^{k-3},\alpha,\alphabar),\eqtag3041
$$
such that
$$\eqalignno{
[&\partial_\Psi^kh](\psi^{k+2},\psibar^{k-
2};\alpha,\beta,\alphabar,\betabar)\aleqtag305
&=<\psi,\alpha><\psi,\beta>A(\psi^{k},\psibar^{k-2}\alphabar\betabar)
+<\psibar,\alphabar><\psibar,\betabar>B(\psi^{k+2}\alpha\beta,\psibar^{k-
4})\cr\eqskip
&+<\psi,\alpha><\alphabar,\psibar>W(\psi^{k+1},\psibar^{k-
3},\beta,\betabar) \ + <\psi,\beta><\betabar,\psibar>
W(\psi^{k+1},\psibar^{k-3},\alpha,\alphabar).}
$$
The spinor $A$ is symmetric in its first $k$ and last $k$ arguments; the
spinor $B$ is symmetric in its first $k+4$ and last $k-4$ arguments; and
the spinor $W$ is symmetric in its first $k+1$ and following $k-3$
arguments.  With these symmetries, the spinors $A, B, W$ are uniquely
determined by $\partial_\Psi^kh$.  When $k=3$, \eq{305} is valid with
$B=0$ and $W=W(\psi^{4},\alpha,\alphabar)$.  When $k=2$, \eq{305}
holds with $B=0$ and $W=0$.

Let us remark that \eq{305} contains the algebraic form of the generalized
diffeomorphism symmetry. Indeed, if
$$
X^{\ss A}_{\ss A^\prime}=X^{\ss A}_{\ss
A^\prime}(\Psi^2,\Psibar^2,\dots,\Psi^{k-1},\Psibar^{k-1})
$$
is the spinor form of a natural vector field of order $k-1$, and we let
$$
d^{\ss A\kern 0.5pt B}_{\ss A^{\kern -.8pt\prime}\kern -1.8pt B^{\kern -
.8pt\prime}}=\nabla_{\ss A^\prime}^{\ss A}
X^{\ss B}_{\ss B^\prime}+\nabla_{\ss B^\prime}^{\ss B}X^{\ss A}_{\ss
A^\prime},
$$
then, by \eq{3011},
$$
\eqalign{
[\partial_\Psi^kd](\psi^{k+2},\psibar^{k-2};\alpha,\beta,\alphabar,\betabar)
=&<\psi,\alpha><\alphabar,\psibar>[\partial_\Psi^{k-
1}X](\psi^{k+1},\psibar^{k-3};\beta,\betabar)\cr +
&<\psi,\beta><\betabar,\psibar> [\partial_\Psi^{k-
1}X](\psi^{k+1},\psibar^{k-3};\alpha,\alphabar).}\eqtag306
$$
We observe that with $W=\partial_\Psi^{k-1}X$ the right hand side of
\eq{306} coincides with the expression involving $W$ in \eq{305}.  In
\S3C we shall prove $W$ satisfies integrability conditions that imply
$W=\partial_\Psi^{k-1}X$.

There is an analogous decomposition for $\partial^k_\Psibar h$.

\proclaim Proposition{\statementtag304 }.
The spinor $[\partial^k_\Psibar h](\psi^{k-
2},\psibar^{k+2};\alpha,\beta,\alphabar,\betabar)$ satisfies the symmetry
conditions \eq{304} if and only if there are natural spinors,
$$
D=D(\psibar^{k},\psi^{k}),\quad
E=E(\psibar^{k+4},\psi^{k-4}),\quad
U=U(\psibar^{k+1},\psi^{k-3},\alpha,\alphabar),\eqtag3211
$$
such that
$$\eqalignno{
[&\partial^k_\Psibar h](\psi^{k-
2},\psibar^{k+2};\alpha,\beta,\alphabar,\betabar)\aleqtag322
&=<\psibar,\alphabar><\psibar,\betabar>D(\psibar^{k},\psi^{k-
2}\alpha\beta)
+<\psi,\alpha><\psi,\beta>E(\psibar^{k+2}\alphabar\betabar,\psi^{k-
4})\cr\eqskip
+&<\psibar,\alphabar><\alpha,\psi>U(\psibar^{k+1},\psi^{k-
3},\beta,\betabar) + <\psibar,\betabar><\beta,\psi>
U(\psibar^{k+1},\psi^{k-3},\alpha,\alphabar).}
$$
The spinor $D$ is symmetric in its first $k$ and last $k$ arguments; the
spinor $E$ is symmetric in its first $k+4$ and last $k-4$ arguments; and
the spinor $U$ is symmetric in its first $k+1$ and following $k-3$
arguments.  With these symmetries the spinors $D, E, U$ are unique.
When $k=3$, \eq{322} is valid with $E=0$ and
$U=U(\psibar^{4},\alpha,\alphabar)$.  When $k=2$, \eq{322} holds with
$E=0$ and $U=0$.
\vskip0.2truein
\noindent{\bf 3B.  The $\Psi^{k+1}\Psi^{k+1}$,
$\Psi^{k+1}\Psibar^{k+1}$, and $\Psibar^{k+1}\Psibar^{k+1}$ Analysis.}

In this step we prove that if $\hab$ is a natural generalized symmetry of
order $k$, then $\hab$ must be linear in the highest derivatives $\Psi^k$
and $\Psibar^k$.  To begin, we use the commutation rules \eq{3011} and
\eq{3012} to find that
$$
\eqalign{
&(\partial_\Psi^{k+1}\partial_\Psi^{k+1}\nabla^{\ss C}_{\ss
C^\prime}\nabla^{\ss D}_{\ss D^\prime}\hab)(\chi^{k+3},\chibar^{k-
1};\psi^{k+3},\psibar^{k-1})\cr
\eqskip
&=\big[\partial_\Psi^{k+1}\{\psi^{\ss C}\psibar_{\ss
C^\prime}(\partial_\Psi^k
\nabla^{\ss D}_{\ss D^\prime}\hab)
(\psi^{k+2},\psibar^{k-2})
+\nabla^{\ss C}_{\ss C^\prime}(\partial_\Psi^{k+1}\nabla^{\ss D}_{\ss
D^\prime}\hab)(\psi^{k+3},\psibar^{k-1})\}\big](\chi^{k+3},\chibar^{k-
1})\cr\eqskip
&=\big[\partial_\Psi^{k+1}\{\psi^{\ss C}\psibar_{\ss
C^\prime}(\partial_\Psi^k
\nabla^{\ss D}_{\ss D^\prime}\hab)
(\psi^{k+2},\psibar^{k-2})
+\psi^{\ss D}\psibar_{\ss D^\prime}\nabla^{\ss C}_{\ss
C^\prime}(\partial_\Psi^{k}\hab)(\psi^{k+2},\psibar^{k-
2})\}\big](\chi^{k+3},\chibar^{k-1})\cr\eqskip
&=\big(\psi^{\ss C}\psibar_{\ss C^\prime}\chi^{\ss D}\chibar_{\ss
D^\prime}
+\psi^{\ss D}\psibar_{\ss D^\prime}\chi^{\ss C}\chibar_{\ss
C^\prime}\big)
(\partial_\Psi^{k}\partial_\Psi^{k}\hab)
(\psi^{k+2},\psibar^{k-2};\chi^{k+2},\chibar^{k-2}).}\eqtag323
$$

We differentiate the symmetry equation \lineq\ twice with respect to
$\Psi^{k+1}$ and use \eq{323}; after some elementary simplifications we
obtain
$$
\eqalign{
-&2<\psi,\chi><\psibar,\chibar>(\partial_\Psi^k\partial^k_\Psi h)
(\psi^{k+2},\psibar^{k-2};\chi^{k+2},\chibar^{k-
2};\alpha,\beta,\alphabar,\betabar)\cr
&+<\psi,\beta><\psibar,\betabar>(\partial_\Psi^k\partial^k_\Psi h)
(\psi^{k+2},\psibar^{k-2};\chi^{k+2},\chibar^{k-
2};\alpha,\chi,\chibar,\alphabar)\cr
&+<\chi,\beta><\chibar,\betabar>(\partial_\Psi^k\partial^k_\Psi h)
(\psi^{k+2},\psibar^{k-2};\chi^{k+2},\chibar^{k-
2};\alpha,\psi,\psibar,\alphabar)\cr
&+<\psi,\alpha><\psibar,\alphabar>(\partial_\Psi^k\partial^k_\Psi h)
(\psi^{k+2},\psibar^{k-2};\chi^{k+2},\chibar^{k-
2};\beta,\chi,\chibar,\betabar)
\cr
&+<\chi,\alpha><\chibar,\alphabar>(\partial_\Psi^k\partial^k_\Psi h)
(\psi^{k+2},\psibar^{k-2};\chi^{k+2},\chibar^{k-
2};\beta,\psi,\psibar,\betabar)
=0.}
$$
In the notation of equation \eq{3000} this is the condition $\gamma=0$.
Using Proposition \statement{302}, we immediately find that this equation
simplifies to
$$
(\partial_\Psi^k\partial^k_\Psi h)
(\psi^{k+2},\psibar^{k-2};\chi^{k+2},\chibar^{k-
2};\alpha,\beta,\alphabar,\betabar)=0.\eqtag324
$$
This proves that $\hab$ is at most linear in the variables $\Psi^k$.
Likewise, if we take the second derivative of the linearized equations
\lineq\ with respect to $\Psibar^{k+1}$ and use Proposition
\statement{302}, we obtain
$$
(\partial^k_\Psibar\partial^k_\Psibar h)
(\psi^{k-2},\psibar^{k+2};\chi^{k-
2},\chibar^{k+2};\alpha,\beta,\alphabar,\betabar)=0,\eqtag325
$$
which implies that $\hab$ is linear in the variables $\Psibar^k$.
Finally, differentiation of the symmetry condition \lineq\ with respect to
$\Psibar^{k+1}$ and $\Psi^{k+1}$, followed by use of Proposition
\statement{302}, leads to
$$
(\partial^k_\Psibar\partial_\Psi^k h)
(\psi^{k-2},\psibar^{k+2};\chi^{k+2},\chibar^{k-
2};\alpha,\beta,\alphabar,\betabar)=0.\eqtag326
$$
Together, equations \eq{324}, \eq{325}, and \eq{326} prove the following
proposition.

\proclaim Proposition\statementtag305 .
Let
$$
\hab
=\hab(\Psi^2,\Psibar^2,\dots,\Psi^k,\Psibar^k)
$$
be a generalized symmetry of the vacuum Einstein equations. Then $\hab$
is at most linear in the top-order Penrose fields $\Psi^k$ and $\Psibar^k$.

\proclaim Corollary\statementtag306 .
The spinors $A, B,W$ and $D,E,U$ in equations \eq{305} and \eq{322}
are at most of order $k-1$.

\proof
This corollary follows from Proposition \statement{305} and the fact that
the spinors $A, B,W$ and $D,E,U$ in the decompositions \eq{305} and
\eq{322} are unique.\endEx

At this point we are able to prove that there are no generalized symmetries
of the Einstein equations of differential order two and three in the metric,
aside from the scaling symmetry \equationlabel{2}{311}{}.

\proclaim Corollary\statementtag3061 .
Let $\hab(\Psi^2,\Psibar^2) $ be a natural generalized symmetry of the
vacuum Einstein equations of order 2.  Then
$$
\hab = c\ \epsilon_{\ss A^\prime B^\prime}\epsilon^{\ss AB},
$$
where $c$ is a constant.

\proof
According to Proposition \statement{303} and Proposition
\statement{304}, we have that
$$
[\partial_\Psi^2h](\psi^4;\alpha,\beta,\alphabar,\betabar)
=<\psi,\alpha><\psi,\beta> A(\psi^2,\alphabar\betabar),
$$
and
$$[\partial^2_\Psibar h](\psibar^4;\alpha,\beta,\alphabar,\betabar)
=<\psibar,\alphabar><\psibar,\betabar> D(\psibar^2,\alpha\beta).
$$
Proposition \statement{305} implies that the spinors $A$ and $D$ are
independent of the Penrose fields $\Psi^2$ and $\Psibar^2$.  Because $h$ is
$SL(2,{\bf C})$ invariant, $A$ and $D$ are $SL(2,{\bf C})$ invariant,
and consequently they are constructed solely from the $\epsilon$-spinors.
It is easy to check that there are no spinors with the rank and symmetries
of $A$ and $D$ built solely from the $\epsilon$-spinors.  Therefore
$A=D=0$.  This implies that $\hab$ is a function only of the $\epsilon$-
spinors from which the corollary follows.\endEx

\proclaim Corollary\statementtag3062 .
Let $\hab(\Psi^2,\Psibar^2,\Psi^3,\Psibar^3) $ be a natural generalized
symmetry of the vacuum Einstein equations of order 3.  Then
$$
\hab = c\ \epsilon_{\ss A^\prime B^\prime}\epsilon^{\ss AB},
$$
where $c$ is a constant.

\proof
According to Proposition \statement{304}, we have that
$$\eqalign{
[\partial_\Psi^3h](\psi^{5},\psibar;\alpha,\beta,\alphabar,\betabar)
=<\psi,\alpha><\psi,\beta>&A(\psi^{3},\psibar\alphabar\betabar)
+<\psi,\alpha><\alphabar,\psibar>W(\psi^{4},\beta,\betabar)\cr
&+ <\psi,\beta><\betabar,\psibar> W(\psi^{4},\alpha,\alphabar),\cr
\noalign{\hbox{and}}
[\partial^3_\Psibar
h](\psi,\psibar^{5};\alpha,\beta,\alphabar,\betabar)=<\psibar,\alphabar>
<\psibar,\betabar>&D(\psibar^{3},\psi\alpha\beta)
+<\psibar,\alphabar><\alpha,\psi>U(\psibar^{4},\beta,\betabar)\cr
& + <\psibar,\betabar><\beta,\psi> U(\psibar^{4},\alpha,\alphabar).}
$$
Proposition \statement{305} implies that $A,D,W,$ and $U$ are functions
of at most $\Psi^2$ and $\Psibar^2$.  However, there are no natural
spinors with the rank and symmtetry of $A,D,W,$ and $U$ built solely
from the undifferentiated Weyl spinors.  To see this, let us focus on the
spinor $A$.  Because $A$ is a natural spinor we have that, for each
$\Lambda,\Omega\in SL(2,{\bf C})$,
$$
A_{\ss A^\prime B^\prime C^\prime}^{\ss
A^{\phantom{\prime}}B^{\phantom{\prime}}
C^{\phantom{\prime}}}(\widetilde\Psi^2,\widetilde{\Psibar}{}^2)
=\Lambda^{\ss A}_{\ss D}\Lambda_{\ss B}^{\ss E}\Lambda_{\ss C}^{\ss
F}\Omega_{\ss A^\prime}^{\ss D^\prime}\Omega_{\ss B^\prime}^{\ss
E^\prime}\Omega_{\ss C^\prime}^{\ss F^\prime}
A_{\ss D^\prime E^\prime F^\prime}^{\ss
D^{\phantom{\prime}}E^{\phantom{\prime}}F^{\phantom{\prime}}}
(\Psi^2,\Psibar^2),\eqtag30611
$$
where we have set
$$
\widetilde\Psi_{\ss ABCD} =\Lambda^{\ss E}_{\ss A}\Lambda_{\ss
B}^{\ss F}\Lambda_{\ss C}^{\ss G}\Lambda_{\ss D}^{\ss H}
\Psi_{\ss EFGH}
\qquad {\rm and}\qquad
\widetilde{\Psibar}{}^{{\ss A^\prime B^\prime C^\prime
D^\prime}}=\Omega^{\ss A^\prime}_{\ss E^\prime}\Omega_{\ss
F^\prime}^{\ss B^\prime}\Omega_{\ss G^\prime}^{\ss
C^\prime}\Omega_{\ss H^\prime}^{\ss D^\prime}
\Psibar^{{\ss E^\prime F^\prime G^\prime H^\prime}}.
$$
As this relation must hold for any $\Lambda,\Omega\in SL(2,{\bf C})$, we
let
$$
\Lambda^{\ss A}_{\ss B} = -\delta^{\ss A}_{\ss B},\qquad{\rm
and}\qquad
\Omega^{\ss A^\prime}_{\ss B^\prime} = \delta^{\ss A^\prime}_{\ss
B^\prime},
$$
in which case
$$
\widetilde\Psi_{\ss ABCD} = \Psi_{\ss ABCD}, \qquad{\rm and}\qquad
\widetilde{\Psibar}{}^{{\ss A^\prime B^\prime C^\prime
D^\prime}}=\Psibar^{{\ss A^\prime B^\prime C^\prime D^\prime}}.
$$
Because there are an odd number of $\Lambda$ matrices on the right hand
side of \eq{30611}, the naturality condition forces $A=0$.  An identical
series of arguments establish that $D=W=U=0$.  We therefore find that
$$
[\partial_\Psi^3h](\psi^5,\psibar^1;\alpha,\beta,\alphabar,\betabar)=0
\qquad {\rm and}\qquad
[\partial_\Psi^3h](\psi^1,\psibar^5;\alpha,\beta,\alphabar,\betabar)=0.
$$
We have reduced the order of $\hab$ by one, and Corollary
\statement{3062} now follows from Corollary \statement{3061}.\endEx

The invariance arguments leading to Corollaries \statement{3061} and
\statement{3062} clearly fail when one allows natural generalized
symmetries of order $k\geq4$.  This reflects the existence, for $k\geq4$,
of generalized diffeomorphism symmetries.  These are analyzed in the next
section.

\vskip0.2truein
\noindent{\bf 3C. The $\Psi^k\Psi^{k+1}$, $\Psibar^k\Psi^{k+1}$,
$\Psi^k\Psibar^{k+1}$, and $\Psibar^k\Psibar^{k+1}$ Analysis.}

In this section we shall prove that $A,B,D$ and $E$ must be of order $k-
2$, and that there exists a natural type $(1,1)$ spinor $X$ of order $k-1$,
$$
X^{\ss A}_{\ss A^\prime}=X^{\ss A}_{\ss
A^\prime}(\Psi^2,\Psibar^2,\dots,\Psi^{k-1},\Psibar^{k-1}),\eqtag3061
$$
such that
$$
\openup-6pt
\eqalignno{
W(\psi^{k+1},\psibar^{k-3},\alpha,\alphabar)&=[\partial_\Psi^{k-1}X]
(\psi^{k+1},\psibar^{k-3};\alpha,\alphabar),\aleqtag3062 \cr
\noalign{\hbox{and}}
U(\psi^{k-3},\psibar^{k+1},\alpha,\alphabar)&=[\partial^{k-1}_\Psibar
X](\psi^{k-3},\psibar^{k+1};\alpha,\alphabar).\aleqtag3063 }
$$
We obtain these results by analyzing the equations arising from the
coefficients of $\Psi^k\Psi^{k+1}$, $\Psi^k\Psibar^{k+1}$,
$\Psibar^k\Psi^{k+1}$, and $\Psibar^k\Psibar^{k+1}$ in the linearized
equations \lineq.

We begin with the $\Psi^k\Psi^{k+1}$ terms.  Because $\hab$ is linear in
the Penrose fields $\Psi^k,\Psibar^{k}$, we can use the commutation rules
in Proposition \statement{301} to deduce that
$$
\eqalign{
[\partial_\Psi^k\partial_\Psi^{k+1}\nabla^{\ss C}_{\ss
C^\prime}&\nabla^{\ss D}_{\ss D^\prime} \hab](\chi^{k+2},\chibar^{k-
2};\psi^{k+3},\psibar^{k-1})\cr\eqskip
&=\psi^{\ss C}\psi^{\ss D}\psibar_{\ss C^\prime}\psibar_{\ss D^\prime}
[\partial_\Psi^{k-1}\partial_\Psi^k\hab](\psi^{k+1},\psibar^{k-
3};\chi^{k+2},\chibar^{k-2})\cr\eqskip
&+\psi^{\ss C}\chi^{\ss D}\psibar_{\ss C^\prime}\chibar_{\ss D^\prime}
[\partial_\Psi^{k-1}\partial_\Psi^k\hab](\chi^{k+1},\chibar^{k-
3};\psi^{k+2},\psibar^{k-2})\cr\eqskip
&+\chi^{\ss C}\psi^{\ss D}\chibar_{\ss C^\prime}\psibar_{\ss D^\prime}
[\partial_\Psi^{k-1}\partial_\Psi^k\hab](\chi^{k+1},\chibar^{k-
3};\psi^{k+2}\psibar^{k-2}).}\eqtag3261
$$
We now apply the operator $\partial_\Psi^k\partial_\Psi^{k+1}$ to the
linearized equations \lineq\ to find, after substituting from \eq{3261} and
simplifying, that
$$
\eqalign{
-2&<\psi,\chi><\psibar,\chibar>[\partial_\Psi^{k-1}\partial_\Psi^kh]
(\chi^{k+1},\chibar^{k-3};\psi^{k+2},\psibar^{k-
2};\alpha,\beta,\alphabar,\betabar)\cr
+&<\psi,\beta><\psibar,\betabar>[\partial_\Psi^{k-1}\partial_\Psi^kh]
(\psi^{k+1},\psibar^{k-3};\chi^{k+2},\chibar^{k-
2};\alpha,\psi,\psibar,\alphabar)\cr
+&<\psi,\alpha><\psibar,\alphabar>[\partial_\Psi^{k-1}\partial_\Psi^kh]
(\psi^{k+1},\psibar^{k-3};\chi^{k+2},\chibar^{k-
2};\beta,\psi,\psibar,\betabar)\cr
+&<\chi,\beta><\chibar,\betabar>[\partial_\Psi^{k-1}\partial_\Psi^kh]
(\chi^{k+1},\chibar^{k-3};\psi^{k+2},\psibar^{k-
2};\alpha,\psi,\psibar,\alphabar)\cr
+&<\chi,\alpha><\chibar,\alphabar>[\partial_\Psi^{k-1}\partial_\Psi^kh]
(\chi^{k+1},\chibar^{k-3};\psi^{k+2},\psibar^{k-
2};\beta,\psi,\psibar,\betabar)\cr
+&<\psi,\beta><\psibar,\betabar>[\partial_\Psi^{k-1}\partial_\Psi^kh]
(\chi^{k+1},\chibar^{k-3};\psi^{k+2},\psibar^{k-
2};\alpha,\chi,\chibar,\alphabar)\cr
+&<\psi,\alpha><\psibar,\alphabar>[\partial_\Psi^{k-1}\partial_\Psi^kh]
(\chi^{k+1},\chibar^{k-3};\psi^{k+2},\psibar^{k-
2};\beta,\chi,\chibar,\betabar)
=0.}\eqtag327
$$
The symmetry condition \eq{303} implies that the coefficients of
$<\chi,\beta><\chibar,\betabar>$ and $<\chi,\alpha><\chibar,\alphabar>$
each vanish, and so we can rewrite equation \eq{327} as
$$
\eqalign{
-2<\psi,\chi>\ <\psibar,\chibar>\ &[\partial_\Psi^{k-1}\partial_\Psi^kh]
(\chi^{k+1},\chibar^{k-3};\psi^{k+2},\psibar^{k-
2};\alpha,\beta,\alphabar,\betabar)\cr
+<\psi,\alpha><\psibar,\alphabar>\{&[\partial_\Psi^{k-1}\partial_\Psi^kh]
(\psi^{k+1},\psibar^{k-3};\chi^{k+2},\chibar^{k-
2};\beta,\psi,\psibar,\betabar)\cr
+&[\partial_\Psi^{k-1}\partial_\Psi^kh]
(\chi^{k+1},\chibar^{k-3};\psi^{k+2},\psibar^{k-
2};\beta,\chi,\chibar,\betabar)\}\cr
+<\psi,\beta><\psibar,\betabar>\{&[\partial_\Psi^{k-1}\partial_\Psi^kh]
(\psi^{k+1},\psibar^{k-3};\chi^{k+2},\chibar^{k-
2};\alpha,\psi,\psibar,\alphabar)\cr
+&[\partial_\Psi^{k-1}\partial_\Psi^kh]
(\chi^{k+1},\chibar^{k-3};\psi^{k+2},\psibar^{k-
2};\alpha,\chi,\chibar,\alphabar)\}=0.}\eqtag328
$$
In this equation we set $\alpha=\beta=\psi$ to arrive at
$$
[\partial_\Psi^{k-1}\partial_\Psi^kh](\chi^{k+1},\chibar^{k-
3};\psi^{k+2},\psibar^{k-2};\psi,\psi,\alphabar,\betabar)=0.
$$
In terms of the decomposition \eq{305} we have that
$$
[\partial_\Psi^kh](\psi^{k+2},\psibar^{k-2};\psi,\psi,\alpha,\betabar)
=<\psibar,\alphabar><\psibar,\betabar>B(\psi^{k+4},\psibar^{k-4}),
$$
and so this equation implies that
$$
[\partial_\Psi^{k-1}B](\chi^{k+1},\chibar^{k-3};\psi^{k+4},\psibar^{k-4})
=0.\eqtag329
$$
In other words, $B$ is independent of the spinor $\Psi^{k-1}$.  Likewise,
by setting $\alphabar=\betabar=\psibar$ in equation \eq{328}, we conclude
that
$$
[\partial_\Psi^{k-1}A](\chi^{k+1},\chibar^{k-3};\psi^{k},\psibar^{k})
=0,\eqtag330
$$
and so $A$ is independent of the spinor $\Psi^{k-1}$.
Together, equations \eq{305}, \eq{329}, and \eq{330} show that
$$
\eqalign{
[\partial_\Psi^{k-1}&\partial_\Psi^kh](\chi^{k+1},\chibar^{k-
3};\psi^{k+2},\psibar^{k-2};\alpha,\beta,\alphabar,\betabar)\cr\eqskip
=&<\psi,\alpha><\alphabar,\psibar>[\partial_\Psi^{k-1}W]
(\chi^{k+1},\chibar^{k-3};\psi^{k+1},\psibar^{k-
3},\beta,\betabar)\cr\eqskip
+&<\psi,\beta><\betabar,\psibar>[\partial_\Psi^{k-1}W]
(\chi^{k+1},\chibar^{k-3};\psi^{k+1},\psibar^{k-
3},\alpha,\alphabar).}\eqtag331
$$

We next set $\alpha=\beta$ and $\alphabar=\betabar$ in \eq{328}, and
substitute from \eq{331} to arrive at
$$
\eqalign{
2 <\psi,\chi><\psibar,\chibar>&[\partial_\Psi^{k-1}W]
(\chi^{k+1},\chibar^{k-3};\psi^{k+1},\psibar^{k-3},\alpha,\alphabar)\cr
=&<\chi,\alpha><\chibar,\psibar>[\partial_\Psi^{k-1}W]
(\psi^{k+1},\psibar^{k-3};\chi^{k+1},\chibar^{k-3},\psi,\alphabar)\cr
+&<\chi,\psi><\chibar,\alphabar>[\partial_\Psi^{k-1}W]
(\psi^{k+1},\psibar^{k-3};\chi^{k+1},\chibar^{k-3},\alpha,\psibar)\cr
+&<\psi,\alpha><\psibar,\chibar>[\partial_\Psi^{k-1}W]
(\chi^{k+1},\chibar^{k-3};\psi^{k+1},\psibar^{k-3},\chi,\alphabar)\cr
+&<\psi,\chi><\psibar,\alphabar>[\partial_\Psi^{k-1}W]
(\chi^{k+1},\chibar^{k-3};\psi^{k+1},\psibar^{k-
3},\alpha,\chibar).}\eqtag3311
$$
The right-hand side of this equation is unchanged by the simultaneous
interchange of $\psi$ with $\chi$ and $\psibar$ with $\chibar$ so we
conclude
$$
[\partial_\Psi^{k-1}W]
(\chi^{k+1},\chibar^{k-3};\psi^{k+1},\psibar^{k-3},\alpha,\alphabar)
=[\partial_\Psi^{k-1}W]
(\psi^{k+1},\psibar^{k-3};\chi^{k+1},\chibar^{k-
3},\alpha,\alphabar).\eqtag332
$$
Equation \eq{332} is necessary and sufficient for equation \eq{3311} to
hold, and is one of the integrability conditions needed to establish equation
\eq{3062}.

In exactly the same fashion we can apply the operator
$\partial^k_\Psibar\,\partial^{k+1}_\Psibar$ to the linearized equations
\lineq\ to show that
$$
\eqalignno{
[\partial^{k-1}_\Psibar D](\psi^{k-
3},\psibar^{k+1};\chibar^{k},\chi^{k})&=0, \aleqtag333 \cr
[\partial^{k-1}_\Psibar E](\psi^{k-
3},\psibar^{k+1};\chibar^{k+4},\chi^{k-4})&=0. \aleqtag334 }
$$
Moreover, we have that
$$
[\partial^{k-1}_\Psibar U]
(\psi^{k-3},\psibar^{k+1};\chibar^{k+1},\chi^{k-3},\alpha,\alphabar)
=[\partial^{k-1}_\Psibar U]
(\chi^{k-3},\chibar^{k+1};\psibar^{k+1},\psi^{k-
3},\alpha,\alphabar).\eqtag335
$$

Before applying the operator $\partial^k_\Psibar\,\partial_\Psi^{k+1}$ to
the linearized equations, we first use the commutation rules of Proposition
\statement{301} and the fact that $\hab$ is linear in $\Psi^k$ and
$\Psibar^k$ to deduce that
$$
\vbox{\eqalignno{
[\partial^k_\Psibar\,\partial_\Psi^{k+1}\nabla^{\ss C}_{\ss
C^\prime}&\nabla^{\ss D}_{\ss D^\prime} \hab](\chi^{k-
2},\chibar^{k+2};\psi^{k+3},\psibar^{k-1})\cr\eqskip
&=\psi^{\ss C}\psi^{\ss D}\psibar_{\ss C^\prime}\psibar_{\ss D^\prime}
[\partial^k_\Psibar\,\partial_\Psi^{k-1}\hab](\chi^{k-
2},\chibar^{k+2};\psi^{k+1},\psibar^{k-3})\cr\eqskip
&+\psi^{\ss C}\chi^{\ss D}\psibar_{\ss C^\prime}\chibar_{\ss D^\prime}
[\partial^{k-1}_\Psibar\partial_\Psi^k\hab](\chi^{k-
3},\chibar^{k+1};\psi^{k+2},\psibar^{k-2})\cr\eqskip
&+\chi^{\ss C}\psi^{\ss D}\chibar_{\ss C^\prime}\psibar_{\ss D^\prime}
[\partial^{k-1}_\Psibar\partial_\Psi^k\hab](\chi^{k-
3},\chibar^{k+1};\psi^{k+2},\psibar^{k-2}).\cr}}
$$
Using this result, if we differentiate \lineq\ with respect to $\Psibar^k$ and
$\Psi^{k+1}$ and take into account the leading order symmetry conditions
of Proposition \statement{302}, we have
$$
\eqalign{
-2<\psi,\chi>\ <\psibar,\chibar>\ &[\partial^{k-1}_\Psibar\partial_\Psi^kh]
(\chi^{k-3},\chibar^{k+1};\psi^{k+2},\psibar^{k-
2};\alpha,\beta,\alphabar,\betabar)\cr
+<\psi,\alpha><\psibar,\alphabar>\{&[\partial_\Psi^{k-1}\partial^k_\Psibar
h]
(\psi^{k+1},\psibar^{k-3};\chi^{k-
2},\chibar^{k+2};\beta,\psi,\psibar,\betabar)\cr
+&[\partial^{k-1}_\Psibar\partial_\Psi^kh]
(\chi^{k-3},\chibar^{k+1};\psi^{k+2},\psibar^{k-
2};\beta,\chi,\chibar,\betabar)\}\cr
+<\psi,\beta><\psibar,\betabar>\{&[\partial_\Psi^{k-1}\partial^k_\Psibar h]
(\psi^{k+1},\psibar^{k-3};\chi^{k-
2},\chibar^{k+2};\alpha,\psi,\psibar,\alphabar)\cr
+&[\partial^{k-1}_\Psibar\partial_\Psi^kh]
(\chi^{k-3},\chibar^{k+1};\psi^{k+2},\psibar^{k-
2};\alpha,\chi,\chibar,\alphabar)\}\cr
&=0.}\eqtag336
$$
With $\alpha=\beta=\psi$, and then with $\alphabar=\betabar=\psibar$,
equation \eq{336} implies
$$
\eqalignno{
[\partial^{k-1}_\Psibar B](\chi^{k+1},\chibar^{k-
3};\psi^{k+4},\psibar^{k-4})
&=0\aleqtag337 \cr
\noalign{\hbox{and}}
[\partial^{k-1}_\Psibar A](\chi^{k+1},\chibar^{k-3};\psi^{k},\psibar^{k})
&=0.\aleqtag338 \cr}
$$
We set $\alpha=\beta$ and $\alphabar=\betabar$ in \eq{336} and take
\eq{337} and \eq{338} into account to find
$$
\eqalign{
2<\psi,\chi>\ <\psibar,\chibar>\ &[\partial^{k-1}_\Psibar\partial_\Psi^kh]
(\chi^{k-3},\chibar^{k+1};\psi^{k+2},\psibar^{k-
2};\alpha,\alpha,\alphabar,\alphabar)
\cr
=\ \ <\psi,\alpha>\ <\psibar,\alphabar>\{&[\partial_\Psi^{k-
1}\partial^k_\Psibar h]
(\psi^{k+1},\psibar^{k-3};\chi^{k-
2},\chibar^{k+2};\alpha,\psi,\psibar,\alphabar)\cr
+&[\partial^{k-1}_\Psibar \partial_\Psi^kh]
(\chi^{k-3},\chibar^{k+1};\psi^{k+2},\psibar^{k-
2};\alpha,\chi,\chibar,\alphabar)
\}.}\eqtag339
$$

Again, in exactly the same manner, the
$\partial_\Psi^k\partial^{k+1}_\Psibar$ derivative of the linearized
equation \lineq\ yields
$$
\eqalignno{
[\partial_\Psi^{k-1}D](\psi^{k+1},\psibar^{k-
3};\chibar^{k},\chi^{k})&=0,\aleqtag340 \cr
[\partial_\Psi^{k-1}E](\psi^{k+1},\psibar^{k-3};\chibar^{k+4}\chi^{k-
4})&=0,\aleqtag341 \cr}
$$
as well as
$$
\eqalign{
2<\psi,\chi>\ <\psibar,\chibar>\ &[\partial_\Psi^{k-1}\partial^k_\Psibar h]
(\chi^{k+1},\chibar^{k-3};\psi^{k-
2},\psibar^{k+2};\alpha,\alpha,\alphabar,\alphabar)
\cr
=\ \ <\psi,\alpha>\ <\psibar,\alphabar>\{&[\partial^{k-
1}_\Psibar\partial_\Psi^kh]
(\psi^{k-3},\psibar^{k+1};\chi^{k+2},\chibar^{k-
2};\alpha,\psi,\psibar,\alphabar)\cr
+&[\partial_\Psi^{k-1}\partial^k_\Psibar h]
(\chi^{k+1},\chibar^{k-3};\psi^{k-
2},\psibar^{k+2};\alpha,\chi,\chibar,\alphabar)
\}.}\eqtag342
$$

Equations \eq{329}, \eq{330}, \eq{333}, \eq{334}, \eq{337}, \eq{338},
\eq{340}, and \eq{341} prove the following proposition.

\proclaim Proposition\statementtag3081 .
Let $\hab$ be a natural generalized symmetry of order $k$.  Then the
spinors $A$, $B$, $D$, $E$ appearing in the decompositions \eq{305} and
\eq{322}
are at most of order $k-2$.

On taking Proposition \statement{3081} into account, the substitution of
\eq{305} and \eq{322} into \eq{339} and \eq{342} gives rise to
$$
\eqalign{
4&<\psi,\chi><\psibar,\chibar>[\partial^{k-1}_\Psibar W]
(\chi^{k-3},\chibar^{k+1};\psi^{k+1},\psibar^{k-3},\alpha,\alphabar)\cr
=&<\chi,\psi><\chibar,\alphabar>[\partial_\Psi^{k-1}U]
(\psi^{k+1},\psibar^{k-3};\chi^{k-3},\chibar^{k+1},\alpha,\psibar)\cr
+&<\chi,\alpha><\chibar,\psibar>[\partial_\Psi^{k-1}U]
(\psi^{k+1},\psibar^{k-3};\chi^{k-3},\chibar^{k+1},\psi,\alphabar)\cr
+&<\psi,\chi><\psibar,\alphabar>[\partial^{k-1}_\Psibar W]
(\chi^{k-3},\chibar^{k+1};\psi^{k+1},\psibar^{k-3},\alpha,\chibar)\cr
+&<\psi,\alpha><\psibar,\chibar>[\partial^{k-1}_\Psibar W]
(\chi^{k-3},\chibar^{k+1};\psi^{k+1},\psibar^{k-
3},\chi,\alphabar),\cr}\eqtag343
$$
along with
$$
\eqalign{
4&<\psi,\chi><\psibar,\chibar>[\partial_\Psi^{k-1}U]
(\chi^{k+1},\chibar^{k-3};\psi^{k-3},\psibar^{k+1},\alpha,\alphabar)\cr
=&<\chi,\alpha><\chibar,\psibar>[\partial^{k-1}_\Psibar W]
(\psi^{k-3},\psibar^{k+1};\chi^{k+1},\chibar^{k-3},\psi,\alphabar)\cr
+&<\chi,\psi><\chibar,\alphabar>[\partial^{k-1}_\Psibar W]
(\psi^{k-3},\psibar^{k+1};\chi^{k+1},\chibar^{k-3},\alpha,\psibar)\cr
+&<\psi,\chi><\psibar,\alphabar>[\partial_\Psi^{k-1}U]
(\chi^{k+1},\chibar^{k-3};\psi^{k-3},\psibar^{k+1},\alpha,\chibar)\cr
+&<\psi,\alpha><\psibar,\chibar>[\partial_\Psi^{k-1}U]
(\chi^{k+1},\chibar^{k-3};\psi^{k-
3},\psibar^{k+1},\chi,\alphabar).\cr}\eqtag344
$$
In this last equation, we simultaneously interchange $\psi$ with $\chi$ and
$\psibar$ with $\chibar$; a comparison with \eq{343} allows us to deduce
that
$$
[\partial^{k-1}_\Psibar W](\chi^{k-
3},\chibar^{k+1};\psi^{k+1},\psibar^{k-3},\alpha,\alphabar)
=
[\partial_\Psi^{k-1}U]
(\psi^{k+1},\psibar^{k-3};\chi^{k-3},\chibar^{k+1},\alpha,\alphabar).
\eqtag345
$$
Equations \eq{332}, \eq{335}, and \eq{345} are the integrability
conditions for \eq{3062} and \eq{3063}.

\proclaim Proposition\statementtag307 .
Let $\hab$ be a generalized symmetry of order $k$.  Then there is a
natural vector field of order $k-1$,
$$
X^{\ss A}_{\ss A^\prime}=X^{\ss A}_{\ss
A^\prime}(\Psi^2,\Psibar^2,\dots,\Psi^{k-1},\Psibar^{k-1}),
$$
such that the spinors $W$ and $U$ in \eq{305} and \eq{322} are the
gradients
$$
\eqalignno{
[\partial_\Psi^{k-1}X]
(\psi^{k+1},\psibar^{k-3};\alpha,\alphabar)
&=W(\psi^{k+1},\psibar^{k-3},\alpha,\alphabar),\aleqtag346 \cr
\noalign{\hbox{and}}
[\partial^{k-1}_\Psibar X](\psi^{k-
3},\psibar^{k+1};\alpha,\alphabar)&=U(\psi^{k-
3},\psibar^{k+1},\alpha,\alphabar).\aleqtag347 }
$$

\proof
We have already seen that the linearized equations \lineq\ imply the
integrability conditions for equations \eq{346} and \eq{347} are satisfied.
It is easy to check that
$$
\eqalign{
X^{\ss A}_{\ss A^\prime}
&=\int_0^1dt\ \Psi^{\ss B_1^\prime\cdots B_{k-3}^\prime}_{\ss B_1\cdots
B_{k-3}\cdots B_{k+1}}W_{\ss B_1^\prime\cdots B_{k-3}^\prime
A^\prime}^{\ss B_1\cdots B_{k-3}\cdots B_{k+1}A}
(\Psi^2,\Psibar^2,\dots,\Psi^{k-2},\Psibar^{k-2},t\Psi^{k-1},t\Psibar^{k-1}
)\cr
&+\int_0^1dt\ \Psibar_{\ss B_1\cdots B_{k-3}}^{\ss B_1^\prime\cdots
B_{k-3}^\prime\cdots B_{k+1}^\prime}U^{\ss B_1\cdots B_{k-3} A}_{\ss
B_1^\prime\cdots B_{k-3}^\prime\cdots B_{k+1}^\prime A^\prime}
(\Psi^2,\Psibar^2,\dots,\Psi^{k-2},\Psibar^{k-2},t\Psi^{k-1},t\Psibar^{k-1}
)}
$$
defines a real, natural vector field that satisfies equations \eq{346} and
\eq{347}.\endEx
\vskip0.2truein
\noindent{\bf 3D. Reduction in Order of $\hab$.}

Let us set
$$
\dab=\nabla^{\ss A}_{\ss A^\prime} X^{\ss B}_{\ss B^\prime}
+\nabla^{\ss B}_{\ss B^\prime} X^{\ss A}_{\ss A^\prime},
$$
where $X^{\ss A}_{\ss A^\prime}$ is defined in Proposition
\statement{307}.
By Proposition \statementlabel{2}{207}, we know that $\dab$ is a solution
to the linearized equations \lineq\ and so defines a generalized symmetry of
the vacuum Einstein equations.  Therefore
$$
\kab=\hab-\dab
$$
is also a generalized symmetry.  Since
$$
\eqalign{
[\partial&^k_\Psi d](\psi^{k+2},\psibar^{k-
2};\alpha,\beta,\alphabar,\betabar)\cr&
=<\psi,\alpha><\alphabar,\psibar>W(\psi^{k+1},\psibar^{k-
3},\beta,\betabar) \ + <\psi,\beta><\betabar,\psibar>
W(\psi^{k+1},\psibar^{k-3},\alpha,\alphabar)}
$$
and
$$
\eqalign{
[&\partial^k_\Psibar d](\psi^{k-
2},\psibar^{k+2};\alpha,\beta,\alphabar,\betabar)\cr&
=<\psibar,\alphabar><\alpha,\psi>U(\psibar^{k+1},\psi^{k-
3},\beta,\betabar) + <\psibar,\betabar><\beta,\psi>
U(\psibar^{k+1},\psi^{k-3},\alpha,\alphabar),}
$$
we have, from our basic decomposition \eq{305} and \eq{322},
$$
\eqalign{
[&\partial_\Psi^kk](\psi^{k+2},\psibar^{k-
2};\alpha,\beta,\alphabar,\betabar)\cr&
=<\psi,\alpha><\psi,\beta>A(\psi^{k},\psibar^{k-2}\alphabar\betabar)
\
+<\psibar,\alphabar><\psibar,\betabar>B(\psi^{k+2}\alpha\beta,\psibar^{k-
4})}\eqtag3491
$$
and
$$
\eqalign{
[&\partial^k_\Psibar k](\psi^{k-
2},\psibar^{k+2};\alpha,\beta,\alphabar,\betabar)\cr&
=<\psibar,\alphabar><\psibar,\betabar>D(\psibar^{k},\psi^{k-
2}\alpha\beta)
+<\psi,\alpha><\psi,\beta>E(\psibar^{k+2}\alphabar\betabar,\psi^{k-4}).}
\eqtag3492
$$
We now show that the linearized equations \lineq\ force
$$
A=B=D=E=0,\eqtag348
$$
and hence
$$
\hab=\nabla^{\ss A}_{\ss A^\prime} X^{\ss B}_{\ss B^\prime}
+\nabla^{\ss B}_{\ss B^\prime} X^{\ss A}_{\ss A^\prime}+\kab,\eqtag349
$$
where $\kab$ is now of order $k-1$.

To prove \eq{348} we differentiate equation \eq{300} one final time with
respect to $\Psi^{k+1}$ and use the leading order symmetry condition
satisfied by $\kab$, namely
$$
[\partial_\Psi^kk](\psi^{k+2},\psibar^{k-2};\psi,
\alpha,\alphabar,\psibar)=0,
$$
to arrive at
$$
\eqalign{
&<\beta,\psi><\betabar,\psibar>\{[\partial_\Psi^{k-
1}k](\psi^{k+1},\psibar^{k-3};\alpha,\psi,\psibar,\alphabar)
+[{\rm Div}\ \partial_\Psi^kk](\psi^{k+2},\psibar^{k-
2};\alpha,\alphabar)\}\cr
+&<\alpha,\psi><\alphabar,\psibar>\{[\partial_\Psi^{k-
1}k](\psi^{k+1},\psibar^{k-3};\beta,\psi,\psibar,\betabar)
+[{\rm Div}\ \partial_\Psi^kk](\psi^{k+2},\psibar^{k-
1};\beta,\betabar)\}\cr
+&[{\rm Grad}\ \partial_\Psi^kk](\psi,\psibar;\psi^{k+2},\psibar^{k-
2};\alpha,\beta,\alphabar,\betabar) = 0,}\eqtag350
$$
where we have introduced the notation
$$
[{\rm Div}\ \partial_\Psi^kk](\psi^{k+2},\psibar^{k-2};\alpha,\alphabar)
=\alpha_{\ss A}\alphabar^{\ss B^\prime}[\nabla_{\ss B}^{\ss
A^\prime}\partial_\Psi^k\kab](\psi^{k+2},\psibar^{k-2}).\eqtag35002
$$
In \eq{350} we now set $\alpha=\beta=\psi$; by virtue of equation
\eq{3491} we then find
$$
[{\rm Grad}\ B](\psi,\psibar;\psi^{k+4},\psibar^{k-4})=0.\eqtag351
$$
Similarly, if we set $\alphabar=\betabar=\psibar$ in \eq{350} and use
\eq{3492} we find that
$$
[{\rm Grad}\  A](\psi,\psibar;\psi^{k},\psibar^{k})=0.\eqtag352
$$
Proposition \statement{308} implies that $A=0$ and $B=0$.

We have thus found that
$$
[\partial_\Psi^kk](\psi^{k+2},\psibar^{k-2};\alpha,\beta,\alphabar,\betabar)
=0.
$$
Likewise, by differentiating the linearized equations \lineq\ with respect to
$\Psibar^{k+1}$ we can show that $D=0$ and $E=0$ so that
$$
[\partial^k_\Psibar\, k](\psi^{k-2},\psibar^{k+2}
\alpha,\beta,\alphabar,\betabar)
=0.
$$
These last two equations prove \eq{349}.

\proclaim Theorem\statementtag309 .
Let
$$
\hab=\hab(\Psi^2,\Psibar^2,\dots,\Psi^k,\Psibar^k)
$$
be a natural generalized symmetry of the vacuum Einstein equations of
order $k$.  Then there exists a natural spinor
$$
X^{\ss A}_{\ss A^\prime}=X^{\ss A}_{\ss
A^\prime}(\Psi^2,\Psibar^2,\dots,\Psi^{k-1},\Psibar^{k-1})
$$
of order $k-1$, and a constant $c$, such that
$$
\hab=c\ \epsilon^{\ss AB}\epsilon_{\ss A^\prime B^\prime}
+ \nabla^{\ss A}_{\ss A^\prime} X^{\ss B}_{\ss B^\prime}
+\nabla^{\ss B}_{\ss B^\prime} X^{\ss A}_{\ss A^\prime}\qquad {\rm
on}\ {\cal E}^k.
$$

\proof
If $k=2,3$ this theorem reduces to Corollaries \statement{3061} and
\statement{3062}.  Let $k>3$.  We have shown that
$$
\hab=\nabla^{\ss A}_{\ss A^\prime} X^{\ss B}_{\ss B^\prime}
+\nabla^{\ss B}_{\ss B^\prime} X^{\ss A}_{\ss A^\prime}+\kab,
$$
where $\kab$ is a natural spinor of order $k-1$.  A straightforward
induction argument now shows that $\kab$ can be reduced to a function of
the Penrose fields $\Psi^2$, $\Psibar^2$, $\Psi^3$, $\Psibar^3$ at the
expense of changing the vector field $X^{\ss A}_{\ss A^\prime}$ (the new
vector field is again denoted $X^{\ss A}_{\ss A^\prime}$).  We apply
Corollary \statement{3062} to the natural generalized symmetry $\kab$ to
show that
$$
\kab=
c\ \epsilon^{\ss AB}\epsilon_{\ss A^\prime B^\prime},
$$
and our classification of the natural generalized symmetries of the vacuum
Einstein equations is complete.\endEx

\vfill\eject

%
\sectionno4
\abovedisplayskip=12pt plus 1pt minus 3pt
\belowdisplayskip=12pt plus 1pt minus 3pt
\def\ss{\scriptscriptstyle}
\def\proof{\par\noindent {\bf Proof:\ }}
\def\eqskip{\noalign{\vskip6pt}}
\def\trace{{\rm tr\,}}
\def\Psibar{\overline\Psi{}}
\def\psibar{\overline\psi{}}
\def\phibar{\overline\phi{}}
\def\betabar{\overline\beta{}}
\def\alphabar{\overline\alpha{}}
\def\chibar{\overline\chi{}}

\def\hab{h^{\ss A\kern 0.5pt B}_{\ss A^{\kern -.8pt\prime}\kern -1.8pt
B^{\kern -.8pt\prime}}}
\def\dab{d^{\ss A\kern 0.5pt B}_{\ss A^{\kern -.8pt\prime}\kern -1.8pt
B^{\kern -.8pt\prime}}}
\def\kab{k^{\ss A\kern 0.5pt B}_{\ss A^{\kern -.8pt\prime}\kern -1.8pt
B^{\kern -.8pt\prime}}}
\overfullrule=0pt
\def\lineq{\equationlabel{2}{2001}{}}
\noindent{\bf 4. First-Order Generalized Symmetries.}

In this section we begin our classification of {\it all} generalized
symmetries of the vacuum Einstein equations by determining all first-order
generalized symmetries.  As mentioned in the introduction, the calculation
of the higher-order generalized symmetries reduces to that of the first-
order generalized symmetries.  While the analysis of the higher-order
symmetries is similar in spirit to that of the natural symmetries, as
presented in the previous section,  the analysis of the first-order
symmetries is rather more complex and merits a separate presentation.

To begin, let
$$
h_{ab}=h_{ab}(x^i,g_{ij},g_{ij},_{k})
$$
be the components of a first-order generalized symmetry.  We emphasize
that the functions $h_{ab}$ are no longer assumed to be the components of
a natural tensor, and hence may depend explicitly upon the coordinates
$x^i$ and the first derivatives of the metric $g_{ij},_k$.  The linearized
equations
$$
\left[-g^{cd}\delta^a_i\delta^b_j-g^{ab}\delta^c_i\delta^d_j
+g^{ac}\left(\delta^b_i\delta^d_j+\delta^b_j\delta^d_i\right)\right]\nabla_c\
nabla_dh_{ab}=0\eqtag401
$$
involve the metric and its first 3 derivatives, and must be satisfied when the
Einstein equations
$$
R_{ab}=0\quad {\rm and}\quad\nabla_cR_{ab}=0\eqtag402
$$
are satisfied.  In accordance with the results of \S2, we write $h_{ab}$ as a
new function
$$
h_{ab}=h_{ab}(x^i,g_{ij},\Gamma^i_{jk})
$$
and express the linearized equations in terms of the jet coordinates
$$
\{x^i, g_{ij}, \Gamma^i_{jk}, \Gamma^i_{jhk}, \Gamma^i_{jhkl},
Q_{ij,kl}, Q_{ij,klm}\}\eqtag4002
$$
for $J^3(\Q)$, which were introduced in \S2 (see
\equationlabel{2}{3102}{} and \equationlabel{2}{312}{}).
The Einstein equations \eq{402} hold if and only if the variables
$Q_{ij,kl}$ and $Q_{ij,klm}$ are completely trace-free.  Consequently,
the linearized equations \eq{401} for the first-order generalized symmetry
must hold identically for all values of
$$
\{x^i, g_{ij}, \Gamma^i_{jk}, \Gamma^i_{jhk}, \Gamma^i_{jhkl},
[Q_{ij,kl}]_{\ss\rm tracefree}, [Q_{ij,klm}]_{\ss\rm tracefree}\}.
$$

In order to determine the dependence of the linearized equations on our
adapted jet coordinates we will need the following structure equations for
the coordinates \eq{4002}:
$$
D_ig_{jk} = g_{jl}\Gamma^l_{ik} + g_{kl}\Gamma^l_{ij},\eqtag100
$$
$$
D_k\Gamma^h_{ij} = \Gamma^h_{ijk} +
{2\over3}Q^{\phantom{k}h}_{k\phantom{h},ij}
+\Gamma^h_{mi}\Gamma^m_{jk}+\Gamma^{h}_{mj}\Gamma^m_{ik},
\eqtag101
$$
$$
D_l\Gamma^h_{ijk}=\Gamma^h_{ijkl} +
{1\over2}Q^{\phantom{l}h}_{l\phantom{j},ijk}
-{2\over3}Q^{\phantom{l}m}_{l\phantom{m},(ij}\Gamma^h_{k)m} +
{4\over3}\Gamma^m_{(ik}R_{j)\phantom{h}lm}^{\phantom{j)}h}
-3\Gamma^m_{(ik}\Gamma^h_{j)ml},\eqtag102
$$
and (see \equationlabel{2}{3101}{})
$$
\nabla_mQ_{ij,kl}=Q_{ij,klm} + {1\over2}(Q_{m(i,j)kl} +
Q_{kl,ijm}).\eqtag103
$$

We will use the following notation.  The derivative of $h_{ab}$ with
respect to the metric $g_{rs}$ and connection variables $\Gamma^t_{rs}$
will be denoted by
$$
\partial^{rs}h_{ab}={\partial h_{ab}\over\partial g_{rs}}\quad {\rm
and}\quad\partial^{rs}_th_{ab} = {\partial
h_{ab}\over\partial\Gamma^t_{rs}}.
$$
Note that these quantities are symmetric in the indices $rs$ and $ab$.    If
$$
X=X^a{\displaystyle {\partial\hfill\over\partial x^a}},
\qquad
Y=Y^a{ {\partial\hfill\over\partial x^a}},\qquad
{\rm and}\qquad\alpha=\alpha_rdx^r,
$$
we let
$$
[\partial_gh](\alpha\alpha;XX)=\alpha_r\alpha_sX^aX^b(\partial^{rs}h_{ab
})
$$
and
$$
[\partial_\Gamma h](\alpha\alpha,Y;XX)=
\alpha_r\alpha_sY^tX^aX^b(\partial^{rs}_th_{ab}).
$$
We denote by $\alpha^\sharp$ the vector field obtained from the 1-form
$\alpha$ by ``raising the index'' with the metric,
$$
\alpha^\sharp=g^{rs}\alpha_s{\partial\hfill\over\partial x^r},
$$
and we denote by $X^\flat$ the 1-form obtained from the vector $X$ by
``lowering the index'' with the metric,
$$
X^\flat=g_{ij}X^idx^j.
$$
The natural pairing of $X$ and $\alpha$ is
$$
<X,\alpha>=X^i\alpha_i.
$$

\proclaim Proposition{\statementtag401 }.
Let $h_{ab}=h_{ab}(x^i,g_{ij},\Gamma_{jk}^i)$ be a first-order
generalized symmetry for the vacuum Einstein equations.  Then there are
zeroth-order quantities
$$
M_{bt}^{\phantom{b}s}=M_{bt}^{\phantom{b}s}(x^i,g_{ij})
$$
such that
$$
\partial^{rs}_th_{ab}=\delta^{(r}_{(a}M_{b)t}^{\phantom{b)}s)}.\eqtag1
$$

\proof
Since
$$
\eqalign{
\nabla_d h_{ab}&=D_dh_{ab}-\Gamma_{ad}^ih_{ib}-
\Gamma_{bd}^ih_{ai}\cr\eqskip
&=(\partial^{rs}_th_{ab})\Gamma^t_{rsd} + \{\star\},}
$$
where $\{\star\}$ denotes terms involving the variables $x^i$, $g_{ij}$,
$\Gamma^i_{jh}$, $Q_{i\phantom{j}hk}^{\phantom{i}j}$, we conclude
using equation \eq{102} and \eq{103} that
$$
\nabla_c\nabla_dh_{ab}=(\partial^{rs}_th_{ab})\Gamma^t_{rscd}+\{\star
\star\},
$$
where $\{\star \star\}$ denotes terms involving the variables
$x^i,g_{ij}$, $\Gamma^i_{jh}$, $\Gamma^i_{hjk}$,
$Q_{i\phantom{j}hk}^{\phantom{i}j}$,
$Q_{i\phantom{j}hkl}^{\phantom{i}j}$.
Hence, by differentiating the linearized equations \eq{401} with respect to
$\Gamma^t_{rscd}$ and contracting the result with
$X^iX^jY^t\alpha_r\alpha_s\alpha_c\alpha_d$, we arrive at
$$
\eqalign{
<\alpha^\sharp,\alpha>[\partial_\Gamma h]&(\alpha\alpha,Y;XX)\cr
&=<X,\alpha>\{-<X,\alpha>[\partial_\Gamma \trace h](\alpha\alpha,Y)
+2\,[\partial_\Gamma h](\alpha\alpha,Y;\alpha^\sharp X)\}.}\eqtag403
$$
Here we have defined the trace of $h_{ab}$ in the usual way:
$$
\trace h = g^{ab}h_{ab}.
$$
When $\alpha$ is a null 1-form, the expression in brackets on the right-
hand side of \eq{403} must vanish.  By Proposition
\statementlabel{2}{2044}, this implies that there are quantities
$M_{bt}^{\phantom{b}s}$ such that
$$
-<X,\alpha>[\partial_\Gamma \trace h](\alpha\alpha,Y)
+2\,[\partial_\Gamma h](\alpha\alpha,Y;\alpha^\sharp X)
=<\alpha^\sharp,\alpha>M(X,Y,\alpha),
$$
where
$$
M(X,Y,\alpha)=M_{bt}^{\phantom{b}s} X^bY^t\alpha_s.
$$
Thus \eq{403} reduces to
$$
[\partial_\Gamma h](\alpha\alpha,Y;XX)
=<X,\alpha>M(X,Y,\alpha).\eqtag404
$$
We have shown that equation \eq{404} is necessary for \eq{403} to hold.
It is also sufficient.  This is easily verified if we observe that \eq{404}
implies
$$
 [\partial_\Gamma h](\alpha\alpha,Y;\alpha^\sharp X)
={1\over2}\left(<\alpha^\sharp,\alpha>M(X,Y,\alpha)
+<X,\alpha>M(\alpha^\sharp,Y,\alpha)\right)
$$
and
$$
[\partial_\Gamma\trace h](\alpha\alpha;Y)=M(\alpha^\sharp,Y,\alpha).
$$

It remains to prove that $M_{bt}^{\phantom{b}s}$ is independent of the
connection variables $\Gamma_{jk}^i$.  To this end we first differentiate
equation \eq{404} with respect to $\Gamma_{jk}^i$ to obtain
$$
[\partial_\Gamma\partial_\Gamma h](\beta\beta,Z;\alpha\alpha,Y;XX)
=<X,\alpha>[\partial_\Gamma M](\beta\beta,Z;X,Y,\alpha).\eqtag407
$$
The left-hand side of this equation is symmetric under interchange of
$(\beta,Z)$ with $(\alpha,Y)$, and therefore
$$
<X,\alpha>[\partial_\Gamma M](\beta\beta,Z;X,Y,\alpha)
=<X,\beta>[\partial_\Gamma M](\alpha\alpha,Y;X,Z,\beta).
$$
Using Proposition \statementlabel{2}{2041} we conclude that
$[\partial_\Gamma M]$ takes the form
$$
[\partial_\Gamma M](\beta\beta,Z;X,Y,\alpha)
=<X,\beta>W(\alpha,\beta,Y,Z),\eqtag408
$$
where $W$ has the symmetry property
$$
W(\alpha,\beta,Y,Z)=W(\beta,\alpha,Z,Y).
$$
Equation \eq{407} becomes
$$
[\partial_\Gamma\partial_\Gamma h](\beta\beta,Z;\alpha\alpha,Y;XX)
=<X,\alpha><X,\beta>W(\alpha,\beta,Y,Z).\eqtag4071
$$

Next we observe that the structure equations \eq{100}--\eq{103} imply
$$
\nabla_c\nabla_dh_{ab}=(\partial^{uv}_w\partial^{rs}_th_{ab})\Gamma^t
_{rsd}\Gamma^w_{uvc}+\{\star\},$$
where $\{\star\}$ denotes terms that are at most linear in the coordinates
$\Gamma^i_{jhk}$.  Using this equation, we now differentiate the
linearized equations with respect to $\Gamma^t_{rsd}$ and
$\Gamma^w_{uvc}$ to find that
$$
\eqalign{
<\beta^\sharp&,\alpha>[\partial_\Gamma\partial_\Gamma h]
(\beta\beta,Z;\alpha\alpha,Y;XX)
+<X,\beta><X,\alpha>[\partial_\Gamma\partial_\Gamma \trace h]
(\beta\beta,Z;\alpha\alpha,Y)\cr
&=<X,\beta>[\partial_\Gamma\partial_\Gamma h]
(\beta\beta,Z;\alpha\alpha,Y;X\alpha^\sharp)
+<X,\alpha>[\partial_\Gamma\partial_\Gamma h]
(\beta\beta,Z;\alpha\alpha,Y;X\beta^\sharp).}
$$
Into this equation we substitute from equation \eq{4071} to deduce that
$$
\eqalign{
[<\beta^\sharp,\alpha><X,\alpha><X,\beta>-
{1\over2}<X,\beta>^2<\alpha^\sharp,\alpha>&-
{1\over2}<X,\alpha>^2<\beta^\sharp,\beta>]\cr
&\times W(\alpha,\beta,Y,Z)=0.}
$$
Because the expression in square brackets is not identically zero, this
equation implies that $W=0$ and therefore $\partial_\Gamma M=0$, as
claimed.\endEx

Next we turn to an analysis of the terms involving $Q_{ij,hkl}$ in the
linearized equations \eq{401}.  In the following proposition we let
$$
M^{\phantom{a}sr}_a=M_{at}^{\phantom{b}s}g^{rt}\quad{\rm
and}\quad
M^{asr}=g^{ab}M_{bt}^{\phantom{b}s}g^{rt}.
$$

\proclaim Proposition{\statementtag402 }.
If $h_{ab}=h_{ab}(x^i,g_{ij},\Gamma^i_{hk})$ is a first-order
generalized symmetry of the vacuum Einstein equations, then there are
quantities
$$
\eqalignno{
V^a&=V^a(x^i,g_{ij})\cr
\noalign{\hbox{such that}}
M^{\phantom{a}[sr]}_a&=\delta_a^{[s}V^{r]}.\aleqtag40801 }
$$

\proof
Because
$$
\nabla_dh_{ab}={2\over3}(\partial^{rs}_th_{ab})
Q^{\phantom{d}t}_{d\phantom{h},rs}
+(\partial^{rs}_th_{ab})\Gamma^t_{rsd}
+\{\star\},
$$
where $\{\star\}$ denotes terms involving the variables $x^i$, $g_{ij}$,
$\Gamma^i_{jk}$,
we can show
$$
\nabla_c\nabla_dh_{ab}
=
{2\over3}(\partial^{rs}_hh_{ab})Q^{\phantom{c}h}_{d\phantom{h},rs|c}
+{1\over2}(\partial^{rs}_hh_{ab})Q^{\phantom{c}h}_{c\phantom{h},rsd
}
+\{\star\star\}
$$
where $\{\star\star\}$ now indicates terms involving the variables
$x^i,g_{ij}$, $\Gamma_{ij}^k$, $\Gamma_{ijh}^k$, $Q_{ij,kl}$.
Therefore, for the linearized equations to hold we must have that
$$
[-g^{cd}\delta^a_i\delta^b_j  -g^{ab}\delta^c_i\delta^d_j +
g^{ac}\delta^b_i\delta^d_j + g^{bc}\delta^a_j\delta^d_i]
[{2\over3}(\partial^{rs}_hh_{ab})Q^{\phantom{c}h}_{d\phantom{h},rs|c
}
+{1\over2}(\partial^{rs}_hh_{ab})Q^{\phantom{c}h}_{c\phantom{h},rsd
}]
=0\eqtag409
$$
for all $Q^{\phantom{c}h}_{c\phantom{h},rs|c}$ and
$Q^{\phantom{c}h}_{c\phantom{h},rsd}$ that are completely trace-free.
We multiply \eq{409} by $X^iX^j$ and substitute for
$\partial^{rs}_hh_{ab}$ from Proposition \statement{401} and for
$Q^{\phantom{c}h}_{c\phantom{h},rsd}$ and
$Q^{\phantom{d}h}_{d\phantom{h},rs}$, from \equationlabel{2}{312}{}
to obtain
$$
\eqalign{
&[-M^{bsh}X^cX^d+M^{csh}X^bX^d]\cr
&\times[{1\over12}(R_{bhcs|d}+R_{dhcb|s}+R_{shcd|b}
+R_{bhcd|s} + R_{shcb|d} + R_{dhcs|b})
+{1\over3}(R_{bhds|c}+R_{shdb|c})]=0.}
$$
By using the algebraic curvature symmetries and the Bianchi identities,
every term in this equation may be expressed as either a multiple of
$M^{bsh}X^cX^dR_{dhbc|s}$ or $M^{bsh}X^cX^dR_{shbc|d}$.  The
coefficient of the former term vanishes, while that of the latter term is one.
Thus \eq{409} holds if and only if
$$
M^{bsh}X^cX^d[R_{shbc|d}]_{\ss\rm tracefree}=0.\eqtag410
$$

To analyze this condition it is convenient to revert to spinors. We set
$$
M^{\ss BB^\prime AA^\prime HH^\prime}=M^{bst}\sigma_b^{\ss
BB^\prime}\sigma_s^{\ss AA^\prime}\sigma_t^{\ss HH^\prime},
$$
and use \equationlabel{2}{34}{} and \equationlabel{2}{39}{} to write
$$
[R_{shbc|d}]_{\ss\rm tracefree}\longleftrightarrow
\epsilon_{\ss SH}\epsilon_{\ss BC}\Psibar_{\ss S^\prime H^\prime
B^\prime C^\prime D^\prime D}
+\epsilon_{\ss S^\prime H^\prime }\epsilon_{\ss B^\prime C^\prime
}\Psi_{\ss SHBCDD^\prime},
$$
so that the condition \eq{410} is equivalent to
$$
X^{\ss CC^\prime}X^{\ss DD^\prime}M^{\ss BB^\prime SS^\prime
HH^\prime}[\epsilon_{\ss SH}\epsilon_{\ss BC}\Psibar_{\ss S^\prime
H^\prime B^\prime C^\prime D^\prime D}
+\epsilon_{\ss S^\prime H^\prime }\epsilon_{\ss B^\prime C^\prime
}\Psi_{\ss SHBCDD^\prime}]=0\eqtag411
$$
for all Penrose spinors $\Psi^3$ and $\Psibar^3$.  We differentiate this
expression with respect to $\Psi_{\ss SHBCDD^\prime}$ and multiply the
resulting equation by $\psi_{\ss S} \psi_{\ss H} \psi_{\ss B} \psi_{\ss C}
\psi_{\ss D} \psibar_{\ss D^\prime}$ to conclude
$$
\epsilon_{\ss A^\prime H^\prime}\psi_{\ss A}\psi_{\ss H}\psi_{\ss B}
M^{\ss BB^\prime AA^\prime HH^\prime}=0.\eqtag412
$$
Similarly, differentiation of \eq{411} with respect to $\Psibar_{\ss
S^\prime H^\prime B^\prime C^\prime D^\prime D}$ leads to
$$
\epsilon_{\ss A H}\psibar_{\ss A^\prime}\psibar_{\ss
H^\prime}\psibar_{\ss B^\prime}
M^{\ss BB^\prime AA^\prime HH^\prime}=0.\eqtag413
$$

To solve equations \eq{412} and \eq{413} we decompose $M$ as
$$
M^{\ss BB^\prime AA^\prime HH^\prime}
=P^{\ss BB^\prime AA^\prime HH^\prime}
+S^{\ss BB^\prime}\epsilon^{\ss AH}\epsilon^{\ss A^\prime H^\prime }
+T^{\ss BB^\prime A^\prime H^\prime}\epsilon^{\ss AH}
+\overline T^{\ss BB^\prime AH}\epsilon^{\ss A^\prime
H^\prime},\eqtag414
$$
where the spinors $P,T,\overline T$ are each symmetric in the indices
$AH$ and $A^\prime H^\prime$.  Note that the spinors $T$ and $\overline
T$ correspond to the skew symmetric part of $M$ in \eq{40801}.
Equations \eq{412} and \eq{413} now imply that
$$
\eqalignno{
\psi_{\ss A}\psi_{\ss H}\psi_{\ss B}\overline T^{\ss BB^\prime AH
}&=0\cr
\noalign{\hbox{and}}
\psibar_{\ss A^\prime}\psibar_{\ss H^\prime}\psibar_{\ss B^\prime}T^{\ss
BB^\prime A^\prime H^\prime}&=0.}
$$
These equations can be analyzed using Proposition
\statementlabel{2}{202}; we find that there must exist quantities $V^{\ss
AA^\prime}$ such that
$$
\eqalignno{
T^{\ss BB^\prime A^\prime H^\prime}&=\epsilon^{\ss A^\prime B^\prime
} V^{\ss BH^\prime }+\epsilon^{\ss H^\prime B^\prime } V^{\ss
BA^\prime },\aleqtag4150
\noalign{\hbox{and}}
\overline T^{\ss BB^\prime A H}&=\epsilon^{\ss AB}\overline V^{\ss
B^\prime H}+\epsilon^{\ss HB}\overline V^{\ss B^\prime A}. \aleqtag4151
}
$$
We insert \eq{4150} and \eq{4151} into \eq{414};  note that only the real
part of $V^{\ss AA^\prime}$ appears. We then write the resulting equation
in tensor form to complete the proof.\endEx

We now turn to an analysis of the conditions arising from the
$\Gamma^r_{stu}\Gamma^m_{pq}$ terms in the linearized equation.  This
analysis will enable us to prove that every first-order generalized
symmetry is, modulo a generalized diffeomorphism symmetry, an
evolutionary zeroth-order symmetry.

\proclaim Proposition{\statementtag403 }.
Let $h_{ab}=h_{ab}(x^i,g_{ij},\Gamma_{ij}^k)$ be a first-order
generalized symmetry of the vacuum Einstein equations.  Then there are
zeroth-order quantities $V_i=V_i(x^i,g_{ij})$ and $\widehat
h_{ab}=\widehat h_{ab}(x^i,g_{ij})$ such that
$$
h_{ab}=\widehat h_{ab}+\nabla_aV_b+\nabla_bV_a.
$$

\proof
Let
$$
\widehat h_{ab}=h_{ab}-(\nabla_aV_b + \nabla_bV_a),
$$
where $V_a$ is defined by Proposition \statement{402}.  Then $\widehat
h_{ab}$ is a first-order generalized symmetry and therefore, by
Proposition \statement{401}, there exist zeroth-order quantities $\widehat
M_{at}^{\phantom{b}s}=\widehat M_{at}^{\phantom{b}s}(x^i,g_{ij})$
such that
$$
\partial^{rs}_t\widehat h_{ab}=\delta^{(r}_{(a}\widehat
M_{b)t}^{\phantom{b)}s)}.\eqtag417
$$
Moreover, by construction, $\widehat M$ will satisfy Proposition
\statement{402} with $V^i=0$, and hence
$$
\widehat M^{bst}=\widehat M^{bts}.\eqtag418
$$
This symmetry condition will allow us to prove, from the coefficient of
$\Gamma^r_{stu}\Gamma^m_{pq}$ in the linearized equations, that
$\widehat M^{bst}=0$, that is,
$$
\widehat h_{ab}=\widehat h_{ab}(x^i,g_{ij}).
$$

The derivation of the condition arising from the coefficient of
$\Gamma^r_{stu}\Gamma^m_{pq}$ in the linearized equations is the
longest single calculation in this paper.  To begin we first compute
$$
\eqalign{
\alpha_s\alpha_t\alpha_u\partial^{stu}_r(\nabla_c\nabla_d\widehat h_{ab})
=&\alpha_s\alpha_t\alpha_u[D_c(\partial_r^{st}\widehat h_{ab})\delta^u_d
+\delta^u_c\partial^{st}_r\nabla_d\widehat h_{ab}
-3\delta^u_c\Gamma_{ij}^t\delta^{(s}_d(\partial_r^{ij)}\widehat
h_{ab})\cr
\eqskip
&-\Gamma_{cd}^s(\partial^{tu}_r\widehat h_{ab})
-\Gamma_{ac}^l\delta^s_d(\partial_r^{tu}\widehat h_{lb})
-\Gamma_{bc}^l\delta^s_d(\partial_r^{tu}\widehat h_{la})].}\eqtag419
$$
The second term on the right-hand side of this equation is found to be
$$
\eqalign{
\alpha_s\alpha_t\partial^{st}_r\nabla_d\widehat h_{ab}
=&\alpha_s\alpha_t\big[D_d(\partial^{st}_r\widehat
h_{ab})+2g_{jr}\delta^t_d(\partial^{sj}\widehat h_{ab})
+2\Gamma^t_{jd}(\partial^{sj}_r\widehat
h_{ab})+2\delta^t_d\Gamma^h_{ri}(\partial^{is}_h\widehat h_{ab})\cr
\eqskip&-\Gamma_{ad}^l(\partial^{st}_r\widehat h_{lb})
-\Gamma_{bd}^l(\partial^{st}_r\widehat h_{la})
-\delta^t_d\delta^s_{a}\widehat h_{rb}
-\delta^t_d\delta^s_{b}\widehat h_{ra}\big].}\eqtag420
$$
Together, equations \eq{419} and \eq{420} imply that
$$
\eqalign{
&X^rY^m\alpha_s\alpha_t\alpha_u\beta_p\beta_q[\partial^{stu}_r\partial^{
pq}_m(\nabla_c\nabla_d\widehat h_{ab})]\cr\eqskip
&=4\beta_{(c}\alpha_{d)}[\partial_g\partial_\Gamma\widehat
h_{ab}](\beta Y^\flat;\alpha\alpha,X)
+2\alpha_c\alpha_d[\partial_g\partial_\Gamma\widehat h_{ab}](\alpha
X^\flat;\beta\beta,Y)\cr\eqskip
&-\alpha_c\beta_a\beta_dY^m[\partial_\Gamma\widehat
h_{mb}](\alpha\alpha,X)
-\alpha_c\beta_b\beta_dY^m[\partial_\Gamma\widehat
h_{ma}](\alpha\alpha,X)
-\alpha_a\alpha_c\alpha_dX^m[\partial_\Gamma\widehat
h_{mb}](\beta\beta,Y)\cr\eqskip
&-\alpha_b\alpha_c\alpha_dX^m[\partial_\Gamma\widehat
h_{ma}](\beta\beta,Y)
-\beta_a\beta_c\alpha_dY^m[\partial_\Gamma\widehat
h_{mb}](\alpha\alpha,X)
-\beta_b\beta_c\alpha_dY^m[\partial_\Gamma\widehat
h_{ma}](\alpha\alpha,X)\cr\eqskip
&+2\alpha_c\alpha_d<X,\beta>[\partial_\Gamma\widehat
h_{ab}](\alpha\beta,Y)
-\alpha_c\alpha_d<Y,\alpha>[\partial_\Gamma\widehat
h_{ab}](\beta\beta,X)\cr\eqskip
&-\beta_c\beta_d<Y,\alpha>[\partial_\Gamma\widehat
h_{ab}](\alpha\alpha,X).}
$$
We substititute this equation into the linearized equations \eq{401}
multiplied by $Z^iZ^j$ and use \eq{417} to obtain, after considerable
algebraic simplifications,
$$
\eqalignno{
&2<Z,\alpha>^2\{[\partial_g\widehat M](\beta
Y^\flat;\beta^\sharp,X,\alpha)
-\partial_g\widehat M(\alpha X^\flat;\beta^\sharp,Y,\beta)\}\cr
+&2<Z,\alpha><Z,\beta>\ \{[\partial_g\widehat M](\alpha
X^\flat;\alpha^\sharp,Y,\beta)
-\partial_g\widehat M(\beta Y^\flat;\alpha^\sharp,X,\alpha)\}\cr
+&2<Z,\alpha><\alpha^\sharp,\beta>\{[\partial_g\widehat M](\alpha
X^\flat;Z,Y,\beta)
-\partial_g\widehat M(\alpha X^\flat;Z,Y,\beta)\}\cr
+&2<\alpha^\sharp,\alpha><Z,\beta>\{[\partial_g\widehat M](\beta
Y^\flat;Z,X,\alpha)
-\partial_g\widehat M(Z^\flat X^\flat;Z,X,\alpha)\}\cr
&-<Z,\alpha>^2<\beta^\sharp,\beta>\widehat M(Y,X,\alpha)
-<Z,\beta>^2<\alpha^\sharp,\alpha>\widehat M(Y,X,\alpha)\aleqtag421
&+<Z,\alpha>^2<\alpha,Y>\widehat M(\beta^\sharp,X,\beta)
-<Z,\alpha>^2<X,\beta>\widehat M(\beta^\sharp,Y,\alpha)\cr
&+\big[<\alpha^\sharp,\alpha><Y,\alpha><Z,\beta>
+<Z,\alpha><Y,\alpha><\alpha^\sharp,\beta>\big]\widehat M(Z,X,\beta)\cr
&+\big[<Z,\alpha><X,\beta><\alpha^\sharp,\beta>
-<\alpha^\sharp,\alpha><X,\beta><Z,\beta>\big]\widehat M(Z,Y,\alpha)\cr
&-<Z,\alpha><Y,\alpha><Z,\beta>\widehat M(\alpha^\sharp,X,\beta)
+<Z,\alpha><X,\beta><Z,\beta>\widehat M(\alpha^\sharp,Y,\alpha)
\cr
&+2<Z,\alpha><Z,\beta><\alpha^\sharp,\beta>\widehat M(Y,X,\alpha)=0.}
$$
As a check of the accuracy of this equation, we used {\sl Maple} to verify
that the diffeomorphism symmetry, for which
$$
\widehat M(\alpha,X,Z)=2[\partial_gV](Z^\flat\alpha;X)
-<X,\alpha>V(Z),
$$
and $V_i=V_i(x^i,g_{kl})$, provides a solution to \eq{421}.

In order to simplify equation \eq{421} using \eq{418} we set
$$
\widehat N_a^{\phantom{a}sr}=\widehat M_{at}^{\phantom{at}s}g^{rt},
$$
and
$$
\widehat N(Z,\beta,\alpha)=\widehat
N_a^{\phantom{a}sr}Z^a\beta_s\alpha_r,
$$
and observe that
$$
\eqalign{
[\partial_g\widehat M](\beta\gamma;Z,X,\alpha)
=&[\partial_g\widehat N](\beta\gamma;Z,X^\flat,\alpha)\cr
&+{1\over2}<X,\beta>\widehat N(Z,\gamma,\alpha)
+{1\over2}<X,\gamma>\widehat N(Z,\beta,\alpha).}
$$
We substitute this equation into \eq{421} and use the fact that
$$
\widehat N(Z,\alpha,\beta)=\widehat N(Z,\beta,\alpha)\eqtag422
$$
to deduce, again after lengthy algebraic simplifications, that
$$
\eqalign{
<Z,\alpha>&^2K(\beta,Y,\beta,\alpha,X)
+<Z,\alpha><Z,\beta>K(\alpha,X,\alpha,\beta,Y)\cr
&+[<Z,\alpha><\alpha^\sharp,\beta>-
<Z,\beta><\alpha^\sharp,\alpha>]K(\alpha,X,Z^\flat,\beta,Y)=0,}\eqtag423
$$
where
$$
\eqalign{
K(\alpha,X,Z,\beta,Y)=\ &[\partial_g\widehat N](\alpha
X^\flat;Z,Y^\flat,\beta)
-[\partial_g\widehat N](\beta Y^\flat;Z,X^\flat,\alpha)\cr
&+{1\over2}[\partial_g\widehat N](\beta Z^\flat;Y,\alpha,X^\flat)
-{1\over2}<Z,\alpha>\widehat N(X,\beta,Y^\flat).}\eqtag424
$$
Equation \eq{423} implies that $K(\alpha,X,Z^\flat,\beta,Y)=0$ whenever
$<Z,\alpha>=0$.  Therefore, by Proposition \statementlabel{2}{2041},
there exist quantities $L$ such that
$$
K(\alpha,X,Z^\flat,\beta,Y)=<Z,\alpha>L(X,\beta,Y).
$$
Substituting this expression back into \eq{423} and simplifying the result,
we find
$$
<\beta^\sharp,\beta>L(Y,\alpha,X)
+<\alpha^\sharp,\beta>L(X,\beta,Y)=0.
$$
In this equation we set $\alpha=\beta$ to conclude that $L=0$ and hence
$K=0$.

On account of the symmetry \eq{422} of $\widehat N$, the condition
$K=0$ implies that
$$
<Z,\alpha>\widehat N(X,\beta,Y^\flat)=<Z,X^\flat>\widehat
N(\alpha^\sharp,\beta,Y^\flat),\eqtag425
$$
which easily implies that $\widehat N=0$ and thus $\widehat
M=0$.\qed\medskip

We are now ready to complete our classification of first-order generalized
symmetries.

\proclaim Theorem{\statementtag404 }.
Let $h_{ab}=h_{ab}(x^i,g_{ij},\Gamma_{ij}^k)$ be a first-order
generalized symmetry of the vacuum Einstein equations.  Then there is a
constant $c$ and zeroth-order quantities $V_i=V_i(x^i,g_{ij})$ such that
$$
h_{ab} = c g_{ab} +\nabla_aV_b + \nabla_bV_a.
$$

\proof
Proposition \statement{403} reduces the proof to showing that the zeroth-
order symmetry $\widehat h_{ab}$ is in fact a constant times the metric.
This follows from the classification of the point symmetries of the Einstein
equations \refto{Ibragimov1985}.  We include the proof here for
completeness.

Let us begin with the conditions placed on $\widehat h_{ab}$ by the
vanishing of the terms in the linearized equations involving
$\Gamma^a_{bcd}$.  From the structure equations \eq{100}--\eq{102} it
is a straightforward matter to show that
$$
\nabla_c\nabla_d\widehat h_{ab}=2 {\partial \widehat h_{ab}\over\partial
g_{mn}}
g_{mp}
[\Gamma^p_{ncd}+{1\over2}Q^{\phantom{c}p}_{c\phantom{p}nd}]
-\widehat
h_{pa}[\Gamma^p_{bdc}+{1\over2}Q^{\phantom{c}p}_{c\phantom{p}db
}]
-\widehat
h_{pb}[\Gamma^p_{adc}+{1\over2}Q^{\phantom{c}p}_{c\phantom{p}da
}]
+\{\star\},\eqtag426
$$
where $\{\star\}$ denotes terms depending only on the variables $x^i$,
$g_{ij}$, $\Gamma_{ij}^k$.
We multiply the linearized equations by $X^iX^j$ and differentiate them
with respect to $\Gamma^a_{bcd}$.  The result, after multiplying by
$\alpha_b\alpha_c\alpha_dZ^a$ and simplifying, is given by
$$
<\alpha^\sharp,\alpha>[\partial_g\widehat h](Z^\flat\alpha;XX)
=<\alpha,X>\{2[\partial_g\widehat h](Z^\flat\alpha;\alpha^\sharp X)
-<\alpha,X>[\partial_g{\rm tr}\widehat h](Z^\flat\alpha)\}.\eqtag427
$$
Proposition \statementlabel{2}{2041} now implies that there exist zeroth-
order quantities $A$ such that
$$
[\partial_g\widehat h](Z^\flat\alpha;XX)=<\alpha,X>A(Z^\flat,X).
$$
The symmetry of $(\partial_g\widehat h)$ in $Z^\flat\alpha$ implies that
$$
<\alpha,X>A(Z^\flat,X)=<Z^\flat,X>A(\alpha,X),
$$
and therefore, by Proposition \statementlabel{2}{2041}, there exists a
zeroth-order function $F=F(x^i,g_{ij})$ such that
$$
A(\alpha,X)=<\alpha,X>F.
$$
We have therefore found that
$$
[\partial_g\widehat h](\alpha\alpha;XX)=<\alpha,X>^2 F.\eqtag428
$$
It is easily verified that this equation is necessary and sufficient for
\eq{427} to hold.  Next, we differentiate \eq{428} with respect to $g_{ij}$
to obtain
$$
[\partial_g\partial_g\widehat h](\beta\beta;\alpha\alpha;XX)=<\alpha,X>^2
[\partial_gF](\beta\beta).
$$
The left-hand side of this equation is symmetric under interchange of
$\alpha$ and $\beta$, and we therefore have
$$
<\alpha,X>^2 [\partial_gF](\beta\beta)=<\beta,X>^2
[\partial_gF](\alpha\alpha).
$$
{}From Proposition \statementlabel{2}{2041} it is easily seen that this
equation implies
$$
[\partial_gF](\alpha\alpha)=0.\eqtag429
$$
Equations \eq{428}, \eq{429} imply that $\widehat h_{ab}$ is of the form
$$
\widehat h_{ab}= F(x^i)g_{ab} + k_{ab}(x^i).\eqtag430
$$

Now we turn to the conditions on $\widehat h_{ab}$ arising from the
terms in the linearized equations depending on $Q_{ab,cd}$.  It is
straightforward to show, using \eq{426}, that this condition takes the form
$$
Q_{ij,kl}[2X^iX^cg^{rk}{\partial\widehat h_{rc}\over\partial g_{jl}}
-X^iX^kg^{bc}{\partial\widehat h_{bc}\over\partial g_{jl}}
-{3\over2}X^iX^j\widehat h^{kl}]=0,
$$
when $Q_{ij,kl}$ is completely trace-free.
If we substitute from \eq{430} the first and second terms vanish leaving us
with
$$
X^iX^j k_{ab}g^{ak}g^{bl}[Q_{ij,kl}]_{\ss\rm tracefree}=0.
$$
Because $k_{ab}$ is independent of the metric, this equation implies that
$k_{ab}=0.$

We have reduced $\widehat h_{ab}$ to the form
$$
\widehat h_{ab}= F(x^i)g_{ab}.
$$
We now substitute this equation for $\widehat h_{ab}$ into the linearized
equations to find
$$
-g_{ij}\nabla^a\nabla_aF-2\nabla_i\nabla_jF=0.
$$
We differentiate this equation with respect to $\Gamma^r_{st}$ and obtain
$$
[g_{ih}g^{st}+2\delta_{(i}^{(s}\delta_{j)}^{t)}]{\partial F\over\partial
x^r}=0,
$$
which implies that $\displaystyle{{\partial F\over\partial x^r}}=0$, and
thus $F$ is a constant.\endEx
\vfill\eject

%
\sectionno5
\abovedisplayskip=12pt plus 1pt minus 3pt
\belowdisplayskip=12pt plus 1pt minus 3pt
\def\ss{\scriptscriptstyle}
\def\gammabar{\overline\gamma}
\def\proof{\par\noindent {\bf Proof:\ }}
\def\eqskip{\noalign{\vskip6pt}}
\def\Psibar{{\overline\Psi}{}}
\def\psibar{\overline\psi{}}
\def\phibar{\overline\phi{}}
\def\betabar{\overline\beta{}}
\def\alphabar{\overline\alpha{}}
\def\chibar{\overline\chi{}}

\def\hab{h^{\ss A\kern 0.5pt B}_{\ss A^{\kern -.8pt\prime}\kern -1.8pt
B^{\kern -.8pt\prime}}}
\def\dab{d^{\ss A\kern 0.5pt B}_{\ss A^{\kern -.8pt\prime}\kern -1.8pt
B^{\kern -.8pt\prime}}}
\def\kab{k^{\ss A\kern 0.5pt B}_{\ss A^{\kern -.8pt\prime}\kern -1.8pt
B^{\kern -.8pt\prime}}}
\overfullrule=0pt
\def\lineq{\eq{300}}
\noindent{\bf 5.  Complete Classification of Generalized Symmetries of the
Vacuum Einstein Equations.}

We now turn to the computation of all generalized symmetries of the
Einstein equations.  Let
$$\hab=\hab(x,\sigma,\Gamma^1,\Gamma^2,
\Psi^2,\Psibar^2,\ldots,\Gamma^{l},\Psi^k,\Psibar^k)\eqtag502
$$
be the components of a generalized symmetry of the Einstein equations.
Initially, we have $l=k$, so the generalized symmetry is of order $k$.
The repeated covariant derivative of $\hab$ can be given schematically by
$$
\nabla\nabla h = DDh + \gamma\cdot D h + (D\gamma)\cdot h +
\gamma\cdot\gamma\cdot h,
$$
where $\gamma\cdot D h$ is a sum of products of spin connections
$\gamma_{\ss AA^\prime}^{\ss BC}$ and $\overline\gamma_{\ss
AA^\prime}^{\ss B^\prime C^\prime }$ and total derivatives $D_{\ss
C^\prime}^{\ss C}\hab$, and so on.  The linearized equation,
$$
\eqalign{
[-\epsilon_{\ss CD}\epsilon^{\ss C^\prime D^\prime}\alpha_{\ss A}
\beta_{\ss B}\alphabar^{\ss A^\prime}\betabar^{\ss B^\prime}
&+\epsilon_{\ss BC}\epsilon^{\ss A^\prime C^\prime}\alpha_{\ss A}
\beta_{\ss D}\alphabar^{\ss B^\prime}\betabar^{\ss D^\prime}\cr
&+\epsilon_{\ss BC}\epsilon^{\ss A^\prime C^\prime}\alpha_{\ss D}
\beta_{\ss A}\alphabar^{\ss D^\prime}\betabar^{\ss B^\prime}]
\nabla_{\ss C^\prime}^{\ss C}\nabla_{\ss D^\prime}^{\ss D}\hab
=0\qquad{\rm on}\  {\cal E}^{k+2},}\eqtag44
$$
is an $SL(2,{\bf C})$ invariant identity depending on the variables
$x^i$, $\sigma_{a{\ss AA^\prime}}$, $\sigma_{a{\ss AA^\prime},b}$,
$\sigma_{a{\ss AA^\prime},bc}$, $\Gamma^1$, $\Gamma^2$, $\Psi^2$,
$\Psibar^2$, \dots, $\Gamma^{l+2}$, $\Psi^{k+2}$, $\Psibar^{k+2}$.  On
the Einstein equation manifold ${\cal E}^{k+2}$ there are relationships
between $\sigma_{a{\ss AA^\prime},bc}$ and $\Gamma^2$, $\Psi^2$,
$\Psibar^2$, but in what follows we are careful only to consider terms
involving $\Psi^l$ and $\Psibar^l$ for $l\geq3$.  The rather complicated
lower-derivative analysis was performed in \S4.

In order to analyze the dependence of this equation on our adapted jet
coordinates, we need the following structure equations on ${\cal
E}^{k+1}$:
$$
\eqalign{
D_{j_{k+1}}\Gamma^i_{j_0j_1\cdots j_k}&
=\Gamma^i_{j_0j_1\cdots j_{k+1}}
+ A^i_{j_0j_1\cdots j_{k+1}}(\sigma,\Psi^{k+1},\Psibar^{k+1})
+ B^i_{j_0j_1\cdots j_{k+1}}(\Gamma^1,\Gamma^k)\cr
&+ C^i_{j_0j_1\cdots j_{k+1}}(\sigma,\Gamma^1,\Psi^k,\Psibar^k)
+ E^i_{j_0j_1\cdots j_{k+1}}(\sigma,\Gamma^1,\ldots,\Gamma^{k-
1},\Psi^2,\Psibar^2,\ldots,\Psi^{k-1},\Psibar^{k-1}).}
\eqtag215
$$
Here $A^{\cdots}_{\cdots}$ is linear in $\Psi^k$ and $\Psibar^k$,
$B^{\cdots}_{\cdots}$ is bilinear in its arguments, $C^{\cdots}_{\cdots}$
is linear in $\Psi^k$ and $\Psibar^k$ with coefficients depending on
$\sigma$ and $\Gamma^1$.

We also have (see \equationlabel{2}{39}{})
$$
\eqalign{
D_{\ss A}^{\ss A^\prime}\Psi_{\ss J_1\cdots J_{k+2}}^{\ss J_1^\prime
\cdots J_{k-2}^\prime}&=\Psi_{\ss A\ J_1\cdots J_{k+2}}^{\ss A^\prime
J_1^\prime \cdots J_{k-2}^\prime}
+M_{\ss A\ J_1\cdots J_{k+2}}^{\ss A^\prime J_1^\prime \cdots J_{k-
2}^\prime}(\gamma,\gammabar,\Psi^k)
+N_{\ss A\ J_1\cdots J_{k+2}}^{\ss A^\prime J_1^\prime \cdots J_{k-
2}^\prime}(\Psi^2,\Psibar^2,\dots,\Psi^{k-1},\Psibar^{k-1}),}
\eqtag37
$$
where $M^{\cdots}_{\cdots}$ is linear in $\Psi^k$.  There is an analogous
formula for the total derivative of $\Psibar^k$.

Let
$$
f(\sigma, \Gamma^1, \Gamma^2, \Psi^2, \Psibar^2, \ldots, \Gamma^l,
\Psi^k, \Psibar^k)
$$
be a smooth function.  We retain the notation
$$
[\partial^m_\Psi f](\psi^{m+2}, \psibar^{m-2})\quad{\rm and}\quad
[\partial^m_\Psibar f](\psi^{m-2}, \psibar^{m+2})
$$
introduced in \S3 for the derivatives of $f$ with respect to $\Psi^m$ and
$\Psibar^m$, and we define
$$
[\partial^m_\Gamma f](Y, \omega^{m+1})
={\partial f\over\partial \Gamma^i_{j_0j_1\dots j_m}}
Y^i\omega_{j_0}\omega_{j_1}\cdots \omega_{j_m}.
$$
In many of our subsequent formulas the spinor components
$$
\omega_{\ss A^\prime}^{\ss A}=\sigma^{j{\ss A}}_{\ss
A^\prime}\omega_j
$$
of the covector $\omega$ will appear.  In addition, we will use $\omega$
as a bilinear map
$$
\omega(\alpha,\betabar)=\omega^{\ss A}_{\ss A^\prime}\alpha_{\ss
A}\betabar^{\ss A^\prime}.
$$
Finally, we write
$$
h(\alpha,\omega,\alphabar)=\hab\alpha_{\ss A}\omega_{\ss B}^{\ss
A^\prime}\alphabar^{\ss B^\prime}.
$$

{}From the structure equations \eq{215}--\eq{37} we readily derive the
following commutation rules. For $l\geq2$ we have
$$
[\partial^{l+1}_\Gamma D_{\ss A^\prime}^{\ss A}f](Y,\omega^{l+2})
=\omega_{\ss A^\prime}^{\ss A}[\partial^l_\Gamma
f](Y,\omega^{l+1})\eqtag5001
$$
and
$$
[\partial^{l}_\Gamma D_{\ss A^\prime}^{\ss A}f](Y,\omega^{l+1})
=\omega_{\ss A^\prime}^{\ss A}[\partial^{l-1}_\Gamma
f](Y,\omega^{l})+\left(D_{\ss A^\prime}^{\ss A}[\partial^l_\Gamma
f]\right)(Y,\omega^{l+1})+ [\Gamma^1\cdot\partial^l_\Gamma f]_{\ss
A^\prime}^{\ss A}(Y,\omega^{l+1}),\eqtag5002
$$
while for $l<k$ we find that
$$
[\partial^{k+1}_\Psi D_{\ss A^\prime}^{\ss A}f](\psi^{k+3},\psibar^{k-
1})
=\psi^{\ss A}\psibar_{\ss A^\prime}[\partial^k_\Psi
f](\psi^{k+2},\psibar^{k-2})\eqtag5003
$$
and
$$
\eqalign{
[\partial^{k}_\Psi D_{\ss A^\prime}^{\ss A}f](\psi^{k+2},\psibar^{k-2})
=&\psi^{\ss A}\psibar_{\ss A^\prime}[\partial^{k-1}_\Psi
f](\psi^{k+1},\psibar^{k-3})
+\left(D_{\ss A^\prime}^{\ss A}[\partial^k_\Psi
f]\right)(\psi^{k+2},\psibar^{k-2})\cr
&+[\Gamma^2\cdot\partial^k_\Psi f]_{\ss A^\prime}^{\ss
A}(\psi^{k+2},\psibar^{k-2}) + [\partial^{k-1}_\Gamma f]_{\ss
A^\prime}^{\ss A}(\psi^{k+2},\psibar^{k-2}).}\eqtag5004
$$

The analysis of \eq{44} now proceeds along lines very similar to those
presented in \S3.  Accordingly, we shall not provide all the details of the
many calculations involved in the lengthy analysis, but rather simply list
the various steps and the conclusions obtained in each.

\vskip0.2truein
\noindent{\bf 5A. The $\Gamma^{l+2}$ Analysis, $l\geq k-1$, $k\geq2$.}

When we differentiate \eq{44} with respect to $\Gamma^{l+2}$, we find
that
$$
\eqalign{
<\omega,\omega>[\partial^{l}_\Gamma
h](Y,\omega^{l+1};\alpha,\alphabar,\beta,\betabar)
+&\omega(\beta,\betabar)[\partial^{l}_\Gamma
h](Y,\omega^{l+1};\alpha,\omega,\alphabar)\cr
+&\omega(\alpha,\alphabar)[\partial^{l}_\Gamma
h](Y,\omega^{l+1};\beta,\omega,\betabar)=0.}\eqtag5005
$$
In this equation, set $\omega^{\ss A}_{\ss A^\prime}=\psi^{\ss
A}\psibar_{\ss A^\prime}$ to conclude that
$$
[\partial^{l}_\Gamma h](Y,\omega^{l+1};\alpha,\omega,\alphabar)=0
$$
whenever $\omega$ is a null vector.  By Proposition
\statementlabel{2}{2044}{} this implies there is a spinor
$$
P=P(Y,\omega^{l},\alpha,\alphabar)
$$
such that
$$
[\partial^{l}_\Gamma h](Y,\omega^{l+1};\alpha,\omega,\alphabar)
=-{1\over2}<\omega,\omega>P(Y,\omega^{l},\alpha,\alphabar).
$$
This fact allows us to use \eq{5005} to show that the highest $\Gamma$
derivative of $h$ has the algebraic form
$$
[\partial^{l}_\Gamma h](Y,\omega^{l+1};\alpha,\alphabar,\beta,\betabar)
={1\over2}\omega(\alpha,\alphabar)P(Y,\omega^{l},\beta,\betabar) +
{1\over2}\omega(\beta,\betabar)P(Y,\omega^{l},\alpha,\alphabar)
\eqtag504
$$
Note that the commutativity of the partial derivatives
$\partial^{l}_\Gamma\partial^{l}_\Gamma$ implies, using equation
\eq{504} with $\beta=\alpha$ and $\betabar=\alphabar$, that
$$
\omega(\alpha,\alphabar)[\partial^{l}_\Gamma
P](Z,\eta^{l+1};Y,\omega^{l},\alpha,\alphabar)
=\eta(\alpha,\alphabar)[\partial^{l}_\Gamma
P](Y,\omega^{l+1};Z,\eta^{l},\alpha,\alphabar).\eqtag506
$$

\vskip0.2truein
\noindent{\bf 5B. The $\Gamma^{l+1}\Gamma^{l+1}$ Analysis, $l\geq k-
1$, $k\geq2$.}

The repeated derivative of \eq{44} with respect to $\Gamma^{l+1}$
becomes, with $\beta=\alpha$ and $\betabar=\alphabar$,
$$
\eqalign{
<\omega,\eta>&[\partial^{l}_\Gamma\partial^{l}_\Gamma h]
(Y,\omega^{l+1};Z,\eta^{l+1};\alpha,\alpha,\alphabar,\alphabar)\cr\eqskip
+\eta(\alpha,\alphabar)&[\partial^{l}_\Gamma\partial^{l}_\Gamma
h](Y,\omega^{l+1};Z,\eta^{l+1};\alpha,\omega,\alphabar)\cr\eqskip
+\omega(\alpha,\alphabar)&[\partial^{l}_\Gamma\partial^{l}_\Gamma
h](Y,\omega^{l+1};Z,\eta^{l+1};\alpha,\eta,\alphabar)=0.}\eqtag505
$$
We now substitute into \eq{505} from \eq{504}, multiply by
$\eta(\alpha,\alphabar)$, and use \eq{506} to deduce that
$$
\eqalign{
[<\omega,\omega>\eta^2(\alpha,\alphabar) +
&<\eta,\eta>\omega^2(\alpha,\alphabar)-
2<\omega,\eta>\omega(\alpha,\alphabar)\eta(\alpha,\alphabar)]\cr
\times&[\partial^{l}_\Gamma
P](Z,\eta^{l+1};Y,\omega^{l},\alpha,\alphabar)=0.}
$$
Because the first spinor in brackets is not identically zero, we find that
$$
[\partial^{l}_\Gamma
P](Z,\eta^{l+1};Y,\omega^{l},\alpha,\alphabar)=0,\eqtag507
$$
and thus $\hab$ is at most linear in the variables $\Gamma^{l}$.

\vskip0.2truein
\noindent{\bf 5C. The $\Psi^{k+2}\Gamma^{l}$ and
$\Psibar^{k+2}\Gamma^{l}$ Analysis, $l\geq k-1$, $k\geq2$.}

The commutation rules \eq{5001}--\eq{5004} do not allow us to
immediately differentiate with respect to $\Psi^{k+2}$ and
$\Psibar^{k+2}$ to arrive at the equations \equationlabel{3}{303}{} and
\equationlabel{3}{304}{}, which were the basic starting equations for the
analysis of natural generalized symmetries.  Nevertheless, if we use the
linearity of $\hab$ in the variables $\Gamma^{l}$, we can differentiate
\eq{44} with respect to $\Psi^{k+2}$ and $\Gamma^{l}$ to find that
$$
[\partial^{l}_\Gamma\partial^k_\Psi
h](Y,\omega^{l+1};\psi^{k+2},\psibar^{k-
2};\psi,\alpha,\alphabar,\psibar)=0,
\eqtag508
$$
and
$$
[\partial^{l}_\Gamma \partial^k_\Psibar h](Y,\omega^{l+1};\psi^{k-
2},\psibar^{k+2};\psi,\alpha,\alphabar,\psibar)=0.\eqtag509
$$

\vskip0.2truein
\noindent{\bf 5D. The $\Gamma^{l+1}\Psi^{k+1}$,
$\Gamma^{l+1}\Psibar^{k+1}$ Analysis, $l\geq k-1$, $k\geq2$.}

Here we find, in a very straightforward manner, that
$$
[\partial^k_\Psi\partial^{l}_\Gamma h](\psi^{k+2},\psibar^{k-
2};Y,\omega^{l+1};\alpha,\beta,\alphabar,\betabar)=0,\eqtag511
$$
and
$$
[\partial^k_{\Psibar}\partial^{l}_\Gamma h](\psi^{k-
2},\psibar^{k+2};Y,\omega^{l+1};\alpha,\beta,\alphabar,\betabar)=0.\eqtag
5110
$$
In deriving these equations we used \eq{508} and \eq{509}.

\vskip0.2truein
\noindent{\bf 5E. The $\Gamma^{l+1}\Gamma^{l}$ Analysis, $l\geq k-1$,
$k\geq3$ and $l=2$, $k=2$.}

We differentiate \eq{44} with respect to $\Gamma^{l}$ and
$\Gamma^{l+1}$.  In the resulting equation we set $\beta=\alpha$,
$\betabar=\alphabar$ and substitute from \eq{504} to obtain
$$
\eqalign{
&<\omega,\omega>\eta(\alpha,\alphabar>\{[\partial^{l-1}_\Gamma P]
(Y,\omega^{l};Z,\eta^{l},\alpha,\alphabar)
-[\partial^{l-1}_\Gamma P]
(Z,\eta^{l};Y,\omega^{l},\alpha,\alphabar)\}\cr\eqskip
&+2\omega(\alpha,\alphabar)\{<\omega,\eta>
[\partial^{l-1}_\Gamma P]
(Z,\eta^{l};Y,\omega^{l},\alpha,\alphabar)
+[\partial^{l}_\Gamma\partial^{l-1}_\Gamma h]
(Z,\eta^{l+1};Y,\omega^{l};\alpha,\omega,\alphabar)\cr\eqskip
&+[\partial^{l-1}_\Gamma\partial^{l}_\Gamma h]
(Z,\eta^{l};Y,\omega^{l+1};\alpha,\eta,\alphabar)\}=0.}
$$
We multiply this equation by $\eta(\alpha,\alphabar)$ and subtract from it
the product of $\omega(\alpha,\alphabar)$ with the result of interchanging
$(Z,\eta)$ with $(Y,\omega)$ to deduce that
$$
[\partial^{l-1}_\Gamma P](Z,\eta^{l};Y,\omega^{l},\alpha,\alphabar)
=[\partial^{l-1}_\Gamma
P](Y,\omega^{l};Z,\eta^{l},\alpha,\alphabar).\eqtag510
$$

\vskip0.2truein
\noindent{\bf 5F. A Partial Reduction in Order.}

Equations \eq{507}, \eq{511}, \eq{5110}, and \eq{510} show that there is
a vector field
$$
X^{\ss A^\prime}_{\ss A}=X^{\ss A^\prime}_{\ss A}
(x,\sigma,\Gamma^1,\dots,\Gamma^{l-1},\Psi^{k-1},\Psibar^{k-1})
$$
such that
$$
[\partial^{l-1}_\Gamma
X](Y,\omega^{l};\alpha,\alphabar)={1\over2}P(Y,\omega^{l};\alpha,\alpha
bar).
$$
Hence the generalized symmetry
$$
\widetilde\hab=\hab-(\nabla^{\ss A^\prime}_{\ss A}X^{\ss B^\prime}_{\ss
B}+\nabla^{\ss B^\prime}_{\ss B}X^{\ss A^\prime}_{\ss A})
$$
is independent of the variables $\Gamma^{l}$, and accordingly we may
now assume that the original generalized symmetry \eq{502} is of the type
$$
\hab=\hab(x,\sigma,
\Gamma^1,\Gamma^2,\Psi^2,\Psibar^2,\dots,\Gamma^{k-
1},\Psi^k,\Psibar^k).\eqtag512
$$
This partial reduction in the order of $\hab$ is important because it enables
us to repeat, almost without modification, the arguments of \S3.

\vskip0.2truein
\noindent{\bf 5G. Repetition of steps A through E and the natural
symmetry analysis, $l=k-1$, $k\geq3$.}

We now repeat steps A through E assuming $\hab$ to be of the form
\eq{512}, that is, with the $\Gamma$ derivative-dependence reduced by
one order.  We can also repeat steps A and B of \S3 to conclude that now
$$\eqalignno{
&[\partial^k_\Psi h](\psi^{k+2},\psibar^{k-
2};\alpha,\beta,\alphabar,\betabar)\cr\eqskip
&\quad=<\psi,\alpha><\psi,\beta>A(\psi^{k},\psibar^{k-
2}\alphabar\betabar)
+<\psibar,\alphabar><\psibar,\betabar>B(\psi^{k+2}\alpha\beta,\psibar^{k-
4})\aleqtag513
\eqskip
&\quad+<\psi,\alpha><\alphabar,\psibar>W(\psi^{k+1},\psibar^{k-
3},\beta,\betabar) \ + <\psi,\beta><\betabar,\psibar>
W(\psi^{k+1},\psibar^{k-3},\alpha,\alphabar),\cr\eqskip
&[\partial^k_{\Psibar} h](\psi^{k-
2},\psibar^{k+2};\alpha,\beta,\alphabar,\betabar)\cr\eqskip
&\quad=<\psibar,\alphabar><\psibar,\betabar>D(\psibar^{k},\psi^{k-
2}\alpha\beta)
+<\psi,\alpha><\psi,\beta>E(\psibar^{k+2}\alphabar\betabar,\psi^{k-
4})\aleqtag514 \eqskip
&\quad+<\psibar,\alphabar><\alpha,\psi>U(\psibar^{k+1},\psi^{k-
3},\beta,\betabar) + <\psibar,\betabar><\beta,\psi>
U(\psibar^{k+1},\psi^{k-3},\alpha,\alphabar),}
$$
and
$$
[\partial^{k-1}_\Gamma h](Y,\omega^{k};\alpha,\omega,\alphabar)
=-{1\over2}<\omega,\omega>P(Y,\omega^{k-
1},\alpha,\alphabar).\eqtag515
$$
The coefficients $A$, $B$, $W$, $D$, $E$, $U$, and $P$ are functions of
the variables $x$, $\sigma$, \dots, $\Gamma^{k-2}$, $\Psi^{k-1}$,
$\Psibar^{k-1}$.  Note that steps \S3A and \S3B are valid even when
$k=2$.

Next we repeat step C of \S3 to find that $A$, $B$, $D$, $E$ are
independent of the variables $\Psi^{k-1}$ and $\Psibar^{k-1}$.  We also
arrive at the integrability conditions \equationlabel{3}{332}{},
\equationlabel{3}{335}{} and \equationlabel{3}{345}{}.  Note that step
\S3C is valid even when $k=2$.

\vskip0.2truein
\noindent{\bf 5H. The $\Gamma^{k-1}\Psi^{k+1}$,  $\Gamma^{k-
1}\Psibar^{k+1}$, $\Gamma^{k}\Psibar^{k}$ and $\Gamma^{k}\Psi^{k}$
Analysis, $k\geq3$.}

The derivative of the linearized equation with respect to $\Gamma^{k-1}$
and $\Psi^{k+1}$ gives, after taking into account
\equationlabel{3}{303}{},
$$
\eqalign{
2\omega(\psi,\psibar)&[\partial^{k-2}_\Gamma \partial^k_\Psi h]
(Y,\omega^{k-1};\psi^{k+2},\psibar^{k-
2};\alpha,\beta,\alphabar,\betabar)\cr
+<\alpha,\psi><\alphabar,\psibar>&[\partial^{k-2}_\Gamma \partial^k_\Psi
h](Y,\omega^{k-1};\psi^{k+2},\psibar^{k-2};\beta,\omega,\betabar)\cr
+<\beta,\psi><\betabar,\psibar>&[\partial^{k-2}_\Gamma \partial^k_\Psi
h](Y,\omega^{k-1};\psi^{k+2},\psibar^{k-2};\alpha,\omega,\alphabar)\cr
+<\alpha,\psi><\alphabar,\psibar>&[\partial^{k-1}_\Psi \partial^{k-
1}_\Gamma h](\psi^{k+1},\psibar^{k-
3};Y,\omega^{k};\beta,\psi,\psibar,\betabar)\cr
+<\beta,\psi><\betabar,\psibar>&[\partial^{k-1}_\Psi \partial^{k-
1}_\Gamma h](\psi^{k+1},\psibar^{k-
3};Y,\omega^{k};\alpha,\psi,\psibar,\alphabar)=0.}\eqtag516
$$
In this equation we set $\alpha=\beta=\psi$ and then
$\alphabar=\betabar=\psibar$ to deduce, in light of \eq{513}, that
$$
[\partial^{k-2}_\Gamma B](Y,\omega^{k-1};\psi^{k+4},\psibar^{k-4})=0
\qquad{\rm and}\qquad
[\partial^{k-2}_\Gamma A](Y,\omega^{k-
1};\psi^{k},\psibar^{k})=0.\eqtag1
$$
Now we set $\beta=\alpha$ and $\betabar=\alphabar$ in \eq{516}; after
substituting from \eq{513} and \eq{515} we find that
$$
\eqalign{
{1\over2}\omega(\alpha,\psibar)[\partial^{k-1}_\Psi P]
(&\psi^{k+1},\psibar^{k-3};Y,\omega^{k-1},\psi,\alphabar)
+{1\over2}\omega(\psi,\alphabar)[\partial^{k-1}_\Psi P]
(\psi^{k+1},\psibar^{k-3};Y,\omega^{k-1},\alpha,\psibar)\cr
-&<\psi,\alpha>[\partial^{k-2}_\Gamma W]
(Y,\omega^{k-1};\psi^{k+1},\psibar^{k-
3},\psibar\cdot\omega,\alphabar)\cr
-&<\psibar,\alphabar>[\partial^{k-2}_\Gamma W]
(Y,\omega^{k-1};\psi^{k+1},\psibar^{k-3},\alpha,\psi\cdot\omega)\cr
=&2\omega(\psi,\psibar)[\partial^{k-2}_\Gamma W]
(Y,\omega^{k-1};\psi^{k+1},\psibar^{k-3},\alpha,\alphabar).}\eqtag517
$$
In this equation we have defined
$$
(\psi\cdot\omega)^{\ss A^\prime}=\omega^{\ss A^\prime}_{\ss A}\psi^{\ss
A}\qquad{\rm and}\qquad
(\psibar\cdot\omega)_{\ss A}=\omega^{\ss A^\prime}_{\ss A}\psibar_{\ss
A^\prime}.
$$
Next we differentiate the linearized equation with respect to
$\Gamma^{k}$ and $\Psi^k$, then set $\alpha=\beta$ and
$\alphabar=\betabar$, and substitute from \eq{513} and \eq{515} to find
$$
\eqalignno{
&\{\omega(\psi,\psibar)\omega(\alpha,\alphabar)
-{1\over2}<\omega,\omega><\psi,\alpha><\psibar,\alphabar>\}[\partial^{k-
1}_\Psi P](\psi^{k+1},\psibar^{k-3};Y,\omega^{k-1},\alpha,\alphabar)\cr
&+<\omega,\omega><\psi,\alpha><\psibar,\alphabar>[\partial^{k-
2}_\Gamma W](Y,\omega^{k-1};\psi^{k+1},\psibar^{k-
3},\alpha,\alphabar)\cr
&-\omega(\alpha,\alphabar)\{{1\over2}\omega(\alpha,\psibar)[\partial^{k-
1}_\Psi P](\psi^{k+1},\psibar^{k-3};Y,\omega^{k-
1},\psi,\alphabar)\aleqtag518
&+{1\over2}\omega(\psi,\alphabar)
[\partial^{k-1}_\Psi P](\psi^{k+1},\psibar^{k-3};Y,\omega^{k-
1},\alpha,\psibar)\cr
&-<\psi,\alpha>[\partial^{k-2}_\Gamma W](Y,\omega^{k-1};\psi^{k+1},
\psibar^{k-3},\psibar\cdot\omega,\alphabar)\cr
&-<\psibar,\alphabar>[\partial^{k-2}_\Gamma W](Y,\omega^{k-
1};\psi^{k+1}, \psibar^{k-3},\alpha, \psi\cdot\omega)\}
=0.}$$
The last four terms in this equation are precisely the four terms on the left-
hand side of \eq{517}.  Therefore, equations \eq{517} and \eq{518} lead
to the integrability condition
$$
{1\over2}[\partial^{k-1}_\Psi P](\psi^{k+1},\psibar^{k-3};Y,\omega^{k-
1},\alpha,\alphabar)
=[\partial^{k-2}_\Gamma W](Y,\omega^{k-1};\psi^{k+1},\psibar^{k-
3},\alpha,\alphabar).\eqtag519
$$
Similarly, an analysis of the $\Gamma^{k-1}\Psibar^{k-1}$ and
$\Gamma^{k}\Psibar^k$ conditions proves that
$$
[\partial_\Gamma^{k-2}D](Y,\omega^{k-1};\psibar^{k},\psi^{k})=0
\quad{\rm and}\quad [\partial_\Gamma^{k-2} E](Y,\omega^{k-
1};\psibar^{k+4},\psi^{k-4})=0,\eqtag2
$$
and
$$
{1\over2}[\partial^{k-1}_{\Psibar} P](\psi^{k-
3},\psibar^{k+1};Y,\omega^{k-1},\alpha,\alphabar)
= [\partial_\Gamma^{k-2} U](Y,\omega^{k-1};\psibar^{k+1},\psi^{k-
3},\alpha,\alphabar).\eqtag520
$$

\vskip0.2truein
\noindent{\bf 5I. Reduction in Order, $k\geq3$.}

The integrability conditions \equationlabel{3}{332}{},
\equationlabel{3}{335}{}, \equationlabel{3}{345}{}, \eq{519}, and
\eq{520} show that there is a vector field
$$
X_{\ss A^\prime}^{\ss A}
=X_{\ss A^\prime}^{\ss A}(x, \sigma, \dots, \Gamma^{k-2},\Psi^{k-
1},\Psibar^{k-1})
$$
such that
$$
\eqalign{
W(\psi^{k+1},\psibar^{k-3},\alpha,\alphabar)
&=[\partial^{k-1}_\Psi X](\psi^{k+1},\psibar^{k-3};\alpha,\alphabar)\cr
U(\psi^{k-3},\psibar^{k+1},\alpha,\alphabar)
&=[\partial^{k-1}_\Psibar X](\psi^{k-
3},\psibar^{k+1};\alpha,\alphabar)\cr
{1\over2}P(Y,\omega^{k-1},\alpha,\alphabar)
&=[\partial^{k-2}_\Gamma X](Y,\omega^{k-1};\alpha,\alphabar).}
$$
Just as in \S3, we set
$$
\kab=\hab - (\nabla^{\ss A}_{\ss A^\prime}X^{\ss B}_{\ss B^\prime}
+\nabla^{\ss B}_{\ss B^\prime}X^{\ss A}_{\ss A^\prime}).\eqtag521
$$
Then
$$
\kab=\kab(x,\sigma,\Gamma^1,\Gamma^2,\Psi^2,\Psibar^2,\ldots,
\Gamma^{k-2},\Psi^k,\Psibar^k),
$$
and
$$
\eqalignno{
&[\partial^k_\Psi k](\psi^{k+2},\psibar^{k-
2};\alpha,\beta,\alphabar,\betabar)\cr\eqskip
&=<\psi,\alpha><\psi,\beta>A(\psi^{k},\psibar^{k-2}\alphabar\betabar)
+<\psibar,\alphabar><\psibar,\betabar>B(\psi^{k+2}\alpha\beta,\psibar^{k-
4})\aleqtag522 \cr
&[\partial^k_\Psibar k](\psi^{k-
2},\psibar^{k+2};\alpha,\beta,\alphabar,\betabar)\cr\eqskip
&=<\psibar,\alphabar><\psibar,\betabar>D(\psibar^{k},\psi^{k-
2}\alpha\beta)
+<\psi,\alpha><\psi,\beta>E(\psibar^{k+2}\alphabar\betabar,\psi^{k-
4}.\aleqtag523 }
$$

Finally, we analyze the terms in the linearized equations involving
$\Psi^{k+1}$ and $\Psibar^{k+1}$.  To this end, it is convenient to set
$$
\eqalign{
R&(\psi^{k+2},\psibar^{k-2},\alpha,\beta,\alphabar,\betabar)\cr
&=<\psi,\alpha><\psi,\beta>A(\psi^{k},\psibar^{k-2}\alphabar\betabar)
+<\psibar,\alphabar><\psibar,\betabar>B(\psi^{k+2}\alpha\beta,\psibar^{k-
4}),}
$$
and
$$
\eqalign{
S&(\psi^{k-2},\psibar^{k+2},\alpha,\beta,\alphabar,\betabar)\cr
&=<\psibar,\alphabar><\psibar,\betabar>D(\psibar^{k},\psi^{k-
2}\alpha\beta)+<\psi,\alpha><\psi,\beta>E(\psibar^{k+2}\alphabar\betabar,\
psi^{k-4}).}
$$
Then equations \eq{521}--\eq{523} imply that
$$
k=R\cdot\Psi^k + S\cdot\Psibar^k + \widetilde k,
$$
where
$$
\widetilde k=\widetilde k(x,\sigma,\dots,\Gamma^{k-2},\Psi^{k-
1},\Psibar^{k-1}).
$$
The repeated covariant derivative of $k$ thus takes the form
$$
\eqalign{
\nabla^{\ss A}_{\ss A^\prime}\nabla^{\ss B}_{\ss B^\prime}k
=&(\nabla^{\ss A}_{\ss A^\prime}\nabla^{\ss B}_{\ss
B^\prime}R)\cdot\Psi^k
+[(\nabla^{\ss A}_{\ss A^\prime}R)\cdot\nabla^{\ss B}_{\ss
B^\prime}\Psi^k
+(\nabla^{\ss B}_{\ss B^\prime}R)\cdot\nabla^{\ss A}_{\ss
A^\prime}\Psi^k]\cr
&+R\cdot(\nabla^{\ss A}_{\ss A^\prime}\nabla^{\ss B}_{\ss
B^\prime}\Psi^k) + \nabla^{\ss A}_{\ss A^\prime}\nabla^{\ss B}_{\ss
B^\prime}\widetilde k + \{\star\},}
$$
where $\{\star\}$ denotes similar terms derived from $S\cdot\Psibar^k$.
By \eq{1} and \eq{2}, $R$ and $S$ depend upon $x$, $\sigma$, \dots,
$\Gamma^{k-3}$, $\Psi^{k-2}$, $\Psibar^{k-2}$, and hence the
derivatives $\nabla^{\ss A}_{\ss A^\prime}\nabla^{\ss B}_{\ss
B^\prime}R$ and $\nabla^{\ss A}_{\ss A^\prime}\nabla^{\ss B}_{\ss
B^\prime}S$ are independent of the variables $\Psi^{k+1}$ and
$\Psibar^{k+1}$.  Moreover, we have that
$$
R\cdot\nabla^{\ss A}_{\ss A^\prime}\nabla^{\ss B}_{\ss B^\prime}\Psi^k
=R\cdot\Psi^{k+2} + \{\star\star\},
$$
where $\{\star\star\}$ denotes terms of order $k$ in the Penrose fields.
Hence $R\cdot\nabla^{\ss A}_{\ss A^\prime}\nabla^{\ss B}_{\ss
B^\prime}\Psi^k$ does not contain $\Psi^{k+1}$ and $\Psibar^{k+1}$.
Consequently, if we differentiate the linearized equations for $\kab$ with
respect to $\Psi^{k+1}$ and set $\alpha=\beta$ and $\alphabar=\betabar$,
we obtain
$$
\eqalign{
&({\rm Grad\ }R)(\psi,\psibar;\psi^{k+2},\psibar^{k-
2};\alpha,\alpha,\alphabar,\alphabar)\cr
&+2<\alpha,\psi><\alphabar,\psibar>[({\rm Div\
}R)(\psi^{k+2},\psibar^{k-2},\alpha,\alphabar)+(\partial^{k-
1}_\Psi\widetilde k)(\psi^{k+1},\psibar^{k-
3},\alpha,\psi,\psibar,\alphabar)]=0,}\eqtag52301
$$
where the covariant derivative operators Grad and Div are given by
\equationlabel{3}{35001}{} and \equationlabel{3}{35002}{}.  With
$\alpha=\psi$ and $\alphabar=\psibar$, we deduce from this equation the
covariant constancy conditions
$$
({\rm Grad\ }A)(\psi,\psibar;\psi^{k},\psibar^{k})=0,\eqtag524
$$
and
$$
({\rm Grad\ }B)(\psi,\psibar;\psi^{k+4},\psibar^{k-4})=0.\eqtag525
$$
Just as in Proposition \statementlabel{3}{308}, equation \eq{524} implies
that $A$ is independent of all the $\Gamma$, $\Psi$, and $\Psibar$
variables, that is,
$$
A=A(x,\sigma).
$$
But now, the covariant derivative of $A$ takes the general form
$$
\nabla^{\ss C}_{\ss C^\prime}A^{\dots}_{\dots}
=D^{\ss C}_{\ss C^\prime}A^{\dots}_{\dots}
+\gamma^{\ss C\dots}_{\ss C^\prime\dots}A^{\dots}_{\dots}
=\sigma^{a{\ss C}}_{\phantom{a}\ss C^\prime}({\partial A\over\partial
x^a}
+{\partial A \over \partial\sigma_{b{\ss BB^\prime}}}
\sigma_{b{\ss BB^\prime}},_a )
+ \gamma^{\ss C\dots}_{\ss C^\prime\dots}A^{\dots}_{\dots}.
$$
Since
$$
\sigma_{b{\ss BB^\prime}},_a
=\Gamma^e_{ba}\sigma_{e{\ss BB^\prime}}
+\gamma_{{\ss B}a}^{\ss C}\sigma_{b{\ss CB^\prime}}
+\gammabar_{{\ss B^\prime}a}^{\ss C^\prime}\sigma_{b{\ss
BC^\prime}}
$$
we find that
$$
\nabla^{\ss C}_{\ss C^\prime}A^{\dots}_{\dots}
=\Gamma^e_{ba}({\partial A \over \partial\sigma_{b{\ss BB^\prime}}}
\sigma_{b{\ss BB^\prime}}\sigma^{a{\ss C}}_{\phantom{a}\ss
C^\prime}) +\{\star\},
$$
where $\{\star\}$ indicates terms involving $x$, $\sigma$, and the spin
connections $\gamma$ and $\gammabar$.  It is now a simple matter to
differentiate \eq{524} with respect to $\Gamma^i_{jk}$, keeping in mind
that $\Gamma^i_{jk}$ is independent of the spin connections, to arrive at
$$
{\partial A \over \partial\sigma_{b{\ss BB^\prime}}}=0.
$$
At this point we can continue, as in the proof of Proposition
\statementlabel{3}{308}, to deduce that $A=0$.  Similarly, $B$, $D$, and
$E$ satisfy covariant constancy conditions that imply they too vanish.

We have now shown that a generalized symmetry of order $k\geq3$ is
equivalent, up to a generalized diffeomorphism symmetry, to a generalized
symmetry of order $k-1$ depending on $x$, $\sigma$, $\Gamma^i$,
$i=1,\ldots,k-2$ and $\Psi^j$, $\Psibar^j$, $j=2,\ldots,k-1$.  A
straightforward induction argument then implies that any generalized
symmetry of order $k\geq3$ is, up to a generalized diffeomorphism
symmetry, given by a generalized symmetry of order $2$ depending on
$x$, $\sigma$, $\Gamma^1$, $\Psi^2$, and $\Psibar^2$.  If the order of
the original symmetry is $k=2$, then by repeating steps \S5A through \S5F
the symmetry is again equivalent, modulo a diffeomorphism symmetry, to
a symmetry of order $2$ depending on $x$, $\sigma$, $\Gamma^1$,
$\Psi^2$, and $\Psibar^2$.

\vskip0.2truein
\noindent{\bf 5J. Reduction to First-Order Generalized Symmetries.}

The induction argument of \S5I shows that, modulo the generalized
diffeomorphism symmetry, any generalized symmetry of order $k\geq2$ is
equivalent to a symmetry $h$ with the functional dependence
$$
h=h(x,\sigma,\Gamma^1,\Psi^2,\Psibar^2).
$$
Steps \S5A through \S5D, with $l=1$ and $k=2$, show that $h$ takes the
schematic form
$$
h=P(x,\sigma)\cdot \Gamma^1+h_0(x,\sigma,\Psi^2,\Psibar^2).
$$
Steps \S3A, \S3B, and \S3C show that
$$
h=P(x,\sigma)\cdot\Gamma^1+A(x,\sigma)\cdot\Psi^2+D(x,\sigma)
\cdot\Psibar^2+k(x,\sigma).
$$
The derivative of the linearized equations with respect to $\Psi^3$ gives an
equation similar to \eq{52301}, which we write symbolically as
$$
{\rm Grad}R +{\rm Div}R + {\cal O}(x,\sigma)=0.
$$
We can then repeat the arguments at the end of step \S5I to conclude that
$A=0$.   A similar analysis of the terms involving $\Psibar^3$ in the
linearized equations leads to $D=0$.  Thus we reduce our analysis to first-
order generalized symmetries, which were classified in \S4 (see Theorem
\statementlabel{4}{404}).  We have now proven our main result.

\proclaim Theorem{\statementtag1 }.
Let $$h_{ab}=h_{ab}(x^i,g_{ij},g_{ij,h_1},\ldots,g_{ij,h_1\cdots h_k})$$
be the components of a $k^{th}$-order generalized symmetry of the
vacuum Einstein equations $R_{ij}=0$ in four spacetime dimensions.
Then there is a constant $c$ and a generalized vector field
$$X^i=X^i(x^i,g_{ij},g_{ij,h_1},\ldots,g_{ij,h_1\cdots h_{k-1}})$$ such
that, modulo the Einstein equations,
$$
h_{ab}=c g_{ab} + \nabla_a X_b + \nabla_b X_a.
$$

\vfill\eject

\noindent{\bf Acknowledgments}
This work was supported by NSF grant DMS-9100674 (I.M.A) and a grant
from the Utah State University office of research (C.G.T).

%
\references

\refis{Olver1993}{P. Olver, {\it Applications of Lie Groups to
Differential Equations}, (Springer-Verlag, New York 1993).}

\refis{Fokas1987}{A. Fokas, \journal Stud. Appl. Math., 77, 253, 1987.}

\refis{Mikhailov1991}{A. Mikhailov, A. Shabat, and V. Sokolov in {\it
What is Integrability?}, ed. V. Zakharov (Springer-Verlag, New York
1991).}

\refis{Belinsky1979}{V. Belinsky and V. Zakharov, \journal Sov. Phys.
JETP, 50, 1, 1979.};

\refis{Hauser1981}{I. Hauser and F. Ernst, \jmp 22, 1051, 1981.}

\refis{Hauser1993a}{I. Hauser and F. Ernst, \prl 71, 316, 1993.}

\refis{Hauser1993b}{I. Hauser and F. Ernst, \jmp 34, 5252, 1993.}

\refis{Noether1918} {E. Noether, \journal Nachr. Konig. Gesell. Wissen.
Gottinger Math. Phys. Kl., , 235, 1918.}

\refis{Penrose1960}{R. Penrose, \ann 10, 171, 1960.}

\refis{Penrose1976}{R. Penrose, \grg 7, 31, 1976.}

\refis{Penrose1984}{R. Penrose and W. Rindler, {\it Spinors and Space-
Time, Vol. 1}, (Cambridge University Press, Cambridge 1984).}

\refis{Winternitz1989}{C. Boyer and P. Winternitz, \jmp 30, 1081, 1989.}

\refis{Grant1993}{J. Grant, \prd 48, 2606, 1993.}

\refis{CGT1993b}{C. G. Torre, \prd 48, R2373, 1993.}

\refis{Smolin1990}{C. Rovelli and L. Smolin, \np B331, 80, 1990.}

\refis{Gurses1993}{M. Gurses, \prl 70, 367, 1993.}
Hauser - diffeos

\refis{Saunders1989}{D. Saunders, {\it The Geometry of Jet Bundles},
(Cambridge University Press, Cambridge 1989).}

\refis{CGT1993}{C. G. Torre  and I. M. Anderson, \prl 70, 3525, 1993.}

\refis{CGT1994a}{I. M. Anderson and C. G. Torre, {\it Two Component
Spinors and Natural Coordinates for the Prolonged Einstein Equation
Manifolds}, {\sl Utah State University Technical Report}, 1994.}

\refis{Tsujishita1982}{T. Tsujishita, {\it Osaka J. Math.}, {\bf 19}, 311
(1982).}

\refis{Tsujishita1989}{T. Tsujishita, {\it Sugaku Exposition}, {\bf 2}, 1
(1989).}

\refis{Ibragimov1985}{N. Ibragimov, {\it Transformation Groups
Applied to Mathematical Physics}, (D. Reidel, Boston 1985).}

\refis{Elliot1979}{J. Elliot and P. Dawber, {\it Symmetry in Physics},
(Oxford University Press, New York, 1979).}

\refis{Gourdin1969}{M. Gourdin, {\it Lagrangian Formalism and
Symmetry Laws}, (Gordon and Breach, New York, 1969).}

\refis{Thomas1934}{T. Y. Thomas, {\it Differential Invariants of
Generalized Spaces}, (Cambridge University Press, Cambridge 1934).}

\refis{Capovilla1994}{R. Capovilla, \prd 49, 879, 1994.}

 \refis{Husain1994}{V. Husain, preprint (1994).}

\refis{Bluman1989}{G. Bluman and S. Kumei, {\it Symmetries of
Differential Equations}, (Springer-Verlag, New York 1989).}

\endreferences
\bye